\documentclass[11pt]{article}

\usepackage{multirow}
\usepackage{amssymb}
\usepackage{stmaryrd}
\usepackage{amsmath}
\usepackage{amsfonts}
\usepackage{mathrsfs}
\usepackage{geometry}
\usepackage{amsmath,amssymb,amsfonts}
\usepackage[square, comma, sort&compress, numbers]{natbib}
\usepackage{graphicx}
\usepackage[figuresleft]{rotating}
\usepackage{float}
\usepackage{slashbox}

\allowdisplaybreaks\allowdisplaybreaks[1]
\def\be{\begin{equation}}
\def\ee{\end{equation}}
\def\ba{\begin{eqnarray}}
\def\ea{\end{eqnarray}}
\def\nn{\nonumber}

\title{ Manifestly Gauge Invariant Perturbations of Scalar-Tensor Theories of Gravity}
\author{{ Yu Han}$^1$ $^2$\thanks{qgyuhan@gmail.com}, \quad {Kristina Giesel}$^2$\thanks{kristina.giesel@gravity.fau.de},\quad {Yongge Ma}$^1$\thanks{mayg@bnu.edu.cn}
\\
{$^1$ Department of Physics, Beijing Normal University,}\\
{Beijing 100875, China}\\
{$^2$ Institute for Quantum Gravity (IQG), FAU Erlangen-N\"urnberg,}\\
{Staudtstr. 7, 91058 Erlangen, Germany}}
\date{}
\begin{document}
\maketitle
\begin {abstract}
The general relativistic perturbations of scalar-tensor theories (STT) of gravity are studied in a manifestly gauge invariant Hamiltonian formalism. After the derivation of the Hamiltonian equations of motion in this framework, the gauge invariant formalism is used to compute
the evolution equations of linear perturbations around a general relativistic spacetime
 background in the Jordan frame. These equations are then specialized  to the case of a flat FRW cosmological background. Furthermore, the equivalence between the Jordan frame and the Einstein frame of STT in the manifestly gauge invariant Hamiltonian formalism is analyzed, and it is shown that also in this framework they can be related by a conformal transformation. Finally, the obtained evolution equations for the linear perturbations in our formalism are compared with those in the standard cosmological perturbation theory. It turns out that the perturbation equations in the two different formalisms coincide with each other in a suitable limit.
\end {abstract}
\newpage
\tableofcontents
\newpage
\section{Introduction}

Perturbation theories of gravity have been very broadly used in both theoretical and experimental explorations during the past few decades. In the theoretical research, linear order perturbation theories have been frequently applied in analyzing the stability issues of different kinds of cosmological solutions and black holes \cite{Wald73,Eath76}, and perturbation theories on FRW cosmological background play an important role in explaining the small inhomogeneities of the cosmological microwave background which we have observed today \cite{Mukhanov92,Ma95,Noh04}. Combined with the inflation hypothesis, cosmological perturbation theories enable us to gain a possible deep insight into the mysterious early universe \cite{Lyth99}.

So far most of the perturbation analysis were done in the context of general relativity (GR). From the viewpoints of the solar system experiments, GR is currently the best candidate theory of gravity. Nevertheless, for a variety of reasons \cite{Wesson80}, alternative theories of gravity never cease to exist even from the day GR was proposed and attracted more and more attention in the last twenty years ever since the anomalous galaxy rotation curves and cosmological acceleration were observed \cite{Zwicky37,Riess98}. There are usually two ways to explain these two phenomenons. One is to introduce some unknown ``dark matter" and ``dark energy" components which interact with other fields only through gravity \cite{Navarro96,Copeland06}. The other is to modify the gravity theory. While to add a cosmological constant into GR could explain the cosmological acceleration, the extremely tiny value of this constant suffers from the ``fine tuning problem" \cite{Zlatev99}. To avoid this problem and also due to some other considerations, many people turned to look for modified gravity theories \cite{Nojiri07}. Among all the reasonable attempts, scalar-tensor theories (STT) of gravity are the ones that receive most attention. On one hand, STT provide the great possibility to solve both the anomalous rotation curve and cosmological acceleration problems \cite{Catena04,Elizalde04}. On the other hand, STT can include a lot of modified gravity models as its special sectors, such as $f(R)$ theory, Brans-Dicke theory, induced gravity, etc \cite{Hwang97,Fujii03}.  What is worth mentioning is that, when studying the cosmological perturbations during the slow-roll inflation period of STT, people noticed that the physics there behaves much different from that of GR \cite{Hwang96,Weenink10}. Some predictions  made by $f(R)=R+\alpha R^2$ gravity and certain STT theories can predict the scalar spectral index and the tensor-to-scalar ratio \emph{completely} consistent with the Planck data \cite{Ade13}. A question that cannot be answered in the context of STT is how matter couples to gravity, but is usually treated as an additional freedom of the model under consideration. The two most prominent choices for STT are the Einstein and the Jordan frame, which are related by a conformal transformation, and which will be also chosen in this paper. A systematic approach how alternative actions for gravitational theories can be derived once the information how matter couples to gravity is known, can for instance be found in \cite{Giesel2012}.
\\
\\
GR can be understood as a gauge theory with the gauge group being the diffeomorphisms $Diff(M)$ of the spacetime manifold. Likewise to other gauge theories physically relevant quantities, called observables, are those that are gauge invariant, and in the context of GR this corresponds to diffeomorphism invariant quantities. In the canonical formalism, that is used in this paper, this carries over to the requirement that observables, also denoted as Dirac observables in this context, are tensors on phase space that do (weakly) Poisson commute with all constraints of the system, and particularly for GR in the ADM formalism this means, with the Hamiltonian and spatial diffeomorphism constraints. The elementary variables in the ADM formalism are the ADM 3-metric, given by the pull back of the spacetime metric onto the spatial hypersurfaces, and its conjugate momenta. The canonical Hamiltonian density consists of the sum of the spatial diffeomorphism and Hamiltonian constraints, weighted by the so called shift vector and lapse function respectively, which parametrize possible foliations into spatial hypersurfaces of the four dimensional spacetime manifold. These configuration and momentum variables are no observables, but their Poisson brackets with the canonical Hamiltonian, and hence with the constraints, together with the constraints themselves yield the analogue of Einstein's equations in the canonical framework. As a consequence, when we discuss observables in the context of GR, the natural question arises, how the evolution of those observables can be described. Certainly, it cannot be generated by the canonical Hamiltonian, since this would allow only trivial evolution contradicting what we observe, because we do observe the evolution of physical observables in our everyday life and in experiments. Thus, what we are looking for is a so called physical Hamiltonian having the property that it generates the evolution of the observables with respect to physical time in the canonical setup of GR. The initially missing physical time and along with this the missing physical Hamiltonian is a common feature of any diffeomorphism-invariant theory and is often called the "problem of time" for short. Many attempts have been made to clarify this conceptual issue in the history, see \cite{Ashtekar13} for some examples. One possible way to circumvent the problem of time was introduced by Rovelli \cite{Rovelli91} and mathematically further developed by Dittrich \cite{Dittrich04} in the context of so called relational observables. The main idea of relational observables is to introduce so called reference fields with respect to which the observables and the evolution of the remaining degrees of freedom are formulated. These reference fields could for instance be additional matter fields but also purely gravitational degrees of freedom. In a seminal paper by Brown and Kucha{\v r} \cite{Brown95} they considered to use pressureless dust particles as such reference fields, which were introduced as additional matter fields. The dust particles can be understood as a free falling observer that is dynamically coupled to the system and with respect to which observables for GR are constructed. The dust fields therefore serve as a physical reference system and provide an interpretation of physical spatial and time coordinates. In \cite{Brown95} observables were constructed explicitly only with respect to the spatial diffeomorphism constraint and for the Hamiltonian constraint a corresponding Schr\"odinger-like equation was discussed within the quantum theory. A combination of the methods of Brown and Kucha{\v r}  and the relational observable framework was used in \cite{Giesel10} to derive a manifestly gauge invariant Hamiltonian formulation of GR. The corresponding physical Hamiltonian associated with the dust observer then generates evolution with respect to this particular chosen physical time. As an application in \cite{Giesel10} the manifestly gauge invariant Hamiltonian formalism was used in order to develop a formalism for general relativistic perturbation theory. The obtained results were then in \cite{Giesel10s} applied to the context of cosmological perturbation theory for which a comparison with the results of the standard formalism, in the following denoted as standard cosmological perturbation theory (SCPT), was  analyzed for linear order perturbation theory. It turns out that they are consistent with each other only up to small corrections. The latter occur due to the reason that in the manifestly gauge invariant formalism the observer is dynamically coupled to the system and thus has an influence on the system, whereas this is not the case in SCPT.
\\
The main difference between the manifestly gauge invariant and the standard formalism lies in the way how gauge invariant quantities and thus observables are constructed. In the manifestly gauge invariant formalism first one constructs observables and the associated physical Hamiltonian, which itself is an observable. Then one derives the equations of motion for all observables and afterwards considers the perturbation of these evolution equations. By construction any quantity and their corresponding perturbations, which are involved in these evolution equations are themselves observables and hence gauge invariant. Following SCPT the strategy is different since perturbations of gauge variant quantities, the metric and and their equations of motion, are considered.  Gauge invariant objects, and hence observables, are constructed afterwards and are required to be only gauge invariant up to the corresponding order of perturbation theory one is interested in. The manifestly gauge invariant formalism might be of advantage when higher order perturbation theory is considered because in SCPT one has to start from scratch again when gauge invariant quantities are constructed \cite{Nakamura05}, while using the manifestly gauge invariant framework any object and thus also higher order perturbations are already gauge invariant.
\\
\\
In this paper we extend this manifestly gauge invariant formalism to the case of STT.
This paper is organized as follows. In section two, after introducing the Brown-Kucha{\v r} formalism, relational observables and the notion of a physical Hamiltonian, we construct the physical Hamiltonian in Jordan frame of STT and derive the second order evolution equations of the canonical variables. In section three the latter will be used to derive the evolution equations for linear perturbations in the context of a general relativistic spacetime.  In section four we apply these equations to flat FRW background and get the cosmological perturbation equations. In the first part of section five we formulate the gauge invariant Hamiltonian formalism in Einstein frame of STT and prove it is conformally equivalent to that of Jordan frame. In the second part we extend our results and study the cosmological perturbations in a different reference system. Our results are then compared with the standard cosmological perturbation theory of STT. In the last section we summarize the above results and draw some conclusions. In the appendix the Lagrangian and Hamiltonian analysis of the action of STT in Einstein frame is presented.

\section{Manifestly Gauge Invariant Hamiltonian Formalism of STT}
For the convenience of readers, in the first part of this section, we will first give a brief review of the idea proposed by Brown and Kucha{\v r}. Then we introduce how to combine the Brown-Kucha{\v r}'s dust formalism with the idea of relational observables to build the physical reference system and the gauge invariant Hamiltonian which generates the evolution of every observable in this physical reference system.  We refer to \cite{Giesel10,Giesel07} for the details. In the second part we will apply these ideas to STT of gravity. First we derive the physical Hamiltonian in the Jordan frame of STT and then use it to derive the evolution equations of the 3-metric and gravitational scalar field.

\subsection{The Brown -- Kucha{\v r} Lagrangian and Relational Observables}
\label{BKFormalism}
In this subsection we will briefly review the idea of Brown and Kuch{\v ar} to use dust as reference system for general relativity and discuss how this idea can be embedded in the framework of relational observables. The generalization from general relativity to STT is not difficult and will be discussed in section \ref{EvEqnJordan}.

\subsubsection{Brown -- Kucha{\v r} Lagrangian and Deparametrization of General Relativity}
In \cite{Brown95} Brown and Kucha{\v r} followed the idea that matter fields can be chosen as a physical reference systems. For the reason that they were mainly interested in GR they considered the Einstein-Hilbert action and the following additional matter action
\be
\label{SBKaction}
S_{dust}=-\frac{1}{2}\int\limits_{M}d^4x\sqrt{|\det(g)|}\rho\left(g^{\mu\nu}U_{\mu}U_{\nu}+1\right)
\ee
which, as we will see below, can be interpreted as an action for pressureless dust.
The action $S_{dust}$ is not taken as a functional of the one form $U$ but the latter is expressed in terms of the scalar fields $T,S^j,W_j$ defined through $U=-dT+W_jdS^j$ where we use the notation that Latin letters run from $1$ to $3$ and Greek letters from $0$ to $4$. Hence, the action above is a functional of the fields $\rho, T, S^j, W_j$ and $g_{\mu\nu}$ and hence in addition to GR we have introduced eight more degrees of freedom. As we will see below the system GR+dust has second class constraints and once these are reduced the additional number of degrees of freedom is reduced to four. As discussed in detail in \cite{Brown95} the Euler-Lagrange equations associated with this action show that the vector field $U^\mu=g^{\mu\nu}U_\nu$ satisfies the differential equation of a geodesic in affine parametrization. The fields $W_j$ and $S^j$ are constant along the geodesics and the field $T$ defines proper time along each geodesic. Furthermore, the energy-momentum tensor associated with $S_{dust}$ has the form of the energy-momentum tensor of a pressureless perfect fluid. In order to use the dust matter as a dynamically coupled observer in the canonical framework we have to discuss the Hamiltonian formulation of (\ref{SBKaction}), which is discussed in detail in \cite{Giesel10}. We assume that $M$ is globally hyperbolic and thus we can perform a 3+1-split of $M\simeq \mathbb{R}\times \chi$ into time and space, with $\chi$ being a spatial manifold of arbitrary topology. For this purpose we introduce a family of embeddings $X_t : \chi\to M$ $x \mapsto X_t(x):=X(t,x)$, also called a foliation of $M$, where $\chi_t:=X_t(\chi)$ are called the leaves of the foliation, where we denoted the coordinates on $\chi$ by $x^a$ with $a=1,2,3$. Given the family of embeddings $X_t$ we can construct a family of tangent vectors $X^\mu_{t,a}$ and a co-normal $n_{\mu}$ for each leave. Using the metric we can also work with the future orientated normal $n^\mu$ and can use it to decompose the variation of the embeddings with respect to the parameter $t$ into a tangential and normal part to the leaves $\chi_t$ given by $\partial_t X^\mu_t=nn^\mu+n^a X^{\mu}_{t,a}$ where this decomposition is parametrized by the so called lapse function $n$ and shift vector $n^a$. The spatial three metric $q_{ab}$ intrinsic to $\chi$ can be constructed by pulling back $g_{\mu\nu}$ using  the tangent vectors  yielding $q_{ab}=g_{\mu\nu}X^\mu_{,a}X^{\nu}_{,b}$. In order to derive the Hamiltonian formulation via the Legendre transformation we introduce conjugate momenta $p^{ab},p,p_a,P,P_j,I,I^j$ associated with the configuration variables $q_{ab},n,n^a,T,S^j,\rho,W_j$. Note that in case we would like to couple additional standard model matter to gravity we would need to introduce additional phase space variables for this matter, which we will not display explicitly here, but we will mention below how such additional matter degrees of freedom enter the model. The theory possesses the following primary constraints
\be
z:=p=0,\quad z_a:=p_a=0,\quad Z:=I=0,\quad Z^j:=I^j=0,\quad Z_j:=P_j-PW_j=0.
\ee
Following the Dirac procedure for constrained systems the stability analysis of the primary constraints with respect to the primary Hamiltonian yields the following secondary constraints
\ba
c^{tot} &=& c+c^{dust},\;c^{dust}=\frac{1}{2}\left[\frac{P^2}{\rho\sqrt{\det(q)}}
+\rho\sqrt{\det(q)}(1+q^{ab} U_a U_b)\right],
\nonumber\\
c^{tot}_a &=& c_a+c_a^{dust},\;c_a^{dust}=P[T_{,a}-W_j S^j_{,a}],
\nonumber\\
\tilde{c} &=& \frac{n}{2}\left[-\frac{P^2}{\rho^2\sqrt{\det(q)}}
+\sqrt{\det(q)}(1+q^{ab} U_a U_b)\right],
\ea
here $c$ denotes the gravitational contribution as well as the contribution from other possible standard model matter to the usual Hamiltonian constraint and $c_a$ denotes the contributions from gravity as well as other possible standard model matter to the usual spatial diffeomorphism constraint. Furthermore we have $U_a=-T_{,a}+W_jS^j_{,a}$. A further application of Dirac's constraint algorithm shows that no tertiary constraints occur and thus the set of constraints has completely been determined and is given by $\{c^{tot},c_a^{tot},\tilde{c},z,z_a,Z,Z^j,Z_j\}$. The next step is to classify them into first and second class constraints. The constraints $\{\tilde{c},Z,Z^j,Z_j\}$ form second class constraints. We will solve them explicitly below using the associated Dirac bracket. Furthermore we will also solve the constraints $z,z_a$ by considering the lapse function $n$ and the shift vector $n^a$ as Lagrange multipliers. Now we extend $c^{tot}$ and $c_a^{tot}$ to ${\cal C}^{tot}$ and ${\cal C}_a^{tot}$ by adding terms proportional to the constraints $z,z_a,Z,Z^j,Z_j$ in the case of $c_a^{tot}$ and terms proportional to $Z,Z_j,Z^j$ in the case of $c^{tot}$. These terms are exactly chosen such that ${\cal C}^{tot}$ and ${\cal C}_a^{tot}$ are first class constraints, see \cite{Giesel10} for more details. Since $z_a,z$ are first class constraints as well we end up with the following set of first class constraints $\{{\cal C}^{tot},{\cal C}_a^{tot},z,z_a\}$. Now solving the second class constraints strongly leads to
\be
\label{Sol2nd}
W_j:=\frac{P_j}{P},\quad \rho^2:=\frac{P^2}{\sqrt{q}}\left(q^{ab}U_aU_b+1\right),\quad I:=0,\quad I^j:=0.
\ee
For the reason that $Z,Z^j$ and $Z_j$ do only depend on the dust variables the associated Dirac bracket reduces to the Poisson bracket when applied to the geometrical degrees of freedom $q_{ab}, p^{ab}$ and other possible standard model degrees of freedom. When we introduce the Dirac bracket and also solve the constraints $z,z_a$ by considering the lapse function and shift vector as Lagrange multipliers, as usually done the ADM framework, we work on a reduced phase space where ${\cal C}^{tot}=c^{tot}$ and ${\cal C}_a^{tot}=c_a^{tot}$. Inserting the explicit solutions of the second class constraints from (\ref{Sol2nd}) we end up with the following first class constraints
\ba
c^{tot}&=&c+c^{dust},\quad\quad c^{dust}=-\sqrt{P^2+q^{ab}c_a^{dust}c_b^{dust}}, \\
c_a^{tot}&=&c_a+c_a^{dust},\quad\quad c_a^{dust}=PT_{,a}+P_jS^j_{,a}.
\ea
Note that in principle we have two possible choices for the sign of $\rho$ here but the one chosen by us $\rho<0$ yielding also to $P<0$ is the part of the phase space that involves also flat space solutions since for them $c>0$ is necessary, for a more detailed discussion about this aspect see \cite{Giesel10}.
\\
\\
The Hamiltonian and spatial diffeomorphism constraints satisfy a complicated Poisson algebra also called the hypersurface deformation algebra for the reason that it can also be derived from purely geometrical considerations in the context of deformation of hypersurfaces \cite{Hojman1976}. One of the motivations in \cite{Brown95} to introduce the dust as a reference systems was that using the dust one can write down an equivalent set of first class constraints that has the property that the corresponding constraint algebra becomes Abelian and the final Hamiltonian constraint can be written in deparametrized form, as we will discuss below. The important observation by Brown and Kucha{\v r}, also denoted as the Brown-Kucha{\v r}-mechanism in \cite{Giesel10,Giesel10s}, was that on the constraint surface of the spatial diffeomorphism constraint $c^{tot}_a=0$, we have $c_a=-c_a^{dust}$ and thus we can rewrite the dust contribution $c^{dust}$ as
\be
c^{dust}=-\sqrt{P^2+q^{ab}c_ac_b}.
\ee
As a consequence the only dependence in $c^{tot}$ on the dust variables is via the dust momentum $P$. Thus on the constraint surface of the Hamiltonian constraint we can solve $c^{tot}$ for $P$ and obtain an equivalent form of the Hamiltonian constraint given by
\be
\label{equivHam}
\tilde{c}^{tot}=P+h,\quad\quad h:=\sqrt{c^2-q^{ab}c_ac_b}.
\ee
The fact that $h$ no longer depends on the dust variables is what is called deparametrization and a consequence of this is that we will end up with a time independent physical Hamiltonian as discussed in the next subsection. In order to obtain an Abelian constraint algebra we also solve the spatial diffeomorphism constraint for the momenta $P_j$. For this purpose we have to assume that the matrix $S^j_{,a}$ is everywhere non-degenerate, an assumption similar to the classical restriction $\det(q)>0$, meaning that the inverse matrix exist, which we denote by $S^a_j$. Then on the constraint surface we have $P_j+S^a_j(c_a+PT_{,a})=P_j+S^a_j(c_a-hT_{,a})$ thus we can write down the following equivalent form of the spatial diffeomorphism constraint
\be
\label{equivDiffeo}
\tilde{c}_j^{tot}=P_j+h_j,\quad\quad h_j:=S^a_j(c_a-hT_{,a}).
\ee
We realize in contrast to the Hamiltonian constraint $\tilde{c}^{tot}$ the spatial diffeomorphism constraint $\tilde{c}^{tot}_j$ does not deparametrize. However this is no problem at all because the construction of observables is not restricted to the deparametrized case and can therefore be equally well  applied to the spatial diffeomorphism constraint. It might only be  technically a little bit more involved. It is only at the level of the physical Hamiltonian, for which however only  the form of $\tilde{c}^{tot}$ turns out to be important, where deparametrization yields to simplifications in the sense that the final physical Hamiltonian will be time independent.
\\
Now considering the constraints $\tilde{c}^{tot}$ and $\tilde{c}^{tot}_j$ one can indeed show that they satisfy an Abelian constraint algebra \cite{Brown95}. This follows also immediately from the following abstract argument \cite{Henneaux1992}: The equivalent constraints $\tilde{c}^{tot}$ and $\tilde{c}^{tot}_j$ are still first class. Therefore their Poisson brackets are again linear combinations of constraints. However since all constraints of the system are linear in the momenta $P,P_j$ their Poisson brackets are independent of $P,P_j$. Consequently, we can evaluate the linear combination of constraints that appear in the Poisson bracket computation in particular at $P=-h$ and $P_j=-h_j$. From the Abelian constraint algebra and the explicit form of $h$ and $h_j$ in (\ref{equivHam}) and (\ref{equivDiffeo}) respectively we can conclude that $h(x)$ are mutually Poisson commuting while $h(x)$ does not Poisson commute with $h_j(x)$ nor do the $h_j(x)$ mutually commute. In the next subsection we will introduce the relational framework and use the latter to construct observables in the context of GR.

\subsection{Relational Framework for Constructing Observables}
The main idea of the relational framework is to introduce so called reference fields, often also denoted as clocks, that will then be used to construct observables with respect to the constraints of the system under consideration. Let us assume we have a system with a set of constraints $\{C_I\}$ labeled by an index $I$, which is up to now arbitrary. The aim is to introduce for each constraint $C_I$ a corresponding reference field $T^I$ such that the constraint and the reference field build, at least weakly, a conjugate pair, that is $\{C_I,T^J\}\approx \delta^J_J$ where $\approx$ means equality up to terms that vanish on the constraint surface. Now since for a given set of constraints finding those reference fields might not be a simple task, one uses the freedom that one can always modify the set of constraints as long as the modified set defines the same constraint surface. Suppose we choose a set of reference field $\{T^I\}$, one for each $C_I$ with the property that $\{C_I,T^J\}=:M^{I}_J$ with $M$ being an invertible matrix, then we can define the equivalent set of constraints $\{C'_I\}$ defined through
\be
 C'_I:=\sum\limits_{J}(M^{-1})_I^JC_J .
\ee
One can easily show that for $\{C'_I\}$ we have  $\{C'_I,T^J\}\approx \delta^J_J$. Given these new set of constraints $\{C'_I\}$ we can use the reference fields $\{T^I\}$ to construct observables for a general phase space functions. This will be a particular combination of the original phase space function under considerations and the reference fields. To discuss this construction more in detail we consider the Hamiltonian vector field associated with $C'_I$, which is denoted by $X_I$. As can be shown and will be crucial in the following constructions the $X_I$  mutually weakly commute. Let us introduce a set of up to now arbitrary real numbers $\{\beta_I\}$, again one for each constraint $C'_I$, and consider the following sum of Hamiltonian vector fields
\be
X_\beta:=\sum\limits_{I}\beta^IX_I~.
\ee
Now we consider a function $f$ on phase space and define a map $f\to \alpha_\beta(f)$  on the set of smooth functions on phase space given by
\be
\alpha_\beta(f):=\exp(X_\beta)\cdot f=\sum\limits_{n=1}^\infty \frac{1}{n!}X^n_\beta\cdot f~,
\ee
here $X^n_\beta\cdot f=\{C_\beta,f\}_{(n)}$ where $\{.,.\}_{(n)}$ denotes the iterative Poisson bracket defined through $\{C_\beta,f\}_{(0)}=f$ and $\{C_\beta,f\}_{(n)}=\{C_\beta,\{C_\beta,f\}_{(n-1)}\}$. $\alpha_\beta$ is a Poisson automorphism on the algebra of functions on phase space associated with the Hamiltonian vector field $X_\beta$ of $C_\beta=\beta^IC'_I$.
We will use the map $\alpha_\beta$ as well as the set of reference fields to construct an observable associated with  a phase space function $f$.  A weak Dirac observable has to weakly Poisson commute with all constraints $\{C_I\}$. Now the idea of the relational observables is that although the phase space function $f$ as well as the reference fields $T^I$ have non-vanishing Poisson brackets with the constraints a particular combination of the two involving the map $\alpha_\beta$ has vanishing Poisson brackets with all constraints. We want to construct a map that returns the value of $f$ at those values where the reference fields $T^I$ take the values $\tau^I$. In order to do so let us choose another set of real numbers $\{\tau^I\}$. We are interested in those values of the gauge parameters $\beta^I$ for which $\alpha_\beta(T^I)=\tau_I$. If we apply $\alpha_\beta$ onto the reference fields we obtain $\alpha_\beta(T^I)\approx T^I+\beta^I$, which can easily be solved for $\beta_I$ yielding $\beta_I=\tau^I-T^I$. We will denote this equation for short as $\beta=\tau-T$ suppressing the indices. Using this we can construct the following map for the phase space function $f$
\be
O_f(\tau):=\left[\alpha_\beta(f)\right]_{\beta=\tau-T}~.
\ee
The notation with the square brackets means that only after one has computed the action of $X_\beta$ with $\beta$ treated as a constant on phase space then one sets $\beta=\tau-T$ which becomes then phase space dependent. As has been proven in \cite{Dittrich04,Thiemann2004} $O_f(\tau)$ is indeed a weak Dirac observable, that is for all $I$ we have
\be
\{O_f(\tau),C_I\}\approx 0.
\ee
We realize that we can also understand the map $O_f$ as a map that returns the value of $f$ in the gauge $\beta=\tau-T$.
As also shown in  \cite{Henneaux1992,Thiemann2004} the  multi parameter family of maps $O^{\tau}: f\to O_f(\tau)$  is a homomorphism from the commutative algebra of functions on phase space to the commutative algebra of weak Dirac observables, both with pointwise multiplication,
\be
O_f(\tau)+O_g(\tau)=O_{f+g}(\tau),\quad\quad
O_f(\tau)O_g(\tau)\approx O_{fg}(\tau).
\ee
This will be a particularly useful property when the explicit construction of the observables is considered for the following reason: Let us denote the coordinates on phase space by $(q^A,p_A)$ where the index $A$ is chosen such that all relevant phase space degrees of freedom are considered. Now for a phase space function $f=f(q^A,p_A)$ have we have
\be
O_f(\tau)=f(O_{q_A},O_{p^A})(\tau)
\ee
which has the important consequence that it is sufficient to construct observables for the elementary phase space variables, something we will use below.
Moreover, multi parameter family of maps $O^{\tau}: f\to O_f(\tau)$ is a Poisson homomorphism with respect to the Dirac bracket $\{.,.\}^*$ associated with the system of second class constraints $C_I,T^I$ \cite{Henneaux1992,Thiemann2004}
, this means
\ba
\{O_f(\tau),O_{g}(\tau)\}\approx \{O_f(\tau),O_g(\tau)\}^*
\approx O_{\{f,g\}^*}(\tau)
\ea
where the Dirac bracket is defined as
\be
\{f,g\}^*=\{f,g\}-\{f,C_I\}(M^{-1})^I_J\{T^J,g\}+\{g,C_I\}(M^{-1})^I_J\{T^J,f\}.
\ee
In the following we want to discuss the special case of constraints that are in deparametrized form and understand how this simplifies the construction of the observables $O_f(\tau)$. In the case of deparametrization we can always find canonical coordinates that consists of two sets $(T^I,P_I)$ and $(q^a,P_a)$ such that all constraints $C_I$ of the system can be written in the following form
\be
C_I=P_I+h_I(q^a,p_a),
\ee
and thus do not depend on the configuration variables $T^I$. In practice this is a very special case and most constrained systems, if at all, can only be written in partially deparametrized form, in which only part of the constraints deparametrize. However, for the following discussion let us assume that we consider a fully deparametrized system. Now following the steps of the construction of observables from the discussion above first we observe that
\be
\{C_I,T^J\}=\delta_I^J~.
\ee
Using the notation above this means the equivalent constraints $C'_I$ are identical to $C_I$ and thus the task of inverting a in general complicated matrix $M_I^J$ is no longer necessary. Furthermore as already discussed above if all constraints are linearly in the momenta $P_I$ then the associated constraint algebra is Abelian. For the reason that here also non of the $h_I$ depends on the reference fields $T_I$ we immediately get $\{h_I,h_J\}=0$ from this we can follow $\{h_I,C_J\}=0$ showing that each $h_I$ is already a Dirac observable. Moreover from the Abelian constraint algebra it follows that also the associated Hamiltonian vector fields commute and in this case here not only on the constraint surface but on the entire phase space. As a consequence all weak equalities that we used above can be replaced by strong equalities here.
\\
\\
First let us discuss the construction of the observables for the elementary variables $(q^a,p_a)$. Since $q^a$ and $p_a$ both commute with all momenta $P_J$ we can consider the Hamiltonian vector field associated with the $h_I$'s instead of defining $X_\beta$ via $C'_I$. Moreover for the reason that also $q^a$ and $p_a$ commute with all reference fields $T^I$ we can already, when applying $X_\beta$ to $f$, replace $\beta$ by the corresponding gauge $\tau^I-T^I$ yielding the following form for the observables for a function $f$ that depends only on $(q^a,p_a)$
\be
\label{OfDep}
O_f(\tau)=\sum\limits_{n=0}^\infty\frac{1}{n!}X^n_\tau\cdot f
\ee
where $X_\tau$ is the Hamiltonian vector field of the function
\be
H_{\tau}=(\tau^I-T^I)H_I
\ee
where $H_I:=O_{h_I}(\tau)$ denotes the observables associated with $h_I$. Because $h_I=h_I(q^a,p_a)$ is a function of $q^a$ and $p_a$ only, once the observables for the elementary variables $O_{q^a}(\tau)=:Q^a(\tau)$ and $O_{p_a}(\tau)=:P_a(\tau)$ are constructed we obtain $H_I$ as $H_I=O_{h_I}(\tau)=h_I(Q^a,P_a)(\tau)$ using the homomorphism property of the observable map. In the particular case of deparametrization we have $H_I=h_I$ because $h_I$ is already a Dirac observable as discussed above. Now if we restrict to functions that do only depend on $q^a$ and $p_a$ the Dirac bracket reduces to the Poisson bracket because those $f$ commute with all reference fields $T^I$. In particular for the algebra of the observables $Q^a(\tau)$ and $P_a(\tau)$ we obtain
\be
\{P_a(\tau),Q^b(\tau)\}=\{O_{p_a}(\tau),O_{q^b}(\tau)\}=O_{\{p_a,q^b\}}(\tau)=O_{\delta_a^b}(\tau)=\delta_a^b~,
\ee
showing that the reduced phase space has a very simple symplectic structure in terms of the coordinates $Q^a,P_a$, an important property if the quantization of such systems is considered. Having finished the discussion about the non-reference field degrees of freedom let us discuss now the case of the remaining reference field degrees of freedom. The observable associated to the reference fields $T^I$ is given by
\be
O_{T^I}(\tau)=\left[\alpha_\beta(T^I)\right]_{\beta=\tau^I-T^I}=\tau^I
\ee
and therefore is just a constant function on phase space. Since all momenta $P_I$ Poisson commute with all constraints they are already Dirac observables. In addition they can also be expressed as function of the observables $Q^a(\tau)$ and $P_a(\tau)$, because on the constraint surface we have
\be
P_I=O_{P_I}(\tau)=-O_{h_I}(\tau)=-h_I(Q^a(\tau),P_a(\tau))=-H_I~.
\ee
Hence, what we finally be interested in is the reduced phase space with elementary variables $Q^a(\tau)$ and $P_a(\tau)$.
\\
\\
Let us again consider an observable $O_f(\tau)$ associated with a function that depends only on $q^a$ and $p_a$. How can we formulate the evolution of such observables? Certainly this cannot be generated by the constraints since by construction $O_f(\tau)$ Poisson commutes with all constraints. However, $O_f(\tau)$ gives us the value of $f$ when the reference fields $T^I$ take the values $\tau^I$. As it will be the case for GR and also for STT one of the chosen reference fields will be associated with physical time and let us without loss of generality denote this reference field by $T^0$ and the values that it takes by $\tau^0$. Then time evolution for $O_f(\tau)$ can be described by the derivative of $O_f(\tau)$  with respect to $\tau^0$ since this encodes how $O_f(\tau)$ changes with time $\tau^0$. However, considering the form of $O_f(\tau)$ in (\ref{OfDep}) we can explicitly compute this derivative and as shown in \cite{Thiemann2004} one obtains
\be
\frac{\partial O_f(\tau)}{\partial\tau_0}=\{H_0,O_f(\tau)\}
\ee
where $H_0$ is the observable associated with $h_0$ that occurs in the constraint $C_0:=P_0+h_0$ associated with the reference field $T^0$ that we interpret as a reference field for time. In the following we will call $H_0$ the physical Hamiltonian because in contrast to the constraint $C_0$, that is generating gauge transformations, $H_0$ does not vanish on the constraint surface and can therefore be understood as a true Hamiltonian, which generates evolution with respect to physical time $\tau^0$. Note that because $h_0$ does not depend on $T^0$ (and also not on any other reference field) the final physical Hamiltonian $H_0$ is time independent.

\subsubsection{Observables for GR Using the Brown-Kucha{\v r} Dust}
In this subsection we will discuss how the Brown-Kucha{\v r} dust can be used to construct relational observables. In the case of GR and STT we have four times infinitely many constraints because we have one Hamiltonian and three spatial diffeomorphism constraints per spacetime point. Following the discussion of the last subsection we therefore need to choose 4 times infinitely many $T^I$ making four scalar fields a natural choice for reference fields. These will become exactly the four additional degrees of freedom $(T,S^j)$ which we added to the system by considering the Brown-Kucha{\v r} Lagrangian. The remaining degrees of freedom $(q_{ab},p^{ab})$ and possible other standard model degrees of freedom will be referred to as non-dust degrees of freedom. In the following in order to keep the discussion more simple we will only consider the system of gravity and dust.
In order to define the Hamiltonian vector field $X_\beta$ in this case, we introduce arbitrary functions $\beta^0, \beta^j$ on $\chi$ and  define using the constraints in (\ref{equivHam}) and (\ref{equivDiffeo})
\be
c_\beta^{tot}:=\int\limits_{\chi}d^3x \beta^\mu(x)\tilde{c}_\mu^{tot}(x)
\ee
where $\beta^\mu=(\beta^0,\beta^k)$ and we have defined $\tilde{c}_0^{tot}=\tilde{c}^{tot}$. We denote the Hamiltonian vector field of $c_\beta^{tot}$ by
$X_\beta$ and using it we can define the map $\alpha_\beta$ given by
\be
\alpha_\beta(f):=\exp(X_\beta)\cdot f=\sum\limits_{n=0}^\infty \frac{1}{n!}X_\beta^n\cdot f~.
\ee
We use the notation $T^\mu(x)=(T(x),S^j(x))$ then applying $\alpha_\beta$ onto $T^\mu$ yields $\alpha_\beta(T^\mu(x))=T^\mu(x)+\beta^\mu(x)$. Let us denote the values that the reference fields $T^\mu(x)$ can take by $\tau^\mu(x)$ with $\tau^0(x)=\tau(x)$ and $\tau^j(x)=\sigma^j(x)$ where up to know $\tau^\mu$ are arbitrary functions on $\chi$. Now solving $\alpha_\beta(T^\mu(x))=\tau^\mu(x)$ for $\beta^\mu(x)$ leads to $\beta^{\mu}(x)=\tau^\mu(x)-T^\mu(x)$, which we again write as $\beta=\tau-T$. The observable associated to $f$ reads
\be
O_f(\tau)=\left[\alpha_\beta(f)\right]_{\beta=\tau-T}~.
\ee
Looking at the explicit form of $\tilde{c}^{tot}$ in (\ref{equivHam}) we realize that $\tilde{c}^{tot}(x)$ commutes with $S^j(y)$ and because of this we can construct the observables in two steps. First we reduce with respect to the spatial diffeomorphism constraints $\tilde{c}^{tot}_j$ and afterwards we construct the complete observables that also Poisson commute with the Hamiltonian constraint. Hence, we can rewrite $O_f(\tau)$ as
\be
\label{CompObs}
O_f(\tau,\sigma^j)=\left[\alpha_{\beta^0}(\left[\alpha_{\beta^j}(f)\right]_{\beta^j=\sigma^j-S^j})\right]_{\beta^0=\tau-T} ~.
\ee
Let us first discuss the inner part, that is how spatially diffeomorphism invariant objects are constructed. The reference fields $S^j$ will be used for this and therefore here the remaining degrees of freedom are $(q_{ab},p^{ab},T,P)$ for which observables need to be constructed. As discussed in detail in \cite{Giesel10} for the choice of a constant function $\sigma^j(x)=\sigma^j$ the observable with respect to $\tilde{c}_j^{tot}$ associated with any scalar function $f$ build from the variables $(q_{ab},p^{ab},T,P)$ can be expressed as
\be
\label{tildef}
\tilde{f}(\sigma^j)=\left[\alpha_{\beta^j}(f)\right]_{\beta^j=\sigma^j-S^j}=f(x)\Big|_{S^j(x)=\sigma^j}
\ee
where we denote the partially reduced function as $\tilde{f}$. The interpretation of the formula above is the following: Whatever the value $x$ is at which the function $f$ is evaluated $\tilde{f}(\sigma^j)$ returns $f$ evaluated at the point $x_\sigma$ at which $S^j(x)=\sigma^j$. We call the range of $S^j$ the dust space and denote it by ${\cal S}$. Since by our assumption $S^j_{,a}$ is everywhere invertible it defines a diffeomorphism $S^j:\chi\to {\cal S}$ and hence the value of $x_\sigma$ is unique. Our strategy is therefore to use $S^j$ to construct scalars $f$ on $\chi$ for $(q_{ab},p^{ab},T,P)$ and then apply the formula in (\ref{tildef}) yielding the partially reduced quantities $(\tilde{q}_{ab},\tilde{p}^{ab},\tilde{T},\tilde{P})$, explicitly we get
\be
\label{TildeNonDust}
\tilde{T}=T\quad \tilde{P}=\frac{1}{J}P\quad
\tilde{q}_{jk}=q_{ab}S^a_jS^b_k
\quad
\tilde{p}^{jk}=\frac{1}{J}p^{ab}S^j_{,a}S^k_{,b}
\ee
where $J:=\det\left(\frac{\partial S}{\partial x}\right)$ was used to obtain the correct density weight. Note while these are scalars on $\chi$ there are tensors on the dust space ${\cal S}$ with the same density weight that they have on $\chi$, see also \cite{Giesel10} for more details. We realize that the evaluation of the functions in (\ref{TildeNonDust}) at $x_\sigma$ is nothing else than the pull back of the corresponding fields to ${\cal S}$ under the inverse of the diffeomorphism $S^j:\chi\to {\cal S}$.
\\
Our remaining task is to compute the complete observables in (\ref{CompObs}) that also Poisson commutes with the Hamiltonian constraint. For this purpose the reference field $\tilde{T}$ will be used and hence we need to construct observables for $\tilde{q}_{jk}$ and $\tilde{p}^{jk}$ that we will denote by $Q_{jk}$ and $P^{jk}$ respectively. The constraint $\tilde{c}^{tot}$ in (\ref{equivHam}) is in deparametrized form and thus wen can apply the simplified construction discussed in the last subsection. Let us look at the smeared version of the constraint given by
\be
c^{tot}_\tau:=\int\limits_{\chi} d^3x (\tau-T)(x)\tilde{c}^{tot}(x)
\ee
where we used $\beta^0(x)=(\tau-T)(x)$. Now in order to construct the observables $Q_{jk}$ and $P^{jk}$ we need to ensure that $c_\tau^{tot}$ is already an observable with respect to the spatial diffeomophism constraint $\tilde{c}_j^{tot}$. For the choice of constant $\tau$ the constraint $c^{tot}_\tau$ is an integral over a density of weight one and we can equivalently express it as an integral over the dust space ${\cal S}$ given by
\be
c^{tot}_\tau=\int\limits_{\cal S}d^3\sigma (\tau-\tilde{T})(\sigma)(\tilde{P}+\tilde{h})(\sigma)
\ee
with
\be
\tilde{h}(\sigma)=\sqrt{\tilde{c}^2(\sigma)-\tilde{q}^{jk}(\sigma)\tilde{c}_j(\sigma)\tilde{c}_k(\sigma)}
\ee
where $\tilde{c}(\sigma)$ and $\tilde{c}_j(\sigma)$ are the observables of $c$ and $c_j$ in (\ref{equivHam}) and (\ref{equivDiffeo}) respectively with respect to the spatially diffeomorphism constraint. Now since $c^{tot}_\tau$ deparametrizes we do not need to consider the Hamiltonian vector field of  $c^{tot}_\tau$ but we can work with the Hamiltonian  vector field $X_\tau$ of
\be
H_\tau:=\int\limits_{\cal S}d^3\sigma (\tau-\tilde{T})\tilde{h}(\sigma)~.
\ee
The observables for a function $f$ that depends only on $\tilde{q}_{jk}$ and $\tilde{p}^{jk}$ (and possible other standard model matter degrees of freedom) is then given by
\be
O_f(\tau,\sigma)=\sum\limits_{n=0}^\infty\frac{1}{n!}X^n_\tau\cdot f=\sum\limits_{n=0}^\infty\frac{1}{n!}\{H_\tau,f\}_{(n)}
\ee
where $\{.,.\}_{(n)}$ again denotes the iterated Poisson bracket. Considering the discussion about the physical Hamiltonian above we observe that in case of the dust as reference fields the physical time is given by $\tau$ and the corresponding evolution of the observables is given by
\be
\frac{d O_f(\tau,\sigma)}{d\tau}=\{{\bf H},O_f(\tau,\sigma)\}
\ee
with the physical Hamiltonian
\be
{\bf H}=\int\limits_{\cal S}d^3\sigma H(\sigma)~.
\ee
Here $H(\sigma)$ is the (complete) observable associated to $\tilde{h}(\sigma)$, that is
\be
H(\sigma):=O_{\tilde{h}}(\tau,\sigma)=\tilde{h}(\sigma)
\ee
and is therefore independent of $\tau$ and hence physical time as expected in the deparametrized case. In the following we will continue to use the notation $Q_{jk}$ and $P^{jk}$ for the observables of $q_{ab}$ and $p^{ab}$. Furthermore we will denote the observables associated with $c$ and $c_j$ also by capital letters $C(\tau,\sigma)$ and $C_j(\tau,\sigma)$ respectively. Using this notation we can also rewrite the physical Hamiltonian density as
\be
H(\sigma )=\sqrt{C^2(\tau ,\sigma )-Q^{ij}(\tau ,\sigma )C_i(\tau ,\sigma )C_j(\tau ,\sigma)}.
\ee

\subsection{Evolution Equations in Jordan Frame}
\label{EvEqnJordan}
As is well known, in the classical formulation of STT, there are different choices for basic variables, corresponding to different frames. These frames are related to each other by conformal transformations. There are many debates on which frame should be regarded as the physical one (the word ``physical" has nothing to do with the gauge invariance here) and the question carries over to the discussion what kind of coupling should be chosen. Two prominent choices are either the Jordan frame, where a scalar field is non-minimally coupled to the metric, or the Einstein frame, where a scalar field is minimally coupled as in GR. If one considers the Einstein frame as the relevant one, the corresponding STT would make no difference from GR with a scalar field, which was already presented in \cite{Giesel10}. Hence, to analyze the non-minimally coupled case, in this paper, we consider the Jordan frame and this choice has the consequence that also the evolution equations of the linear perturbations will be formulated using the Jordan frame.

We consider the following action
\ba
S_{STT}=S_{Jordan}+S_{dust},\label{action}\nn
\ea
where the STT action in Jordan frame reads

\ba
S\, _{Jordan}=\int d^4x\sqrt{|\det (g)|}\left[F(\phi)R^{(4)}-K(\phi)g^{\mu\nu}(\triangledown_\mu\phi)\triangledown_\nu\phi-V(\phi)\right].\label{actionjordan}
\ea
Here for simplicity we set $16\pi G=1$,  $F (\phi)$ and $K (\phi)$
are positive coupling functions, and $V(\phi)$ is the potential of the gravitational scalar field $\phi$.
 As mentioned in section 2.1, the dust action reads
 \ba
 S_{dust}=-\frac{1}{2}\int d^4x\sqrt{|\det(g)|}\rho \left[g^{\mu\nu}U_{\mu}U_{\nu}+1\right]
 \ea
with $U_{\mu}=-T_{,\mu}+W_jS^j_{,\mu}$. The Hamiltonian analysis of STT in the generalized Brans-Dicke form is given in Refs. \cite{Zhang11}  in Jordan frame. Using the results there and following the procedures in the last subsection, the physical Hamiltonian density reads
\ba
H(\sigma )=\sqrt{C^2(\tau ,\sigma )-Q^{ij}(\tau ,\sigma )C_i(\tau ,\sigma )C_j(\tau ,\sigma)},\label{phyham}
\ea
with
\ba
C_j(\tau,\sigma )&=&\left[-2Q_{jl}D_kP^{kl}+\Pi D_j\Phi\right](\tau,\sigma) ,\label{Cj}\\
C(\tau, \sigma)&=&\Bigg[\frac{1}{\sqrt{\det Q}}\Big(\frac{\Big(Q_{i m}Q_{j n}-\frac{1}{2}Q_{i j}Q_{mn}\Big)P^{ij}P^{mn}}{F(\Phi)}+\frac{\Big(F'(\Phi )Q_{ij}P^{ij}-F(\Phi)\Pi\Big)^2}{2F(\Phi)\big[3(F'(\Phi))^2+2F(\Phi)K(\Phi)\big]}\Big)\nn\\
&&
+\sqrt{\det Q}\Big(-F(\Phi )R^{(3)}+K(\Phi)Q^{ij}\big(D_i\Phi \big)\big(D_j\Phi\big)+2Q^{ij}D_i D_j F(\Phi )+V(\Phi)\Big)\Bigg](\tau ,\sigma ),\label{C}\nn\\
\ea
and the elementary Poisson bracket read:
\ba
\{P^{ij}(\sigma),Q_{kl}(\tilde{\sigma})\}=\delta ^i_{(k}\delta^j_{l)}\delta(\sigma,\tilde{\sigma}),\qquad\{\Pi(\sigma),\Phi(\tilde{\sigma})\}=\delta (\sigma,\tilde{\sigma}),
\ea
where a prime $'$ in Eq. (\ref{C}) means the derivative with respect to $\Phi$, i.e. $F'(\Phi)=\frac{F(\Phi)}{d\Phi}, F''(\Phi)=\frac{d^2F(\Phi)}{d\Phi^2}$. Here we denote all the objects after gauge completion with capital letters. All quantities are now evaluated on the dust manifold. Note that  Eq. (\ref{C}) is valid in the case of $3(F'(\Phi))^2+2F(\Phi)K(\Phi)\neq0$, which corresponds to the case $\omega\neq -\frac{3}{2}$ in Refs. \cite{Zhang11}, on which we focus in the paper.

Given a function $f$ on phase space, that does not depend on the dust variables, we can construct its associated observable $O_f$ by applying the procedure described in section \ref{BKFormalism}.
 We get the Hamiltonian equations of motion for $O_f$, being first order differential equations with respect to the physical time $\tau$, by simply calculating its Poisson bracket with the physical Hamiltonian denoted by $\textbf{H}$ and given by $\textbf{H}:=\int d\sigma H(\sigma)$:
 \ba
 \dot{O}_f\equiv\frac{dO_f}{d\tau}=\{\textbf{H},O_f\}.
 \ea
 In our case we are interested in the equations of motion for $O_{q_{ab}}=:Q_{jk}, O_{p^{ab}}=:P^{jk}, O_{\phi}=:\Phi, O_{\pi}=:\Pi$ and using the physical Hamiltonian of the dust model we end up with the following Hamiltonian equations:
\ba
\dot{\Phi}&=&-\frac{N}{\sqrt{\det Q}}\frac{F'(\Phi)Q_{i j}P^{i j}-F(\Phi)\Pi}{3(F'(\Phi))^2+2F(\Phi)K(\Phi)}+\mathcal{L}_{\overset{\rightharpoonup }{N}}\Phi  ,\label{phifirst}\\
\dot{\Pi}&=&-\frac{N}{\sqrt{\det Q}}\bigg[\frac{\Big(F'(\Phi)Q_{ij}P^{ij}-F(\Phi)\Pi\Big)^2}{2\big(F(\Phi)\big)^2\big[3(F'(\Phi))^2+2F(\Phi)K(\Phi )\big]^2}\cdot\Big[-F(\Phi)\big(3(F'(\Phi))^2+2F(\Phi)K(\Phi)\big)\Big]'\nn\\
&&\qquad\qquad\quad
+\frac{F'(\Phi)Q_{ij}P^{ij}-F(\Phi)\Pi}{F(\Phi)\big[3(F'(\Phi))^2+2F(\Phi)K(\Phi)\big]}\Big(F''(\Phi)Q_{ij}P^{ij}-F'(\Phi)\Pi\Big) \nn\\
&&\qquad\qquad\quad
-\frac{F'(\Phi)}{(F(\Phi))^2}\Big(Q_{im}Q_{jn}-\frac{1}{2}Q_{ij}Q_{mn}\Big)P^{ij}P^{mn}\bigg]\nn\\
&&
-\sqrt{\det Q}\left[-NF'(\Phi)R^{(3)}+N K'(\Phi)Q^{ij}\big(D_i\Phi\big)D_j\Phi+2F'(\Phi)Q^{ij}D_i D_jN+NV'(\Phi)\right]\nn\\
&&
+2\partial_j\left[N\sqrt{\det Q}K(\Phi )Q^{jk}\Phi _{,k}\right]+\mathcal{L}_{\overset{\rightharpoonup}{N}}\Pi,\label{pifirst}\\
\dot{Q}_{j k}&=&\frac{N}{\sqrt{\det Q}F(\Phi)}\Big(2G_{j k m n}P^{m n}+\frac{(F'(\Phi))^2Q_{m n}P^{mn}-F'(\Phi)F(\Phi)\Pi}{3(F'(\Phi))^2+2F(\Phi )K(\Phi)}Q_{jk}\Big)+\big(\mathcal{L}_{\overset{\rightharpoonup }{N}}Q\big)_{j k} ,\label{qfirst}\\
\dot{P}^{j k}&=&\frac{N}{\sqrt{\det Q}}\left[-\frac{Q_{mn}\Big(2P^{jm}P^{kn}-P^{jk}P^{mn}\Big)}{F(\Phi)}-\frac{(F'(\Phi))^2Q_{mn}P^{m n}-F'(\Phi)F(\Phi)\Pi}{F(\Phi)\big[3(F'(\Phi))^2+2F(\Phi)K(\Phi)\big]}P^{jk}\right]\nn\\
&&
+\sqrt{\det Q}\bigg[N\Big(-K(\Phi)Q^{jk}\Phi^{,m}\Phi_{,m}-V(\Phi)Q^{jk}-Q^{jk}D^iD_i F(\Phi)+K(\Phi)\Phi^{,j}\Phi^{,k}\Big)\nn\\
&&\qquad\qquad\quad
+\left[G^{-1}\right]^{jkmn}\Big(D_m D_n(NF(\Phi))-NF(\Phi)R_{mn}\Big)-2(D^{(j}N)D^{k)}F(\Phi)\nn\\
&&\qquad\qquad\quad
+Q^{jk}\big(D^iN\big)D_i F(\Phi)\bigg]\nn\\
&&
+\frac{N}{2}Q^{j k}C(\tau,\sigma)-\frac{1}{2}H(\tau,\sigma )Q^{j m}Q^{kn}N_m N_n+\big(\mathcal{L}_{\overset{\rightharpoonup }{N}}P\big)^{j k},\label{pfirst}
\ea
where $N:=\frac{C}{H}$, $N_j:=-\frac{C_j}{H}$, $\Phi_{,k}:=\frac{\partial \Phi}{\partial x^k}$, $ G_{jkmn}:=\frac12 (Q_{jm}Q_{nk}+Q_{jn}Q_{mk}-Q_{jk}Q_{mn})$ and its inverse $ [G^{-1}]^{jkmn}:=\frac12(Q^{jm}Q^{nk}+Q^{jn}Q^{mk}-2Q^{jk}Q^{mn})$ satisfying $G_{jkmn}G^{mnrs}=\delta^{r}_{(j}\delta^{s}_{k)}$. All the indices are lowered and raised by the three-metric and its inverse. Note that both $C(\tau,\sigma)$ and $H(\tau,\sigma)$ are non-vanishing because they are no longer constraints by introducing the dust fields as dynamically coupled observers. It is only the total Hamiltonian and diffeomorphism constraints, which involve the dust as well as the gravitational and scalar field contributions, that is still vanishing.
 Moreover, as shown in \cite{Giesel10}, both $H$ and $N_j$ are conserved quantities because their Poisson brackets with the physical Hamiltonian equal zero. Now it is easy to see from Eqs. (\ref{phifirst}) and (\ref{qfirst}) that
\ba
\Pi &=&\frac{1}{F(\Phi)}\Big(F'(\Phi)Q_{ij}P^{ij}+\frac{\sqrt{\det Q}}{N}\big(\dot{\Phi}-\mathcal{L}_{\overset{\rightharpoonup}{N}}\Phi\big)\Big[3 (F'(\Phi))^2+2F(\Phi)K(\Phi)\Big]\Big), \label{Pi}\\
P^{j k}&=&\frac{\sqrt{\det Q}F(\Phi )}{2N}\left[G^{-1}\right]^{jkmn}\Big(\dot{Q}_{mn}-(\mathcal{L}_{\overset{\rightharpoonup }{N}}Q)_{mn}+\frac{F'(\Phi )}{F(\Phi)}\big(\dot{\Phi}-\mathcal{L}_{\overset{\rightharpoonup }{N}}\Phi\big)Q_{mn}\Big).\label{Pjk}
\ea
Taking the second order time derivative of Eqs. (\ref{phifirst}) and (\ref{qfirst}) and substituting Eqs. (\ref{pifirst}), (\ref{pfirst}), (\ref{Pi}), (\ref{Pjk}) into them, we get the following equations:
\ba
\overset{\cdot\cdot}{\Phi}&=&\left[\frac{\dot{N}}{N}-\frac{\Big(\sqrt{\det Q}\Big)^{\cdot}}{\sqrt{\det Q}}+\frac{N}{\sqrt{\det Q}}\Big(\mathcal{L}_{\overset{\rightharpoonup }{N}}(\frac{\sqrt{\det Q}}{N})\Big)\right]\big(\dot{\Phi}
-\mathcal{L}_{\overset{\rightharpoonup }{N}}\Phi \big)+2\big(\mathcal{L}_{\overset{\rightharpoonup }{N}}\dot{\Phi}\big)+\big(\mathcal{L}_{\dot{\overset{\rightharpoonup}{N}}}\Phi\big)-\big(\mathcal{L}_{\overset{\rightharpoonup }{N}}(\mathcal{L}_{\overset{\rightharpoonup}{N}}\Phi)\big) \nn\\
&&
-\frac{3F'(\Phi)F''(\Phi)+F(\Phi)K'(\Phi)}{3(F'(\Phi))^2+2F(\Phi)K(\Phi)}\big(\dot{\Phi}-\mathcal{L}_{\overset{\rightharpoonup}{N}}\Phi\big)^2
-\frac32\frac{N^2}{\sqrt{\det Q}}\frac{F'(\Phi)}{3(F'(\Phi))^2+2F(\Phi)K(\Phi)}C(\tau, \sigma)\nn\\
&&
+\frac{2F(\Phi)}{3(F'(\Phi))^2+2F(\Phi)K(\Phi)}\Bigg[N^2\bigg[K(\Phi)D^iD_i\Phi+K(\Phi)[Q^{jk}]_{,j}\Phi _{,k}-\frac{1}{2}V'(\Phi)\bigg]\nn\\
&&\qquad\qquad\qquad\qquad\qquad\qquad
+K(\Phi)Q^{jk}\Phi_{,k}\Big[\frac{N}{\sqrt{\det Q}}\big(N\sqrt{\det Q}\big)_{,j}\Big]\Bigg]\nn\\
&&
+\frac{F'(\Phi)}{3(F'(\Phi))^2+2F(\Phi)K(\Phi)}\Bigg[\frac{F(\Phi)}{4}\left[G^{-1}\right]^{rstu}\Big(\dot{Q}_{rs}
-(\mathcal{L}_{\overset{\rightharpoonup }{N}}Q)_{rs}\Big)\Big(\dot{Q}_{tu}-(\mathcal{L}_{\overset{\rightharpoonup }{N}}Q)_{tu}\Big)\nn\\
&&\qquad\qquad\qquad\qquad\qquad\qquad\quad
-F'(\Phi)Q^{mn}\Big(\dot{Q}_{mn}-(\mathcal{L}_{\overset{\rightharpoonup }{N}}Q)_{mn}\Big)\big(\dot{\Phi}-\mathcal{L}_{\overset{\rightharpoonup }{N}}\Phi\big)\nn\\
&&\qquad\qquad\qquad\qquad\qquad\qquad\quad
+\Big(2N^2K(\Phi)+\frac{F(\Phi)}{F'(\Phi)}N^2K'(\Phi)\Big)\Phi^{,i}\Phi_{,i}+3N^2V(\Phi)\nn\\
&&\qquad\qquad\qquad\qquad\qquad\qquad\quad
+5N^2D^iD_i F(\Phi)+3N(D^iN)D_i F(\Phi)-N^2F(\Phi)R^{(3)}\nn\\
&&\qquad\qquad\qquad\qquad\qquad\qquad\quad
+\frac{1}{2}\frac{N}{\sqrt{\det Q}}H(\sigma)Q^{mn}N_m N_n\Bigg],\label{phisecond}\\
\overset{\cdot\cdot}{Q}_{jk}&=&\left[\frac{\dot{N}}{N}-\frac{\Big(\sqrt{\det Q}\Big)^{\cdot}}{\sqrt{\det Q}}+\frac{N}{\sqrt{\det Q}}\Big(\mathcal{L}_{\overset{\rightharpoonup}{N}}(\frac{\sqrt{\det Q}}{N})\Big)\right]\Big(\dot{Q}_{jk}-(\mathcal{L}_{\overset{\rightharpoonup }{N}}Q)_{jk}\Big)\nn\\
&&
+Q^{m n}\Big(\dot{Q}_{mj}-(\mathcal{L}_{\overset{\rightharpoonup }{N}}Q)_{mj}\Big)\Big(\dot{Q}_{nk}-(\mathcal{L}_{\overset{\rightharpoonup}{N}}Q)_{nk }\Big)+N^2\Big(\frac{2K(\Phi)}{F(\Phi)}\Phi _{,j}\Phi _{,k}-2R_{jk}\Big)\nn\\
&&
+2ND_j D_k N+2\big(\mathcal{L}_{\overset{\rightharpoonup }{N}}\dot{Q}\big)_{jk}+\big(\mathcal{L}_{\dot{\overset{\rightharpoonup}{N}}}Q \big)_{jk}-\big(\mathcal{L}_{\overset{\rightharpoonup}{N}}(\mathcal{L}_{\overset{\rightharpoonup}{N}}Q)\big)_{jk}
-\frac{F'(\Phi)}{F(\Phi)}\Big(\dot{Q}_{jk}-(\mathcal{L}_{\overset{\rightharpoonup }{N}}Q)_{jk}\Big)\big(\dot{\Phi}
-\mathcal{L}_{\overset{\rightharpoonup }{N}}\Phi\big)\nn\\
&&
+\frac{2N^2}{F(\Phi)}D_j D_k F(\Phi)-\frac{NH(\sigma)}{\sqrt{\det Q}F(\Phi)}G_{jkmn}N^mN^n+\frac{Q_{j k}}{F(\Phi)}\Big[N(D^iN)D_i F(\Phi)+N^2 D^iD_i F(\Phi)+N^2V(\Phi)\Big]\nn\\
&&
+\frac{-2K(\Phi)F''(\Phi)+K'(\Phi)F'(\Phi)}{3(F'(\Phi))^2+2F(\Phi)K(\Phi)}
\big(\dot{\Phi}-\mathcal{L}_{\overset{\rightharpoonup }{N}}\Phi \big)^2Q_{jk}-\frac{N^2}{\sqrt{\det Q}}\frac{K(\Phi)Q_{jk}}{3(F'(\Phi))^2+2F(\Phi)K(\Phi)}C(\tau, \sigma)\nn\\
&&
-\frac{2F'(\Phi)Q_{jk}}{3(F'(\Phi))^2+2F(\Phi)K(\Phi)}\Bigg[N^2\bigg(K(\Phi)D^iD_i\Phi+K(\Phi)[Q^{jk}]_{,j}\Phi _{,k}-\frac{1}{2}V'(\Phi)\bigg)\nn\\
&&\qquad\qquad\qquad\qquad\qquad\qquad\quad
+K(\Phi)Q^{jk}\Phi_{,k}\Big[\frac{N}{\sqrt{\det Q}}\big(N\sqrt{\det Q}\big)_{,j}\Big]\Bigg]\nn\\
&&
-\frac{(F'(\Phi))^2Q_{jk}}{F(\Phi)\big[3(F'(\Phi))^2+2F(\Phi)K(\Phi)\big]}\Bigg[\frac{F(\Phi)}{4}\left[G^{-1}\right]^{rstu}\Big(\dot{Q}_{r s}-(\mathcal{L}_{\overset{\rightharpoonup}{N}}Q)_{rs}\Big)\Big(\dot{Q}_{tu}-(\mathcal{L}_{\overset{\rightharpoonup }{N}}Q)_{tu}\Big)\nn\\
&&\qquad\qquad\qquad\qquad\qquad\qquad\qquad\qquad
-F'(\Phi)Q^{mn}\Big(\dot{Q}_{mn}-(\mathcal{L}_{\overset{\rightharpoonup}{N}}Q)_{mn}\Big)\Big(\dot{\Phi}
-\mathcal{L}_{\overset{\rightharpoonup}{N}}\Phi \Big)\nn\\
&&\qquad\qquad\qquad\qquad\qquad\qquad\qquad\qquad
+\Big(2N^2K(\Phi)+\frac{F(\Phi)}{F'(\Phi)}N^2K'(\Phi)\Big)\Phi ^{,i}\Phi_{,i}+3N^2V(\Phi)\nn\\
&&\qquad\qquad\qquad\qquad\qquad\qquad\qquad\qquad
+5N^2D^iD_i F(\Phi)+3N(D^iN)D_i F(\Phi)-N^2F(\Phi)R^{(3)}\nn\\
&&\qquad\qquad\qquad\qquad\qquad\qquad\qquad\qquad
+\frac{1}{2}\frac{N}{\sqrt{\det Q}}H(\sigma )Q^{mn}N_m N_n\Bigg],\label{qsecond}
\ea
where $C(\tau, \sigma)$ is now expressed as a function depending on the configuration variables and their first order time derivatives given by
\ba
C&=&\frac{\sqrt{\det Q}F(\Phi)}{4N^2}\left[G^{-1}\right]^{rstu}\Big(\dot{Q}_{rs}-(\mathcal{L}_{\overset{\rightharpoonup }{N}}Q)_{r s}\Big)\Big(\dot{Q}_{t u}-(\mathcal{L}_{\overset{\rightharpoonup }{N}}Q)_{tu}\Big)-\sqrt{\det Q}F(\Phi)R^{(3)}+\sqrt{\det Q}K(\Phi )Q^{jk}\Phi _{, j}\Phi _{,k}\nn\\
&&
-\frac{\sqrt{\det Q}}{N^2}F'(\Phi )Q^{rs}\Big(\dot{Q}_{rs}-(\mathcal{L}_{\overset{\rightharpoonup }{N}}Q)_{rs}\Big)\big(\dot{\Phi }-\mathcal{L}_{\overset{\rightharpoonup }{N}}\Phi\big)+\frac{\sqrt{\det Q}}{N^2}K(\Phi)\big(\dot{\Phi}-\mathcal{L}_{\overset{\rightharpoonup }{N}}\Phi \big)^2+2\sqrt{\det Q}Q^{jk}D_j D_k F(\Phi)\nn\\&&+\sqrt{\det Q}V(\Phi).\label{Hamiltonianconstr}
\ea

So far we have derived the evolution equations for the gauge invariant 3-metric $Q_{jk}$ and the scalar field $\Phi$.  Notice that if we choose in the equation of motions shown in  (\ref{phisecond}) and (\ref{qsecond}) $F(\Phi)=1$ and $K(\Phi)=\frac{1}{2}$, all  terms containing $F'(\Phi)$, $F''(\Phi)$, $K'(\Phi)$ naturally vanish and the remaining terms reduce to the evolution equations of the observables for a physical system consisting of dust, gravity and a minimally coupled scalar field. In this special case the evolution equations in  (\ref{phisecond}) and (\ref{qsecond}) agree with the evolution equations for the observables shown in equations (4.23) and (4.24) in \cite{Giesel10}\footnote{Note that the definition of the potential $V(\phi)$ used in \cite{Giesel10} differs by a factor of 2 and this needs to be considered in order to have an exact agreement between the equations.}.

\section{Linear Perturbations of STT on a General Background}
In this section we will use the gauge invariant variables $(\Phi, Q_{jk})$ and their equations of motions, obtained in last section, to derive the evolution equations for their corresponding linearly perturbed variables. For this purpose, we first split the configuration variables into a background and a perturbed part
\ba
\Phi=\overline{\Phi}+\delta \Phi, \qquad Q_{jk}=\overline{Q}_{jk}+\delta Q_{jk},
\ea
where $\overline{\Phi}$ and $\overline{Q}_{jk}$ denote the background variables satisfying the evolution equations shown in (\ref{phisecond}) and (\ref{qsecond}) and are thus solutions of the classical scalar-tensor equations. From now on we denote all occurring background variables with a bar. Introducing the dust as dynamically coupled observers has the consequence, that the shift vector $N_j$ and the lapse function $N$ are no longer arbitrary but become fixed as particular functions on the reduced phase space, spanned by $(Q_{jk},P^{jk},\Phi,\Pi)$. Therefore $N, N_j$ are not treated as independent variables and the same is true for the associated perturbations. Since we consider the perturbations of the second order evolution equations for $Q_{jk}$ and $\Phi$, the lapse function $N$ and the shift vector $N_j$ can be understood as functions of $Q_{jk},\Phi$ and their associated velocities.
The explicit form of their perturbations reads
\ba
N_j=\overline{N}_j+\delta N_j=-\frac{\overline{C}_j}{\overline{H}}+\delta N_j,\qquad N=\overline{N}+\delta N=\frac{\overline{C}}{\overline{H}}-\frac{\overline{N}^j\overline{N}^k}{2\overline{N}}\delta Q_{jk}+\frac{\overline{N}^j}{\overline{N}}\delta N_j.
\ea
Note that in principle $\delta N_j$ can be expressed in terms of $\delta Q_{jk}$ and $\delta \Phi$, but we keep $\delta N_j$  here because as proved in \cite{Giesel10} both $\delta N_j$ and $\delta H$ are conserved quantities, while  a conserved quantity is fixed at an initial time, it remains a constant during the evolution. Keeping $\delta N_j$ here helps to formulate the final evolutions equations for the linear perturbations in more compact form. Likewise we also keep the explicit expression $\delta H$ in the following for the same reason.

Since the evolution equations (\ref{phisecond}) and (\ref{qsecond}) are already very complicated, it is expected that the evolution equations for $\delta \Phi$ and $\delta Q_{jk}$ will be even more complicated. In order to express them in a concise form, we notice that there are three important common terms in Eqs. (\ref{phisecond}) and (\ref{qsecond}), namely $C(\tau,\sigma)$ and all terms inside the last two square brackets. We introduce the following abbreviations, which agree for both equations:
\ba
X&\equiv&N^2\Big[K(\Phi)D^iD_i\Phi+K(\Phi)[Q^{jk}]_{,j}\Phi_{,k}-\frac{1}{2}V'(\Phi)\Big]
+K(\Phi)Q^{jk}\Phi_{,k}\Big[\frac{N}{\sqrt{\det Q}}\Big(N\sqrt{\det Q}\Big)_{,j}\Big],\label{X}\\
Y&\equiv&\frac{F(\Phi)}{4}\left[G^{-1}\right]^{rstu}\Big(\dot{Q}_{r s}-(\mathcal{L}_{\overset{\rightharpoonup}{N}}Q)_{rs}\Big)\Big(\dot{Q}_{tu}-(\mathcal{L}_{\overset{\rightharpoonup }{N}}Q)_{tu}\Big)
-F'(\Phi)Q^{mn}\Big(\dot{Q}_{mn}-(\mathcal{L}_{\overset{\rightharpoonup}{N}}Q)_{mn}\Big)\Big(\dot{\Phi}
-\mathcal{L}_{\overset{\rightharpoonup}{N}}\Phi \Big)\nn\\
&&
+\Big(2N^2K(\Phi)+\frac{F(\Phi)}{F'(\Phi)}N^2K'(\Phi)\Big)\Phi^{,i}\Phi_{,i}+3N^2V(\Phi)
+5N^2D^iD_i F(\Phi)+3N(D^iN)D_i F(\Phi)\nn\\
&&
-N^2F(\Phi)R^{(3)}+\frac{1}{2}\frac{N}{\sqrt{\det Q}}H(\sigma )Q^{mn}N_m N_n.\label{Y}
\ea
The next step is to derive the equations of motions for the linear perturbations. For this purpose we use the following identities:
\ba
\lefteqn{\delta\left[\left[G^{-1}\right]^{rstu}\Big(\dot{Q}_{rs}-(\mathcal{L}_{\overset{\rightharpoonup }{N}}Q)_{rs}\Big)\Big(\dot{Q}_{t u}-(\mathcal{L}_{\overset{\rightharpoonup }{N}}{Q})_{tu}\Big)\right]}\nn\\
&=&2\Big(\dot{\overline{Q}}_{rs}-(\mathcal{L}_{\overset{\rightharpoonup }{\overline{N}}}\overline{Q})_{rs}\Big)\Bigg[\left[\overline{G}^{-1}\right]^{rsjk}(\frac{\partial}{\partial\tau}-\mathcal{L}_{\overset{\rightharpoonup }{\overline{N}}})-\left[\overline{G}^{-1}\right]^{turj}\overline{Q}^{sk}\Big(\dot{\overline{Q}}_{tu}-(\mathcal{L}_{\overset{\rightharpoonup }{\overline{N}}}\overline{Q})_{tu}\Big)\nn\\
&&\qquad\qquad\qquad\qquad\quad
+[\overline{G}^{-1}]^{rsmn}\Big[\overline{Q}^{jt}\overline{N}^k\big[\overline{Q}_{mn}\big]_{,t}+2\overline{Q}_{tn}\frac{\partial}{\partial x^m}\Big(\overline{Q}^{jt}\overline{N}^k\Big)\Big]\Bigg]\delta{Q}_{jk}\nn\\
&&
-2\Big(\dot{\overline{Q}}_{rs}-(\mathcal{L}_{\overset{\rightharpoonup }{\overline{N}}}\overline{Q})_{rs}\Big)\left[\overline{G}^{-1}\right]^{rsjn}\left[\overline{Q}^{mt}\left[\overline{Q}_{jn}\right]_{,t}
+2\overline{Q}_{tj}\frac{\partial}{\partial x^n}\Big(\overline{Q}^{mt}\Big)\right]\delta{N}_m,\\
\lefteqn{\delta \Big[Q^{mn}\Big(\dot{Q}_{mn}-(\mathcal{L}_{\overset{\rightharpoonup }{N}}Q)_{mn}\Big)(\dot{\Phi}-\mathcal{L}_{\overset{\rightharpoonup}{N}}\Phi)\Big]}\nn\\
&=&\Bigg[-\overline{Q}^{jm}\overline{Q}^{kn}\Big(\dot{\overline{Q}}_{mn}-(\mathcal{L}_{\overset{\rightharpoonup}{N}}\overline{Q})_{m n}\Big)\big(\dot{\overline{\Phi}}-\mathcal{L}_{\overset{\rightharpoonup}{\overline{N}}}\overline{\Phi}\big)+\overline{Q}^{jk}
\big(\dot{\overline{\Phi}}-\mathcal{L}_{\overset{\rightharpoonup}{\overline{N}}}\overline{\Phi}\big)(\frac{\partial}{\partial\tau }-\mathcal{L}_{\overset{\rightharpoonup}{\overline{N}}})\nn\\
&&
-\overline{Q}^{mn}\big(\dot{\overline{\Phi}}-\mathcal{L}_{\overset{\rightharpoonup}{\overline{N}}}\overline{\Phi}\big)\left[-\overline{Q}^{j r}\overline{N}^k[\overline{Q}_{mn}]_{,r}-\overline{Q}_{nr}\frac{\partial}{\partial x^m}(\overline{Q}^{jr}\overline{N}^k)
-\overline{Q}_{mr}\frac{\partial }{\partial x^n}(\overline{Q}^{jr}\overline{N}^k)\right]\nn\\
&&
+\overline{Q}^{mn}\big[\overline{Q}^{jr}\overline{\Phi}_{,r}\overline{N}^k\big]\Big(\dot{\overline{Q}}_{mn}-(\mathcal{L}_{\overset{\rightharpoonup }{N}}\overline{Q})_{mn}\Big)\Bigg]\delta Q_{jk}\nn\\
&&
+\Bigg[-\overline{Q}^{jk}(\dot{\overline{\Phi}}-\mathcal{L}_{\overset{\rightharpoonup}{\overline{N}}}\overline{\Phi})\left[\overline{Q}^{m n}[\overline{Q}_{jk}]_{,n}+\overline{Q}_{kn}\frac{\partial}{\partial x^j}(\overline{Q}^{mn})+\overline{Q}_{jn}\frac{\partial}{\partial x^k}(\overline{Q}^{mn})\right]\nn\\
&&\quad
-\overline{Q}^{jk}[\overline{Q}^{m n}\overline{\Phi}_{,n}]\Big(\dot{\overline{Q}}_{jk}-(\mathcal{L}_{\overset{\rightharpoonup }{\overline{N}}}\overline{Q})_{jk}\Big)\Bigg]\delta N_m\nn\\
&&
+\bigg[\overline{Q}^{mn}\Big(\dot{\overline{Q}}_{mn}-(\mathcal{L}_{\overset{\rightharpoonup }{\overline{N}}}\overline{Q})_{m n}\Big)(\frac{\partial }{\partial\tau}-\mathcal{L}_{\overset{\rightharpoonup}{\overline{N}}})\bigg]\delta\Phi,\\
\lefteqn{\delta\left[D_j D_k F(\Phi)\right]=\left[-\Big(\overline{D}_n F(\overline{\Phi})\Big)\overline{D}_{(k}\overline{Q}^{m n}\right]\delta  Q_{j)m}+\left[\frac{1}{2}\big(\overline{D}_n F(\overline{\Phi})\big)\overline{D}_m\overline{Q}^{m n}\right]\delta Q_{j k}+\left[\overline{D}_j\overline{D}_k F'(\overline{\Phi})\right]\delta\Phi,}\\
\lefteqn{\delta\left[NQ^{jk}(D_j N)D_k F(\Phi)\right]=\left[-\frac{1}{2}\overline{Q}^{m n}\Big(\overline{D}_n F(\overline{\Phi})\Big)\overline{D}_m\big(\overline{N}^j\overline{N}^k\big)-\frac{1}{2}\overline{Q}^{jm}\overline{Q}^{n k}\big(\overline{D}_m\overline{N}^2\big)\Big(\overline{D}_n F(\overline{\Phi})\Big)\right]\delta Q_{jk}}\nn\\
&&\qquad\qquad\qquad\qquad\quad
+\left[\overline{Q}^{jk}\Big(\overline{D}_k F(\overline{\Phi})\Big)\overline{D}_j\overline{N}^m\right]\delta N_m
+\left[\frac{1}{2}\big(\overline{D}_j\overline{N}^2\big)\overline{Q}^{jk}\overline{D}_k F'(\overline{\Phi})\right]\delta\Phi,\\
\lefteqn{\delta\Big(-\frac{N}{\sqrt{\det {Q}}}H G_{jkmn}N^m N^n\Big)=\Big[\frac{1}{2}\frac{\overline{N}}{\sqrt{\det {\overline{Q}}}}(\overline{Q}^{m n}+\frac{\overline{N}^m\overline{N}^n}{\overline{N}^2})\overline{H} \overline{G}_{jkrs}\overline{N}^r \overline{N}^s+\frac{1}{2}\frac{\overline{N}}{\sqrt{\det {\overline{Q}}}}\overline{H}\overline{N}^m\overline{N}^n \overline{Q}_{jk}}\nn\\
&&\qquad\qquad\qquad\qquad\qquad\qquad\quad
+2\frac{\overline{N}}{\sqrt{\det {\overline{Q}}}}\overline{H} \overline{G}_{jkrs} \overline{N}^n \overline{N}^s \overline{Q}^{rm}\Big]\delta Q_{mn}
+\Big[-2\frac{\overline{N}}{\sqrt{\det {\overline{Q}}}}\overline{H} \overline{N}^r \overline{N}^t \overline{Q}^{s m} \overline{G} _{rts(j}\Big]\delta Q_{k)m}\nn\\
&&\qquad\qquad\qquad\qquad\qquad\qquad
+\Big[-\frac{1}{2}\frac{\overline{N}}{\sqrt{\det {\overline{Q}}}}\overline{H} \overline{N}^m \overline{N}^n \overline{Q}_{mn}\Big]\delta Q_{jk}\nn\\
&&\qquad\qquad\qquad\qquad\qquad\qquad
+\Big[-\frac{\overline{N}}{\sqrt{\det {\overline{Q}}}}\frac{\overline{N}^m}{\overline{N}^2}\overline{H} \overline{G}_{jkrs}\overline{N}^r \overline{N}^s-2\frac{\overline{N}}{\sqrt{\det {\overline{Q}}}}\overline{H} \overline{G}_{jkrs}\overline{N}^s \overline{Q}^{rm}\Big]\delta N_{m}\nn\\
&&\qquad\qquad\qquad\qquad\qquad\qquad
+\Big[-\frac{\overline{N}}{\sqrt{\det {\overline{Q}}}}\overline{G}_{jkrs}\overline{N}^r \overline{N}^s\Big]\delta H,
\ea
where the derivatives inside the square bracket act on all terms including the perturbed variables to its right side, while derivatives surrounded by the round bracket like $(\overline{D}_i\overline{D}_j...)$ act only on the elements inside it. With these identities, the perturbation of the expression $X$ in (\ref{X}) can be expressed as
\ba
\delta X&=&\Bigg[\overline{N}^2\big[(\overline{D}^i\overline{D}_i\overline{\Phi})K'(\overline{\Phi})+ K(\overline{\Phi})\overline{D}^i\overline{D}_i+ (\overline{Q}^{jk})_{,j}\overline{\Phi}_{,k}K'(\overline{\Phi})
+K(\overline{\Phi})(\overline{Q}^{mn})_{,n}\frac{\partial}{\partial x^m}-\frac12V''(\overline{\Phi})\big]\nn\\
&&\quad
+K'(\overline{\Phi})\overline{Q}^{jk}\overline{\Phi}_{,k}\Big[\frac{\overline{N}}{\sqrt{\det\overline{Q}}}\big(\overline{N}\sqrt{\det \overline{Q}}\big)_{,j}\Big]
+K(\overline{\Phi})\Big[\frac{\overline{N}}{\sqrt{\det \overline{Q}}}\big(\overline{N}\sqrt{\det \overline{Q}}\big)_{,j}\Big]\overline{Q}^{j k}\frac{\partial}{\partial x^k}\Bigg]\delta \Phi\nn\\
&&
+\Bigg[-\overline{N}^j\overline{N}^k\Big[K(\overline{\Phi})\overline{D}^i\overline{D}_i\overline{\Phi} +K(\overline{\Phi})(\overline{Q}^{m n})_{, m}\overline{\Phi} _{,n}-\frac{1}{2}V'(\overline{\Phi})\Big]
+\overline{N}^2K(\overline{\Phi})\Big[-\frac{\partial}{\partial x^n}(\overline{Q}^{jm}\overline{Q}^{kn}\overline{\Phi}_{,m})\Big]\nn\\
&&\quad
+K(\overline{\Phi})\overline{Q}^{mn}\overline{\Phi}_{,n}\Big[-\frac12\frac{\overline{N}}{\sqrt{\det {\overline{Q}}}}(\overline{Q}^{j k}+\frac{\overline{N}^j\overline{N}^k}{\overline{N}^2})[\overline{N}\sqrt{\det {\overline{Q}}}]_{,m}
+\frac12\frac{\overline{N}}{\sqrt{\det {\overline{Q}}}}\frac{\partial}{\partial x^m}\big(\overline{N}\sqrt{\det {\overline{Q}}}(\overline{Q}^{j k}-\frac{\overline{N}^j\overline{N}^k}{\overline{N}^2})\big)\Big]\nn\\
&&\quad
+K(\overline{\Phi})\Big[\frac{\overline{N}}{\sqrt{\det {\overline{Q}}}}(\overline{N}\sqrt{\det {\overline{Q}}})_{,n}\Big]\Big[-\overline{Q}^{jm}\overline{Q}^{kn}\overline{\Phi}_{,m}\Big]\Bigg]\delta Q_{jk}\nn\\
&&
+\Bigg[2\overline{N}^m\Big(K(\overline{\Phi})\overline{D}^i\overline{D}_i\overline{\Phi} +K(\overline{\Phi})\left[\overline{Q}^{m n}\right]_{, m}\overline{\Phi} _{,n}-\frac{1}{2}V'(\overline{\Phi})\Big)
+K(\overline{\Phi})\overline{Q}^{jk}\overline{\Phi}_{,k}\Big[\frac{\overline{N}}{\sqrt{\det\overline{ Q}}}\frac{\overline{N}^m}{\overline{N}^2}\big(\overline{N}\sqrt{\det \overline{Q}}\big)_{,j}\nn\\
&&\quad
+\frac{\overline{N}}{\sqrt{\det \overline{Q}}}\frac{\partial}{\partial x^j}\big(\frac{\overline{N}^m}{\overline{N}^2}\overline{N}\sqrt{\det{\overline{Q}}}\big)\Big]\Bigg]\delta N_m ,\label{deltaX}
\ea
and the perturbation of the expression $Y$ in (\ref{Y}) yields
\ba
\delta Y&=&\Bigg[\frac{F'(\overline{\Phi})}{4}[\overline{G}^{-1}]^{r s t u}\Big(\dot{\overline{Q}}_{r s}-(\mathcal{L}_{\overset{\rightharpoonup }{\overline{N}}}\overline{Q})_{r s}\Big)\Big(\dot{\overline{Q}}_{t u}-(\mathcal{L}_{\overset{\rightharpoonup }{\overline{N}}}{\overline{Q}})_{t u}\Big)
-F(\overline{\Phi})''(\Phi)\overline{Q}^{m n}\Big(\dot{\overline{Q}}_{m n}-(\mathcal{L}_{\overset{\rightharpoonup }{\overline{N}}}\overline{Q})_{m n}\Big)\Big(\dot{\overline{\Phi}}-\mathcal{L}_{\overset{\rightharpoonup }{\overline{N}}}\overline{\Phi }\Big)\nn\\
&&\quad
-F'(\overline{\Phi})\overline{Q}^{m n}\Big(\dot{\overline{Q}}_{m n}-(\mathcal{L}_{\overset{\rightharpoonup }{\overline{N}}}\overline{Q})_{m n}\Big)(\frac{\partial}{\partial \tau}-\mathcal{L}_{\overset{\rightharpoonup} {\overline{N}}})+\overline{N}^2\overline{Q}^{mn}\overline{\Phi}_{,m}\overline{\Phi}_{,n}\big(2\overline{N}^2K(\overline{\Phi})
+\overline{N}^2\frac{F(\overline{\Phi})}{F'(\overline{\Phi})}K'(\overline{\Phi})\big)'
\nn\\
&&\quad
+2\big(2K(\overline{\Phi})+\frac{F(\overline{\Phi})}{F'(\overline{\Phi})}K'(\overline{\Phi})\big)
\overline{Q}^{mn}\overline{\Phi}_{,m}\frac{\partial}{\partial x^n}+3\overline{N}^2V'(\overline{\Phi})
+5\overline{N}^2\overline{D}^i\overline{D}_i F'(\overline{\Phi})+3\overline{N}(\overline{D}^i\overline{N})\overline{D}_i F'(\overline{\Phi})\nn\\
&&\quad
-\overline{N}^2\overline{R}^{(3)}F'(\overline{\Phi})\Bigg]\delta\Phi\nn\\
&&
+\Bigg[\frac{F(\overline{\Phi})}{2}\Big(\dot{\overline{Q}}_{r s}-(\mathcal{L}_{\overset{\rightharpoonup }{\overline{N}}}\overline{Q})_{r s}\Big)
\bigg[[\overline{G}^{-1}]^{rsmn}(\frac{\partial}{\partial \tau} - \mathcal{L}_{\overset{\rightharpoonup} {\overline{N}}})
-[\overline{G}^{-1}]^{turm}\overline{Q}^{sn}\Big(\dot{\overline{Q}}_{tu}-\mathcal{L}_{\overset{\rightharpoonup} {\overline{N}}}\overline{Q}_{tu}\Big)\nn\\
&&\qquad
+[\overline{G}^{-1}]^{rsvw}\Big(\overline{Q}^{mt}\overline{N}^n[\overline{Q}_{vw}]_{,t}+2\overline{Q}_{tw}\frac{\partial}{\partial x^v}(\overline{Q}^{mt}\overline{N}^n)\Big)\bigg]\nn\\
&&\quad
-F'(\overline{\Phi})\bigg[-\overline{Q}^{v m}\overline{Q}^{nw}\Big(\dot{\overline{Q}}_{vw}-(\mathcal{L}_{\overset{\rightharpoonup }{\overline{N}}}\overline{Q})_{vw}\Big)\big(\dot{\overline{\Phi}}-\mathcal{L}_{\overset{\rightharpoonup }{\overline{N}}}\overline{\Phi}\big)
+\overline{Q}^{mn}\big(\dot{\overline{\Phi}}-\mathcal{L}_{\overset{\rightharpoonup}{\overline{N}}}\overline{\Phi}\big)(\frac{\partial}{\partial\tau}- \mathcal{L}_{\overset{\rightharpoonup}{\overline{N}}})\nn\\
&&\qquad
+ \overline{Q}^{vw}\Big(\dot{\overline{Q}}_{vw}-(\mathcal{L}_{\overset{\rightharpoonup }{\overline{N}}}\overline{Q})_{vw}\Big)[\overline{Q}^{mr}\overline{\Phi}_{,r}N^n]\nn\\
&&\qquad
- \overline{Q}^{vw}\big(\dot{\overline{\Phi}}-\mathcal{L}_{\overset{\rightharpoonup }{\overline{N}}}\overline{\Phi}\big)\Big[-\overline{Q}^{mr}\overline{N}^n(\overline{Q}_{vw})_{,r}-\overline{Q}_{wr}\frac{\partial}{\partial x^v}(\overline{Q}^{mr}\overline{N}^n)-\overline{Q}_{vr}\frac{\partial}{\partial x^w}(\overline{Q}^{mr}\overline{N}^n)\Big]\bigg]\nn\\
&&\quad
-\overline{N}^m\overline{N}^n\big(2K(\overline{\Phi})+\frac{F(\overline{\Phi})}{F'(\overline{\Phi})}K'(\overline{\Phi})\big)\overline{\Phi}^{,i}
\overline{\Phi}_{,i}-\overline{N}^2\big(2K(\overline{\Phi})+\frac{F(\overline{\Phi})}{F'(\overline{\Phi})}K'(\overline{\Phi})\big)\overline{\Phi}^{,m}
\overline{\Phi}^{,n}\nn\\
&&\quad
-3V(\overline{\Phi})\overline{N}^m\overline{N}^n+5\Big[-\overline{N}^m\overline{N}^n\overline{Q}^{vw}(\overline{D}_v \overline{D}_w F(\overline{\Phi}))+\frac12\overline{N}^2\overline{Q}^{mn}(\overline{D}_wF(\overline{\Phi}))\overline{D}_v\overline{Q}^{vw}
\nn\\
&&\qquad
-\overline{N}^2\overline{Q}^{mr}(\overline{D}_wF(\overline{\Phi}))\overline{D}_r\overline{Q}^{nw}-\overline{N}^2\overline{Q}^{mv}
\overline{Q}^{nw}(\overline{D}_v\overline{D}_wF(\overline{\Phi}))\Big]\nn\\
&&\quad
+3\Big[-\frac12\overline{Q}^{vw}(\overline{D}_vF(\overline{\Phi}))\overline{D}_w(\overline{N}^m\overline{N}^n)-\frac12\overline{Q}^{mv}\overline{Q}^{nw}
(\overline{D}_v\overline{N}^2)(\overline{D}_wF(\overline{\Phi}))\Big]\nn\\
&&\quad
+F(\overline{\Phi})\overline{R}^{(3)}\overline{N}^m\overline{N}^n-\overline{N}^2F(\overline{\Phi})\Big[[\overline{G}^{-1}]^{mnvw}\overline{D}_v \overline{D}_w-\overline{R}^{mn}\Big]\nn\\
&&\quad
+\frac12\overline{H}(\sigma)\overline{Q}^{vw}\overline{N}_v \overline{N}_w\Big[-\frac12\frac{\overline{N}}{\sqrt{\det {\overline{Q}}}}(\overline{Q}^{mn}+\frac{\overline{N}^m\overline{N}^n}{\overline{N}^2})\Big]-\frac12\frac{\overline{N}}{\sqrt{\det {\overline{Q}}}}\overline{H}(\sigma)\overline{N}^m\overline{N}^n\Bigg]\delta Q_{mn}\nn\\
&&
+\Bigg[-\frac{F(\overline{\Phi})}{2}\Big(\dot{\overline{Q}}_{r s}-(\mathcal{L}_{\overset{\rightharpoonup }{\overline{N}}}\overline{Q})_{r s}\Big)[\overline{G}^{-1}]^{rsvn}\Big(\overline{Q}^{mt}[\overline{Q}_{vn}]_{,t}
+2\overline{Q}_{tv}\frac{\partial}{\partial x^n}(\overline{Q}^{mt})\Big)\nn\\
&&\quad
-F'(\overline{\Phi})\bigg[-\overline{Q}^{v w}\big(\dot{\overline{\Phi}}-\mathcal{L}_{\overset{\rightharpoonup }{\overline{N}}}\overline{\Phi}\big)\big[\overline{Q}^{mr}(\overline{Q}_{vw})_{,r}+\overline{Q}_{w r}\frac{\partial}{\partial x^v}(\overline{Q}^{mr})+\overline{Q}_{vr}\frac{\partial}{\partial x^w}(\overline{Q}^{mr})\big]\nn\\
&&\qquad
-\overline{Q}^{vw}\Big(\dot{\overline{Q}}_{vw}-(\mathcal{L}_{\overset{\rightharpoonup }{\overline{N}}}\overline{Q})_{vw}\Big)[\overline{Q}^{mn}\overline{\Phi}_{,n}]\bigg]\nn\\
&&\quad
+2\overline{N}^m\big(2K(\overline{\Phi})+\frac{F(\overline{\Phi})}{F'(\overline{\Phi})}K'(\overline{\Phi})\big)\overline{\Phi}^{,i}\overline{\Phi}_{,i}
+6V(\overline{\Phi})\overline{N}^m\nn\\
&&\quad
+10\big[\overline{N}^m\overline{Q}^{vw}(\overline{D}_v\overline{D}_wF(\overline{\Phi}))\big]+3\big[\overline{Q}^{vw}(\overline{D}_wF(\overline{\Phi}))
\overline{D}_v\overline{N}^m\big]
-2F(\overline{\Phi})\overline{R}^{(3)}\overline{N}^m\nn\\
&&\quad
+\frac{1}{2}\frac{\overline{N}}{\sqrt{\det \overline{Q}}}\overline{H}(\sigma )\overline{Q}^{rs}\overline{N}_r \overline{N}_s(\frac{\overline{N}^m}{\overline{N}^2})+\frac{\overline{N}}{\sqrt{\det \overline{Q}}}\overline{H}(\sigma )\overline{Q}^{mn}\overline{N}_n\Bigg]\delta N_m\nn\\
&&
+\Bigg[\frac12\frac{\overline{N}}{\sqrt{\det \overline{Q}}}\overline{Q}^{mn}\overline{N}_m\overline{N}_n\Bigg]\delta H.\label{deltaY}
\ea
Furthermore, we will need the perturbation of $C(\tau, \sigma)$ that is given by
\ba
\delta C&=&\Bigg[\frac{1}{4}\frac{\sqrt{\det \overline{Q}}}{\overline{N}^2} F(\overline{\Phi})' \left[\overline{G}^{-1}\right]^{r s t u}\Big(\dot{\overline{Q}}_{r s}-(\mathcal{L}_{\overset{\rightharpoonup }{\overline{N}}}\overline{Q})_{r s}\Big)\Big(\dot{\overline{Q}}_{t u}-(\mathcal{L}_{\overset{\rightharpoonup }{\overline{N}}}{\overline{Q}})_{t u}\Big)-\sqrt{\det \overline{Q}} \overline{R}^{(3)}F'(\overline{\Phi})\nn\\
&&\quad
+\sqrt{\det \overline{Q}} K'(\overline{\Phi})\overline{Q}^{jk}\overline{\Phi}_{,j}\overline{\Phi}_{,k}+2 \sqrt{\det \overline{Q}} K(\overline{\Phi}) \overline{Q}^{jk} \overline{\Phi}_{,j} \frac{\partial}{\partial x^k}- F'(\overline{\Phi}) \overline{Q}^{m n}\Big(\dot{\overline{Q}}_{m n}-(\mathcal{L}_{\overset{\rightharpoonup }{\overline{N}}}\overline{Q})_{m n}\Big)(\frac{\partial}{\partial \tau} - \mathcal{L}_{\overset{\rightharpoonup}{\overline{N}}})\nn\\
&&\quad
-F''(\overline{\Phi})\overline{Q}^{m n}\Big(\dot{\overline{Q}}_{m n}-(\mathcal{L}_{\overset{\rightharpoonup}{\overline{N}}}\overline{Q})_{m n}\Big)\big(\dot{\overline{\Phi}}-\mathcal{L}_{\overset{\rightharpoonup }{\overline{N}}}\overline{\Phi}\big)+ \frac{\sqrt{\det \overline{Q}}}{\overline{N}^2}K'(\overline{\Phi})\big(\dot{\overline{\Phi}}-\mathcal{L}_{\overset{\rightharpoonup }{\overline{N}}}\overline{\Phi}\big)^2
\nn\\
&&\quad
+2\frac{\sqrt{\det \overline{Q}}}{\overline{N}^2} K(\overline{\Phi})\big(\dot{\overline{\Phi}}-\mathcal{L}_{\overset{\rightharpoonup }{\overline{N}}}\overline{\Phi}\big)(\frac{\partial}{\partial \tau} - \mathcal{L}_{\overset{\rightharpoonup} {\overline{N}}})+2 \sqrt{\det\overline{Q}}\overline{Q}^{jk} \overline{D}_{j} \overline{D}_{k} F'(\overline{\Phi})+ \sqrt{\det \overline{Q}} V'(\overline{\Phi})\Bigg]\delta\Phi\nn\\
&&
+\Bigg[\frac12 \overline{Q}^{jk}\overline{C}+\frac{\overline{N}^j\overline{N}^k}{\overline{N}^2}\bigg(\frac{\sqrt{\det \overline{Q}}F(\overline{\Phi} )}{4\overline{N}^2}\left[\overline{G}^{-1}\right]^{rstu}\Big(\dot{\overline{Q}}_{rs}-(\mathcal{L}_{\overset{\rightharpoonup}{\overline{N}}}\overline{Q})_{r s}\Big)\Big(\dot{\overline{Q}}_{tu}-(\mathcal{L}_{\overset{\rightharpoonup }{\overline{N}}}\overline{Q})_{tu}\Big)\nn\\
&&\quad
-\frac{\sqrt{\det \overline{Q}}}{\overline{N}^2}F'(\overline{\Phi})\overline{Q}^{rs}\Big(\dot{\overline{Q}}_{r s}-(\mathcal{L}_{\overset{\rightharpoonup}{\overline{N}}}\overline{Q})_{r s}\Big)\big(\dot{\overline{\Phi}}-\mathcal{L}_{\overset{\rightharpoonup }{\overline{N}}}\overline{\Phi}\big)+\frac{\sqrt{\det\overline{Q}}}{\overline{N}^2}K(\overline{\Phi} )(\dot{\overline{\Phi}}-\mathcal{L}_{\overset{\rightharpoonup }{\overline{N}}}\overline{\Phi})^2\bigg)\nn\\
&&\quad
+\frac{\sqrt{\det \overline{Q}}}{\overline{N}^2}\frac{F(\overline{\Phi})}{2}\Big(\dot{\overline{Q}}_{rs}-(\mathcal{L}_{\overset{\rightharpoonup }{\overline{N}}}\overline{Q})_{r s}\Big)
\Big[[\overline{G}^{-1}]^{rsjk}(\frac{\partial}{\partial\tau}-\mathcal{L}_{\overset{\rightharpoonup}{\overline{N}}})
-[\overline{G}^{-1}]^{turj}\overline{Q}^{sk}(\dot{\overline{Q}}_{tu}-\mathcal{L}_{\overset{\rightharpoonup}{\overline{N}}}\overline{Q}_{tu})\nn\\
&&\qquad
+[\overline{G}^{-1}]^{rsmn}\Big(\overline{Q}^{jt}\overline{N}^k[\overline{Q}_{mn}]_{,t}+2\overline{Q}_{tn}\frac{\partial}{\partial x^m}(\overline{Q}^{jt}\overline{N}^k)\Big)\Big]-\sqrt{\det \overline{Q}}F(\overline{\Phi})\big [[\overline{G}^{-1}]^{jkmn}\overline{D}_m \overline{D}_n-\overline{R}^{jk}]\nn\\
&&\quad
-\frac{\sqrt{\det\overline{Q}}}{\overline{N}^2}K(\overline{\Phi})\overline{\Phi}^{,j}\overline{\Phi}^{,k}-F'(\overline{\Phi})\Big[-\overline{Q}^{m j}\overline{Q}^{k n}\Big(\dot{\overline{Q}}_{m n}-(\mathcal{L}_{\overset{\rightharpoonup}{\overline{N}}}\overline{Q})_{mn}\Big)\big(\dot{\overline{\Phi}}-\mathcal{L}_{\overset{\rightharpoonup }{\overline{N}}}\overline{\Phi}\big)\nn\\
&&\qquad
+\overline{Q}^{jk}\big(\dot{\overline{\Phi}}-\mathcal{L}_{\overset{\rightharpoonup }{\overline{N}}}\overline{\Phi}\big)(\frac{\partial}{\partial\tau}-\mathcal{L}_{\overset{\rightharpoonup} {\overline{N}}})+\overline{Q}^{m n}\Big(\dot{\overline{Q}}_{mn}-(\mathcal{L}_{\overset{\rightharpoonup }{\overline{N}}}\overline{Q})_{m n}\Big)[\overline{Q}^{jr}\overline{\Phi}_{,r}\overline{N}^k]\nn\\
&&\qquad
- \overline{Q}^{m n}\big(\dot{\overline{\Phi}}-\mathcal{L}_{\overset{\rightharpoonup }{\overline{N}}}\overline{\Phi}\big)\big[-\overline{Q}^{jr}\overline{N}^k(\overline{Q}_{mn})_{,r}-\overline{Q}_{nr}\frac{\partial}{\partial x^m}(\overline{Q}^{jr}\overline{N}^k)-\overline{Q}_{mr}\frac{\partial}{\partial x^n}(\overline{Q}^{jr}\overline{N}^k)\big]\Big]\nn\\
&&\quad
+2\frac{\sqrt{\det \overline{Q}}}{\overline{N}^2}K(\overline{\Phi})\big(\dot{\overline{\Phi}}-\mathcal{L}_{\overset{\rightharpoonup }{\overline{N}}}\overline{\Phi}\big)[\overline{Q}^{jm}\overline{\Phi}_{,m}\overline{N}^k]
+2\Big[\frac{1}{2}\sqrt{\det\overline{Q}}\overline{Q}^{j k}\Big(\overline{D}_nF(\overline{\Phi})\Big)\overline{D}_m\overline{Q}^{m n} \nn\\
&&\qquad
-\sqrt{\det \overline{Q}}\overline{Q}^{j m}\Big(\overline{D}_nF(\overline{\Phi})\Big)\overline{D}_{m}\overline{Q}^{k n} -\overline{Q}^{m j}\overline{Q}^{k n}\sqrt{\det \overline{Q}}\Big(\overline{D}_m\overline{D}_nF(\overline{\Phi})\Big)\Big]\Bigg]\delta Q_{jk}\nn\\
&&
+\Bigg[-2\frac{\overline{N}^j}{\overline{N}^2}\bigg(\frac{\sqrt{\det \overline{Q}}F(\overline{\Phi} )}{4\overline{N}^2}\left[\overline{G}^{-1}\right]^{r s t u}\Big(\dot{\overline{Q}}_{r s}-(\mathcal{L}_{\overset{\rightharpoonup }{\overline{N}}}\overline{Q})_{r s}\Big)\Big(\dot{\overline{Q}}_{t u}-(\mathcal{L}_{\overset{\rightharpoonup }{\overline{N}}}\overline{Q})_{t u}\Big)\nn\\
&&\qquad
-\frac{\sqrt{\det \overline{Q}}}{\overline{N}^2}\overline{F}'(\overline{\Phi})\overline{Q}^{rs}\Big(\dot{\overline{Q}}_{rs}-(\mathcal{L}_{\overset{\rightharpoonup}{\overline{N}}}\overline{Q})_{rs}\Big)
\big(\dot{\overline{\Phi}}-\mathcal{L}_{\overset{\rightharpoonup}{\overline{N}}}\overline{\Phi}\big)+\frac{\sqrt{\det\overline{Q}}}{\overline{N}^2}K(\overline{\Phi})(\dot{\overline{\Phi} }-\mathcal{L}_{\overset{\rightharpoonup}{\overline{N}}}\overline{\Phi})^2\bigg)\nn\\
&&\quad
-\frac{\sqrt{\det \overline{Q}}}{\overline{N}^2}\frac{F(\overline{\Phi})}{2}\Big(\dot{\overline{Q}}_{r s}-(\mathcal{L}_{\overset{\rightharpoonup }{\overline{N}}}\overline{Q})_{r s}\Big)[\overline{G}^{-1}]^{rsjn}\big[\overline{Q}^{mt}[\overline{Q}_{jn}]_{,t}+2\overline{Q}_{tj}\frac{\partial}{\partial x^n}(\overline{Q}^{mt})\big]\nn\\
&&\quad
-\frac{\sqrt{\det \overline{Q}}}{\overline{N}^2}F'(\overline{\Phi})\Big[-\overline{Q}^{jk}\big(\dot{\overline{\Phi}}-\mathcal{L}_{\overset{\rightharpoonup }{\overline{N}}}\overline{\Phi}\big)\big[\overline{Q}^{mn}[\overline{Q}_{jk}]_{,n}+\overline{Q}_{kn}\frac{\partial}{\partial x^j}(\overline{Q}^{mn})+\overline{Q}_{j n}\frac{\partial}{\partial x^k}(\overline{Q}^{mn})\big]\nn\\
&&\qquad
-\overline{Q}^{jk}\Big(\dot{\overline{Q}}_{jk}-(\mathcal{L}_{\overset{\rightharpoonup}{\overline{N}}}\overline{Q})_{jk}\Big)[\overline{Q}^{mn}\overline{\Phi}_{,n}]\Big]
+2\frac{\sqrt{\det \overline{Q}}}{\overline{N}^2}K(\overline{\Phi})\big(\dot{\overline{\Phi}}-\mathcal{L}_{\overset{\rightharpoonup }{\overline{N}}}\overline{\Phi}\big)[-\overline{Q}^{mn}\overline{\Phi}_{,n}]\Bigg]\delta N_m.\label{deltaC}
\ea
With the above formulas, we have all intermediate results available in order to derive the evolution equations for the linear perturbations $\delta \Phi$ and $\delta Q_{jk}$. We will start to discuss the case for $\delta \Phi$. First we write the evolution equation for $\delta\Phi$ in the following form
\ba
\delta \ddot{\Phi}=\textbf{A}\delta\Phi+\textbf{B}^{jk}\delta Q_{jk}+\textbf{C}^m\delta N_m+\textbf{D}\delta H.\nn
\ea
This is always possible with appropriate choices for the coefficients $\textbf{A}$, $\textbf{B}^{jk}$, $\textbf{C}^m$ and $\textbf{D}$. Using the  explicit form of the coefficients $\textbf{A}$, $\textbf{B}^{jk}$, $\textbf{C}^m$, $\textbf{D}$ in the above equation we obtain
\ba
\delta \ddot{\Phi}&=&\Bigg[\bigg[\frac{\dot{\overline{N}}}{\overline{N}}-\frac{\Big(\sqrt{\det \overline{Q}}\Big)^{\cdot}}{\sqrt{\det \overline{Q}}}+\frac{\overline{N}}{\sqrt{\det \overline{Q}}}\Big(\mathcal{L}_{\overset{\rightharpoonup }{\overline{N}}}(\frac{\sqrt{\det \overline{Q}} }{\overline{N}})\Big)\bigg](\frac{\partial}{\partial \tau}
-\mathcal{L}_{\overset{\rightharpoonup}{\overline{N}}})+\Big[\mathcal{L}_{\overset{\rightharpoonup} {\overline{N}}}(\frac{\partial}{\partial \tau}
-\mathcal{L}_{\overset{\rightharpoonup}{\overline{N}}})+\frac{\partial}{\partial \tau}
\mathcal{L}_{\overset{\rightharpoonup}{\overline{N}}}\Big]\nn\\
&&\quad
-\bigg(\frac{3F'(\overline{\Phi})F''(\overline{\Phi})+F(\overline{\Phi})K'(\overline{\Phi})}
{3(F'(\overline{\Phi}))^2+2F(\overline{\Phi})K(\overline{\Phi})}\bigg)'\big(\dot{\overline{\Phi}}-\mathcal{L}_{\overset{\rightharpoonup}{N}}
\overline{\Phi}\big)^2
-2\bigg(\frac{3F'(\overline{\Phi})F''(\overline{\Phi})+F(\overline{\Phi})K'(\overline{\Phi})}{3(F'(\overline{\Phi}))^2
+2F(\overline{\Phi})K(\overline{\Phi})}\bigg)\nn\\
&&\quad
\cdot\big(\dot{\overline{\Phi}}-\mathcal{L}_{\overset{\rightharpoonup}{\overline{N}}}\overline{\Phi}\big)
(\frac{\partial}{\partial \tau}- \mathcal{L}_{\overset{\rightharpoonup}{\overline{N}}})-\frac32\frac{\overline{N}^2}{\sqrt{\det \overline{Q}}}\bigg(\frac{F'(\overline{\Phi})}{3(F'(\overline{\Phi}))^2+2F(\overline{\Phi})K(\overline{\Phi})}\bigg)'\overline{C}(\tau, \sigma)\nn\\
&&\quad
+\bigg(\frac{2F(\overline{\Phi})}{3(F'(\overline{\Phi}))^2+2F(\overline{\Phi})K(\overline{\Phi})}\bigg)'\, \overline{X}+\bigg(\frac{F'(\overline{\Phi})}{3(F'(\overline{\Phi}))^2+2F(\overline{\Phi})K(\overline{\Phi})}\bigg)'\, \overline{Y}\Bigg]\delta \Phi\nn\\
&&
+\Bigg[\big(\dot{\overline{\Phi}}-\mathcal{L}_{\overset{\rightharpoonup }{\overline{N}}}\overline{\Phi}\big)\Big[-\frac12(\frac{\partial}{\partial \tau}-\mathcal{L}_{\overset{\rightharpoonup} {\overline{N}}})(\overline{Q}^{jk}+\frac{\overline{N}^j\overline{N}^k}{\overline{N}^2})-\frac{\overline{N}}{\sqrt{\det {\overline{Q}}}}\frac{\partial}{\partial x^m}(\frac{\sqrt{\det {\overline{Q}}}}{\overline{N}}\overline{Q}^{jm}\overline{N}^k)\Big]\nn\\
&&\quad
+(\overline{Q}^{jm}\overline{\Phi}_{,m}\overline{N}^k)\bigg[\frac{\dot{\overline{N}}}{\overline{N}}-\frac{\Big(\sqrt{\det\overline{Q}}\Big)^{\cdot }}{\sqrt{\det\overline{Q}}}+\frac{\overline{N}}{\sqrt{\det \overline{Q}}}\Big(\mathcal{L}_{\overset{\rightharpoonup }{\overline{N}}}(\frac{ \sqrt{\det \overline{Q}}}{\overline{N}})\Big)\bigg]\nn\\
&&\quad
+\Big[-\overline{Q}^{jm}\overline{Q}^{kn}\overline{N}_m\big(\dot{\overline{\Phi}}-\mathcal{L}_{\overset{\rightharpoonup }{\overline{N}}}\overline{\Phi}\big)_{,n}-(\frac{\partial}{\partial \tau}-\mathcal{L}_{\overset{\rightharpoonup} {\overline{N}}})\Big(\overline{Q}^{jm}\overline{Q}^{kn}\overline{N}_m\overline{\Phi}_{,n}\Big)\Big]\nn\\
&&\quad
-2\bigg(\frac{3F'(\overline{\Phi})F''(\overline{\Phi})+F(\overline{\Phi})K'(\overline{\Phi})}
{3(F'(\overline{\Phi}))^2+2F(\overline{\Phi})K(\overline{\Phi})}\bigg)\big(\dot{\overline{\Phi}}-\mathcal{L}_{\overset{\rightharpoonup }{\overline{N}}}\overline{\Phi}\big)[\overline{Q}^{jr}\overline{\Phi}_{,r}\overline{N}^k]\nn\\
&&\quad
+\frac32\bigg(\frac{F'(\overline{\Phi})}{3(F'(\overline{\Phi}))^2+2F(\overline{\Phi})K(\overline{\Phi})}\bigg)\bigg[\frac{1}{\sqrt{\det \overline{Q}}}\Big(\overline{N}^j\overline{N}^k+\frac{1}{2}\overline{N}^2\overline{Q}^{jk }\Big)\bigg]\overline{C}(\tau, \sigma)\Bigg]\delta Q_{jk}\nn\\
&&
+\Bigg[\big(\dot{\overline{\Phi}}-\mathcal{L}_{\overset{\rightharpoonup}{\overline{N}}}
\overline{\Phi}\big)\Big[(\frac{\partial}{\partial \tau}-\mathcal{L}_{\overset{\rightharpoonup} {\overline{N}}})\frac{\overline{N}^m}{\overline{N}^2}+\frac{\overline{N}}{\sqrt{\det {\overline{Q}}}}\frac{\partial}{\partial x^k}(\frac{\sqrt{\det {\overline{Q}}}}{\overline{N}}\overline{Q}^{mk})\Big]\nn\\
&&\quad
+\left[\frac{\dot{\overline{N}}}{\overline{N}}-\frac{\Big(\sqrt{\det \overline{Q}}\Big)^{\cdot}}{\sqrt{\det \overline{Q}}}+\frac{\overline{N}}{\sqrt{\det \overline{Q}}}\Big(\mathcal{L}_{\overset{\rightharpoonup }{\overline{N}}}(\frac{ \sqrt{\det \overline{Q}} }{\overline{N}})\Big)\right][-\overline{Q}^{mk}\overline{\Phi}_{,k}]\nn\\
&&\quad
+\bigg[\overline{Q}^{mk}\big(\dot{\overline{\Phi}}-\mathcal{L}_{\overset{\rightharpoonup }{\overline{N}}}\overline{\Phi}\big)_{,k}+(\frac{\partial}{\partial \tau} - \mathcal{L}_{\overset{\rightharpoonup} {\overline{N}}})(\overline{Q}^{mk}\overline{\Phi}_{,k})\bigg]\nn\\
&&\quad
+2\bigg(\frac{3F'(\overline{\Phi})F''(\overline{\Phi})+F(\overline{\Phi})K'(\overline{\Phi})}
{3(F'(\overline{\Phi}))^2+2F(\overline{\Phi})K(\overline{\Phi})}\bigg)\big(\dot{\overline{\Phi}}-\mathcal{L}_{\overset{\rightharpoonup }{\overline{N}}}\overline{\Phi}\big)[\overline{Q}^{mn}\overline{\Phi}_{,n}]\nn\\
&&\quad
-\frac{3\overline{N}^m}{\sqrt{\det \overline{Q}}}\frac{F'(\overline{\Phi})}{3(F'(\overline{\Phi}))^2+2F(\overline{\Phi})K(\overline{\Phi})}\overline{C}(\tau, \sigma)\Bigg]\delta N_m\nn\\
&&
-\frac32\frac{\overline{N}^2}{\sqrt{\det \overline{Q}}}\bigg(\frac{F'(\overline{\Phi})}{3(F'(\overline{\Phi}))^2+2F(\overline{\Phi})K(\overline{\Phi})}\bigg)\delta C+\bigg(\frac{2F(\overline{\Phi})}{3(F'(\overline{\Phi}))^2+2F(\overline{\Phi})K(\overline{\Phi})}\bigg)\delta X\nn\\
&&\
+\bigg(\frac{F'(\overline{\Phi})}{3(F'(\overline{\Phi}))^2+2F(\overline{\Phi})K(\overline{\Phi})}\bigg)\delta Y,\label{deltaphi}
\ea
Notice that we did not write down the explicit expressions of $\delta C$, $\delta X$ and $\delta Y$ here in order to show the final equation of motion in more compact form. Now we consider the evolution equations for the linear perturbation $\delta Q_{jk}$. Analogue to the case of $\delta\Phi$ we can also formulate the final evolution equations in the following form:
\ba
\delta Q_{jk}=\textbf{U}_{jk}\delta\Phi+\textbf{V}\delta Q_{jk}+\textbf{W}_{jk}^{mn}\delta Q_{mn}+\textbf{X}^{m}_{(j}\delta Q_{k)m}+\textbf{Y}_{jk}^{m}\delta N_{m}+\textbf{Z}\delta H,\nn
\ea
with appropriate choices for the coefficients $\textbf{U}_{jk}$, $\textbf{V}$, $\textbf{W}_{jk}^{mn}$, $\textbf{X}^{m}_{(j}$, $\textbf{Y}_{jk}^{m}$ and $\textbf{Z}$. Using the explicit form of these coefficients the final evolution equation is given by
\ba
\delta \ddot{Q}_{jk}&=&\Bigg[2\overline{N}^2\overline{\Phi}_{,j}
\overline{\Phi}_{,k}\Big(\frac{K(\overline{\Phi})}{F(\overline{\Phi})}\Big)'+4\overline{N}^2\frac{K(\overline{\Phi})}{F(\overline{\Phi})}
\overline{\Phi}_{,j}\frac{\partial}{\partial x^k}-\Big(\dot{\overline{Q}}_{jk}-(\mathcal{L}_{\overset{\rightharpoonup }{\overline{N}}}\overline{Q})_{jk}\Big)\big(\dot{\overline{\Phi}}-\mathcal{L}_{\overset{\rightharpoonup }{\overline{N}}}\overline{\Phi}\big)\Big(\frac{F'(\overline{\Phi})}{F(\overline{\Phi})}\Big)'\nn\\
&&\quad
-\frac{F'(\overline{\Phi})}{F(\overline{\Phi})}
\Big(\dot{\overline{Q}}_{jk}-(\mathcal{L}_{\overset{\rightharpoonup }{\overline{N}}}\overline{Q})_{jk}\Big)(\frac{\partial}{\partial \tau} - \mathcal{L}_{\overset{\rightharpoonup} {\overline{N}}})-2\overline{N}^2\frac{F'(\overline{\Phi})}{(F(\overline{\Phi}))^2}\big(\overline{D}_j \overline{D}_k F(\overline{\Phi})\big)+\frac{2\overline{N}^2}{F(\overline{\Phi})}\overline{D}_j \overline{D}_k F'(\overline{\Phi})\nn\\
&&\quad
+\frac{\overline{N}F'(\overline{\Phi})}{\sqrt{\det \overline{Q}}F^2(\overline{\Phi})}\overline{H}(\sigma)\overline{G}_{jkmn}\overline{N}^m\overline{N}^n
-\overline{Q}_{jk}\frac{F'(\overline{\Phi})}{(F(\overline{\Phi}))^2}\Big[\overline{N}(\overline{D}^i\overline{N})(\overline{D}_i F(\overline{\Phi}))+\overline{N}^2(\overline{D}^i\overline{D}_i F(\overline{\Phi}))\nn\\
&&\qquad
+\overline{N}^2V(\overline{\Phi})\Big]+\frac{\overline{Q}_{jk}}{F(\overline{\Phi})}\Big[\overline{N}(\overline{D}^i\overline{N})\overline{D}_i F'(\overline{\Phi})+\overline{N}^2 \overline{D}^i\overline{D}_i F'(\overline{\Phi})+\overline{N}^2V'(\overline{\Phi})\Big]\nn\\
&&\quad
+\bigg(\frac{-2K(\overline{\Phi})F''(\overline{\Phi})+K'(\overline{\Phi})F'(\overline{\Phi})}{3(F'(\overline{\Phi}))^2
+2F(\overline{\Phi})K(\overline{\Phi})}\bigg)'
\big(\dot{\overline{\Phi}}-\mathcal{L}_{\overset{\rightharpoonup}{\overline{N}}}\overline{\Phi} \big)^2\overline{Q}_{jk}+2\overline{Q}_{jk}\bigg(\frac{-2K(\overline{\Phi})F''(\overline{\Phi})
+K'(\overline{\Phi})F'(\overline{\Phi})}{3(F'(\overline{\Phi}))^2+2F(\overline{\Phi})K(\overline{\Phi})}\bigg)\nn\\
&&\quad
\cdot\big(\dot{\overline{\Phi}}-\mathcal{L}_{\overset{\rightharpoonup }{\overline{N}}}\overline{\Phi} \big)(\frac{\partial}{\partial \tau} - \mathcal{L}_{\overset{\rightharpoonup} {\overline{N}}})
-\frac{\overline{N}^2}{\sqrt{\det \overline{Q}}}\overline{Q}_{jk}\bigg(\frac{K(\overline{\Phi})}{3(F'(\overline{\Phi}))^2
+2F(\overline{\Phi})K(\overline{\Phi})}\bigg)'\overline{C}(\tau, \sigma)\nn\\
&&\quad
-2\overline{Q}_{jk}\bigg(\frac{F'(\overline{\Phi})}{3(F'(\overline{\Phi}))^2+2F(\overline{\Phi})K(\overline{\Phi})}\bigg)'\, \overline{X}-\overline{Q}_{jk}\bigg(\frac{(F'(\overline{\Phi}))^2}{F(\overline{\Phi})\big[3(F'(\overline{\Phi}))^2
+2F(\overline{\Phi})K(\overline{\Phi})\big]}\bigg)'\, \overline{Y}\Bigg]\delta \Phi\nn\\
&&
+\Bigg[\bigg[\frac{\dot{\overline{N}}}{\overline{N}}-\frac{\Big(\sqrt{\det\overline{ Q}}\Big)^{\cdot}}{\sqrt{\det \overline{Q}}}+\frac{\overline{N}}{\sqrt{\det \overline{Q}}}\Big(\mathcal{L}_{\overset{\rightharpoonup }{\overline{N}}}(\frac{ \sqrt{\det \overline{Q}} }{\overline{N}})\Big)\bigg](\frac{\partial}{\partial\tau}-\mathcal{L}_{\overset{\rightharpoonup} {\overline{N}}})
-2\overline{N}^2[-\frac12\overline{D}_m \overline{D}_n\overline{Q}^{mn}]\nn\\
&&\quad
+\big[\overline{N}(\overline{D}_n\overline{N})\overline{D}_m\overline{Q}^{mn}\big]+\big[\mathcal{L}_{\overset{\rightharpoonup }{\overline{N}}}(\frac{\partial}{\partial \tau}-\mathcal{L}_{\overset{\rightharpoonup} {\overline{N}}})+\frac{\partial}{\partial \tau} \mathcal{L}_{\overset{\rightharpoonup} {\overline{N}}}\big]-\frac{F'(\overline{\Phi})}{F(\overline{\Phi})}\big(\dot{\overline{\Phi}}-\mathcal{L}_{\overset{\rightharpoonup }{\overline{N}}}\overline{\Phi}\big)(\frac{\partial}{\partial \tau}-\mathcal{L}_{\overset{\rightharpoonup}{\overline{N}}})\nn\\
&&\quad
+\frac{2\overline{N}^2}{F(\overline{\Phi})}\Big[\frac12(\overline{D}_n F(\overline{\Phi}))\overline{D}_m\overline{Q}^{mn}\Big]
+\Big[-\frac{1}{2}\frac{\overline{N}}{F(\overline{\Phi})\sqrt{\det {\overline{Q}}} }\overline{H} \overline{N}^m \overline{N}^n \overline{Q}_{mn}\Big]\nn\\
&&\quad
+\frac{1}{F(\overline{\Phi})}\Big[\overline{N}(\overline{D}^i\overline{N})(\overline{D}_iF(\overline{\Phi}))+\overline{N}^2 (\overline{D}^i\overline{D}_iF(\overline{\Phi}))+\overline{N}^2V(\overline{\Phi})\Big]\nn\\
&&\quad
+\bigg(\frac{-2K(\overline{\Phi})F''(\overline{\Phi})+K'(\overline{\Phi})F'(\overline{\Phi})}{3(F'(\overline{\Phi}))^2
+2F(\overline{\Phi})K(\overline{\Phi})}\bigg)
\big(\dot{\overline{\Phi}}-\mathcal{L}_{\overset{\rightharpoonup }{\overline{N}}}\overline{\Phi}\big)^2
-\frac{\overline{N}^2}{\sqrt{\det \overline{Q}}}\bigg(\frac{K(\overline{\Phi})}{3(F'(\overline{\Phi}))^2
+2F(\overline{\Phi})K(\overline{\Phi})}\bigg)\overline{C}(\tau, \sigma)\nn\\
&&\quad
-2\bigg(\frac{F'(\overline{\Phi})}{3(F'(\overline{\Phi}))^2+2F(\overline{\Phi})K(\overline{\Phi})}\bigg)\, \overline{X}-\bigg(\frac{(F'(\overline{\Phi}))^2}{F(\overline{\Phi})\big[3(F'(\overline{\Phi}))^2+2F(\overline{\Phi})K(\overline{\Phi})\big]}\bigg)\, \overline{Y}\Bigg]\delta Q_{jk}\nn\\
&&
+\Bigg[\Big(\dot{\overline{Q}}_{jk}-(\mathcal{L}_{\overset{\rightharpoonup }{\overline{N}}}\overline{Q})_{jk}\Big)\Big[-\frac12(\frac{\partial}{\partial \tau}-\mathcal{L}_{\overset{\rightharpoonup} {\overline{N}}})(\overline{Q}^{mn}+\frac{\overline{N}^m\overline{N}^n}{\overline{N}^2})-\frac{\overline{N}}{\sqrt{\det {\overline{Q}}}}\frac{\partial}{\partial x^r}(\frac{\sqrt{\det {\overline{Q}}}}{\overline{N}}\overline{Q}^{mr}\overline{N}^n)\Big]\nn\\
&&\quad
+\bigg[\frac{\dot{\overline{N}}}{\overline{N}}-\frac{\Big(\sqrt{\det \overline{Q}}\Big)^{\cdot }}{\sqrt{\det \overline{Q}}}+\frac{\overline{N}}{\sqrt{\det \overline{Q}}}\Big(\mathcal{L}_{\overset{\rightharpoonup }{\overline{N}}}(\frac{ \sqrt{\det \overline{Q}} }{\overline{N}})\Big)\bigg]\bigg(-\overline{Q}^{mr}\overline{N}^n(\overline{Q}_{jk})_{,r}-\overline{Q}_{k r}\frac{\partial}{\partial x^j}(\overline{Q}^{mr}\overline{N}^n)\nn\\
&&\qquad
-\overline{Q}_{jr}\frac{\partial}{\partial x^k}(\overline{Q}^{mr}\overline{N}^n)\bigg)+\bigg[-\overline{Q}^{mr}\overline{Q}^{ns}\Big(\dot{\overline{Q}}_{rj}-(\mathcal{L}_{\overset{\rightharpoonup }{\overline{N}}}\overline{Q})_{r j}\Big)\Big(\dot{\overline{Q}}_{sk}-(\mathcal{L}_{\overset{\rightharpoonup }{\overline{N}}}{\overline{Q}})_{sk}\Big)\nn\\
&&\qquad
+(-2\overline{Q}^{tu})\Big(\dot{\overline{Q}}_{t(k}-(\mathcal{L}_{\overset{\rightharpoonup }{\overline{N}}}\overline{Q})_{t(k}\Big)\Big(-\overline{Q}^{mr}\overline{N}^n(\overline{Q}_{j)u})_{,r}-\overline{Q}_{ur}\frac{\partial}{\partial x^{j)}}(\overline{Q}^{mr}\overline{N}^n)-\overline{Q}_{j)r}\frac{\partial}{\partial x^u}(\overline{Q}^{mr}\overline{N}^n)\Big)\bigg]\nn\\
&&\quad
-\overline{N}^m\overline{N}^n
\Big(\frac{2K(\overline{\Phi})}{F(\overline{\Phi})}\overline{\Phi}_{,j}\overline{\Phi}_{,k}-2\overline{R}_{jk}\Big)-2\overline{N}^2
[-\frac12\overline{D}_j\overline{D}_k\overline{Q}^{mn}]
+\Big[-2\frac{\overline{N}^m\overline{N}^n}{\overline{N}^2}\overline{N}(\overline{D}_j\overline{D}_k\overline{N})\nn\\
&&\qquad
-\overline{N}^2
\overline{D}_j\overline{D}_k(\frac{\overline{N}^m\overline{N}^n}{\overline{N}^2})\Big]
+\bigg[-\overline{Q}^{mr}\overline{N}^n\big[\dot{\overline{Q}}_{jk}-(\mathcal{L}_{\overset{\rightharpoonup }{\overline{N}}}\overline{Q})_{jk}\big]_{,r}-\Big(\dot{\overline{Q}}_{kr}-(\mathcal{L}_{\overset{\rightharpoonup }{\overline{N}}}\overline{Q})_{kr}\Big)\frac{\partial}{\partial x^j}(\overline{Q}^{mr}\overline{N}^n)\nn\\
&&\qquad
-\Big(\dot{\overline{Q}}_{jr}-(\mathcal{L}_{\overset{\rightharpoonup }{\overline{N}}}\overline{Q})_{jr}\Big)\frac{\partial}{\partial x^k}(\overline{Q}^{mr}\overline{N}^n)+(\frac{\partial}{\partial\tau}-\mathcal{L}_{\overset{\rightharpoonup} {\overline{N}}})\big[-\overline{Q}^{mr}\overline{N}^n(\overline{Q}_{jk})_{,r}-\overline{Q}_{k r}\frac{\partial}{\partial x^j}(\overline{Q}^{mr}\overline{N}^n)\nn\\
&&\qquad
-\overline{Q}_{jr}\frac{\partial}{\partial x^k}(\overline{Q}^{mr}\overline{N}^n)\big]\bigg]-\frac{F'(\overline{\Phi})}{F(\overline{\Phi})}\big(\dot{\overline{Q}}_{jk}
-(\mathcal{L}_{\overset{\rightharpoonup }{\overline{N}}}\overline{Q})_{jk}\big)[\overline{Q}^{mr}\overline{\Phi}_{,r}\overline{N}^n]-\frac{F'(\overline{\Phi})}{F(\overline{\Phi})
}\big(\dot{\overline{\Phi}}-\mathcal{L}_{\overset{\rightharpoonup }{\overline{N}}}\overline{\Phi}\big)\nn\\
&&\qquad
\cdot\Big[\overline{Q}^{mr}\overline{N}^n[\overline{Q}_{jk}]_{,r}+\overline{Q}_{rk}\frac{\partial}{\partial x^j}(\overline{Q}^{mr}\overline{N}^n)+\overline{Q}_{rj}\frac{\partial}{\partial x^k}(\overline{Q}^{mr}\overline{N}^n)\Big]-\frac{2}{F(\overline{\Phi})}\overline{N}^m\overline{N}^n(\overline{D}_j\overline{D}_kF(\overline{\Phi}))\nn\\
&&\quad
+\Big[\frac{1}{2}\frac{\overline{N}}{F(\overline{\Phi})\sqrt{\det {\overline{Q}}}}(\overline{Q}^{m n}+\frac{\overline{N}^m\overline{N}^n}{\overline{N}^2})\overline{H} \overline{G}_{jkrs}\overline{N}^r \overline{N}^s+\frac{1}{2}\frac{\overline{N}}{F(\overline{\Phi})\sqrt{\det {\overline{Q}}}}\overline{H} \overline{N}^m\overline{N}^n \overline{Q}_{jk}\nn\\
&&\qquad
+2\frac{\overline{N}}{F(\overline{\Phi})\sqrt{\det {\overline{Q}}}}\overline{H} \overline{G}_{jkrs} \overline{N}^n \overline{N}^s \overline{Q}^{rm}\Big]+\frac{\overline{Q}_{jk}}{F(\overline{\Phi})}\Big[-\frac12\overline{Q}^{rs}
(\overline{D}_sF(\overline{\Phi}))\overline{D}_r(\overline{N}^m\overline{N}^n)\nn\\
&&\qquad
-\frac12\overline{Q}^{mr}\overline{Q}^{ns}(\overline{D}_r\overline{N}^2)(\overline{D}_sF(\overline{\Phi}))-\overline{N}^m\overline{N}^n\overline{Q}^{rs}
(\overline{D}_r\overline{D}_sF(\overline{\Phi}))+\frac12\overline{N}^2\overline{Q}^{mn}(\overline{D}_sF(\overline{\Phi}))\overline{D}_r\overline{Q}^{rs}\nn\\
&&\qquad
-\overline{N}^2\overline{Q}^{nr}
(\overline{D}_sF(\overline{\Phi}))\overline{D}_r\overline{Q}^{sm}-\overline{Q}^{mr}\overline{Q}^{sn}\overline{N}^2(\overline{D}_r\overline{D}_s
F(\overline{\Phi}))-\overline{N}^m\overline{N}^nV(\overline{\Phi})\Big]\nn\\
&&\quad
+2\overline{Q}_{jk}\bigg(\frac{-2K(\overline{\Phi})F''(\overline{\Phi})+K'(\overline{\Phi})F'(\overline{\Phi})}{3(F'(\overline{\Phi}))^2
+2F(\overline{\Phi})K(\overline{\Phi})}\bigg)
\big(\dot{\overline{\Phi}}-\mathcal{L}_{\overset{\rightharpoonup}{\overline{N}}}\overline{\Phi}\big)[\overline{Q}^{mr}\overline{\Phi}_{,r}\overline{N}^n]\nn\\
&&\quad
+\frac{K(\overline{\Phi})\overline{Q}_{jk}}{3(F'(\overline{\Phi}))^2+2F(\overline{\Phi})K(\overline{\Phi})}\Big[\frac{1}{\sqrt{\det \overline{Q}}}\Big(\overline{N}^m\overline{N}^n+\frac{1}{2}\overline{N}^2\overline{Q}^{mn }\Big)\Big]\overline{C}(\tau, \sigma)\Bigg]\delta Q_{mn}\nn\\
&&
+\Bigg[\Big[2\overline{Q}^{mn}\Big(\dot{\overline{Q}}_{n(k}-(\mathcal{L}_{\overset{\rightharpoonup }{\overline{N}}}\overline{Q})_{n(k}\Big)(\frac{\partial}{\partial \tau} - \mathcal{L}_{\overset{\rightharpoonup} {\overline{N}}})\Big]-2\overline{N}^2\Big[\overline{D}_n\overline{D}_{(k}\overline{Q}^{mn}\Big]-2\overline{N}(D_n\overline{N})D_{(k}\overline{Q}^{mn}\nn\\
&&\quad
+\frac{2\overline{N}^2}
{F(\overline{\Phi})}\Big[-(\overline{D}_nF(\overline{\Phi}))\overline{D}_{(k}\overline{Q}^{mn}\Big]+\Big[-\frac{2\overline{N}}{F(\overline{\Phi})\sqrt{\det {\overline{Q}}}}\overline{H} \overline{N}^r \overline{N}^t \overline{Q}^{s m} \overline{G} _{r t s (k}\Big]\Bigg]\delta Q_{j)m}\nn\\
&&
+\Bigg[\Big(\dot{\overline{Q}}_{jk}-(\mathcal{L}_{\overset{\rightharpoonup }{\overline{N}}}\overline{Q})_{jk}\Big)\Big[(\frac{\partial}{\partial \tau} - \mathcal{L}_{\overset{\rightharpoonup} {\overline{N}}})\frac{\overline{N}^m}{\overline{N}^2}\Big]+\Big(\dot{\overline{Q}}_{jk}-(\mathcal{L}_{\overset{\rightharpoonup }{\overline{N}}}\overline{Q})_{jk}\Big)\Big[\frac{\overline{N}}{\sqrt{\det \overline{Q}}}\frac{\partial}{\partial x^n}(\frac{\sqrt{\det \overline{Q}}}{\overline{N}}\overline{Q}^{mn})\Big]\nn\\
&&\quad
+\bigg[\frac{\dot{\overline{N}}}{\overline{N}}-\frac{\Big(\sqrt{\det\overline{ Q}}\Big)^{\cdot }}{\sqrt{\det\overline{ Q}}}+\frac{\overline{N}}{\sqrt{\det \overline{Q}}}\Big(\mathcal{L}_{\overset{\rightharpoonup }{\overline{N}}}(\frac{ \sqrt{\det \overline{Q}} }{\overline{N}})\Big)\bigg]\Big[-\overline{Q}^{mr}(\overline{Q}_{jk})_{,r}-\overline{Q}_{k r}\frac{\partial}{\partial x^j}(\overline{Q}^{mr})-\overline{Q}_{jr}\frac{\partial}{\partial x^k}(\overline{Q}^{mr})\Big]\nn\\
&&\quad
+\bigg[\Big(-2\overline{Q}^{tu}\Big(\dot{\overline{Q}}_{t(k}-(\mathcal{L}_{\overset{\rightharpoonup }{\overline{N}}}\overline{Q})_{t(k}\Big)\Big)\Big(\overline{Q}^{mn}(\overline{Q}_{j)u})_{,n}+\overline{Q}_{un}\frac{\partial}{\partial x^{j)}}(\overline{Q}^{mn})+\overline{Q}_{j)n}\frac{\partial}{\partial x^u}(\overline{Q}^{mn})\Big)\bigg]\nn\\
&&\quad
+2\overline{N}^m\Big(\frac{2K(\overline{\Phi})}{F(\overline{\Phi})}\overline{\Phi } _{, j}\overline{\Phi} _{, k}-2\overline{R}_{j k}\Big)+\Big[\frac{\overline{N}^m}{\overline{N}^2}\Big(4\overline{N}(\overline{D}_j\overline{D}_k\overline{N})\Big)+2\overline{N}^2
\overline{D}_j\overline{D}_k(\frac{\overline{N}^m}{\overline{N}^2})\Big]\nn\\
&&\quad
+\bigg[\overline{Q}^{mn}\big[\dot{\overline{Q}}_{jk}-(\mathcal{L}_{\overset{\rightharpoonup }{\overline{N}}}\overline{Q})_{jk}\big]_{,n}+\big(\dot{\overline{Q}}_{kn}-(\mathcal{L}_{\overset{\rightharpoonup }{\overline{N}}}\overline{Q})_{kn}\big)\frac{\partial}{\partial x^j}(\overline{Q}^{mn})+\big(\dot{\overline{Q}}_{jn}-(\mathcal{L}_{\overset{\rightharpoonup }{\overline{N}}}\overline{Q})_{jn}\big)\frac{\partial}{\partial x^k}(\overline{Q}^{mn})\nn\\
&&\qquad
+(\frac{\partial}{\partial \tau} - \mathcal{L}_{\overset{\rightharpoonup} {\overline{N}}})\Big(\overline{Q}^{mn}[\overline{Q}_{jk}]_{,n}+\overline{Q}_{kn}\frac{\partial}{\partial x^j}(\overline{Q}^{mn})+\overline{Q}_{jn}\frac{\partial}{\partial x^k}(\overline{Q}^{mn})\Big)\bigg]\nn\\
&&\quad
+\frac{F'(\overline{\Phi})}{F(\overline{\Phi})}\big(\dot{\overline{\Phi}}-\mathcal{L}_{\overset{\rightharpoonup }{\overline{N}}}\overline{\Phi}\big)\big[\overline{Q}^{mn}[\overline{Q}_{jk}]_{,n}+\overline{Q}_{kn}\frac{\partial}{\partial x^j}(\overline{Q}^{mn})+\overline{Q}_{j n}\frac{\partial}{\partial x^k}(\overline{Q}^{mn})\big]\nn\\
&&\quad
+\frac{F'(\overline{\Phi})}{F(\overline{\Phi})}\Big(\dot{\overline{Q}}_{jk}-(\mathcal{L}_{\overset{\rightharpoonup }{\overline{N}}}\overline{Q})_{jk}\Big)(\overline{Q}^{mr}\overline{\Phi}_{,r})+\frac{4\overline{N}^m}{F(\overline{\Phi})}
(\overline{D}_j\overline{D}_kF(\overline{\Phi}))\nn\\
&&\quad
+\Big[-\frac{\overline{N}}{F(\overline{\Phi})\sqrt{\det {\overline{Q}}}}\frac{\overline{N}^m}{\overline{N}^2}\overline{H}(\sigma) \overline{G}_{jkrs}\overline{N}^r \overline{N}^s
-2\frac{\overline{N}}{F(\overline{\Phi})\sqrt{\det {\overline{Q}}}}\overline{H}(\sigma) \overline{G}_{jkrs}\overline{N}^s \overline{Q}^{rm}\Big]\nn\\
&&\quad
+\frac{\overline{Q}_{jk}}{F(\overline{\Phi})}\Big[2\overline{N}^m\overline{Q}^{rs}(\overline{D}_r\overline{D}_sF(\overline{\Phi}))
+\overline{Q}^{rs}(\overline{D}_sF(\overline{\Phi}))\overline{D}_r\overline{N}^m+2\overline{N}^mV(\overline{\Phi})\Big]\nn\\
&&\quad
-2\overline{Q}_{jk}\bigg(\frac{-2K(\overline{\Phi})F''(\overline{\Phi})+K'(\overline{\Phi})F'(\overline{\Phi})}{3(F'(\overline{\Phi}))^2
+2F(\overline{\Phi})K(\overline{\Phi})}\bigg)
\big(\dot{\overline{\Phi}}-\mathcal{L}_{\overset{\rightharpoonup }{\overline{N}}}\overline{\Phi}\big)\big[\overline{Q}^{mn}\overline{\Phi}_{,n}\big]\nn\\
&&\quad
-2\frac{\overline{N}^m}{\sqrt{\det \overline{Q}}}\frac{K(\overline{\Phi})\overline{Q}_{jk}}{3(F'(\overline{\Phi}))^2+2F(\overline{\Phi})K(\overline{\Phi})}\overline{C}(\tau, \sigma)\Bigg]\delta N_m\nn\\
&&
+\Bigg[-\frac{\overline{N}}{F(\overline{\Phi})\sqrt{\det \overline{Q}}}\overline{N}^n\overline{N}^m\overline{G}_{jkmn}\Bigg]\delta H-\frac{\overline{N}^2}{\sqrt{\det \overline{Q}}}\frac{K(\overline{\Phi})\overline{Q}_{jk}}{3(F'(\overline{\Phi}))^2+2F(\overline{\Phi})K(\overline{\Phi})}\delta C\nn\\
&&
-\frac{2F'(\overline{\Phi})\overline{Q}_{jk}}{3(F'(\overline{\Phi}))^2+2F(\overline{\Phi})K(\overline{\Phi})}\delta X
-\frac{(F'(\overline{\Phi}))^2\overline{Q}_{jk}}{F(\overline{\Phi})\big[3(F'(\overline{\Phi}))^2+2F(\overline{\Phi})K(\overline{\Phi})\big]}\delta Y.\label{deltaQ}
\ea
The long and complicated equations (\ref{deltaphi}) and (\ref{deltaQ}) describe the evolution of the linearly perturbed scalar and
metric fields when the background metric are solutions of the STT equations of motion without any further symmetry requirements. Therefore these equations describe the evolution of the linear perturbations for general, and not only cosmological, spacetimes in the context of STT theories. The evolution equations in (\ref{deltaphi}) and (\ref{deltaQ}) are the generalization of the evolution equations (6.13) and (6.31) in \cite{Giesel10} to the framework of STT theories.
However, for the purpose of practical use, in the next section we will choose a specific background, namely a cosmological one, and study the detailed behavior of the perturbation equations. This also allows to compare the results obtained in this manifestly gauge invariant framework with already existing results in the literature obtained in the standard framework of cosmological perturbation theory.

\section{Cosmological Perturbations on FRW Background }
In this section we specialize the general relativistic equations from the last section to homogenous and isotropic FRW backgrounds and compare them with the already existing results in the literature. For simplicity, we only focus on the spatially flat case.
 The equations are greatly simplified because the spatial derivatives of background variables in Eqs. (\ref{deltaphi}) and (\ref{deltaQ}) vanish, and we have $\overline{N}_j=\frac{\overline{C}_j}{\overline{H}}=0$ and  $\overline{N}=\frac{\overline{C}}{\overline{H}}=\sqrt{1+\overline{Q}^{jk}\overline{N}_j\overline{N}_k}=1$. We denote $\overline{Q}_{jk}\equiv A^2\delta_{jk}$ where $A$ represents the gauge invariant scale factor. Then from Eqs. (\ref{phisecond}) and (\ref{qsecond}) we obtain the following evolution equations for the background variables
\ba
\ddot{\overline{\Phi}}&=&-3\frac{\dot{A}}{A}\dot{\overline{\Phi}}-\Big(\frac{F'(\overline{\Phi})K(\overline{\Phi})+F(\overline{\Phi})K'(\overline{\Phi})
+3F'(\overline{\Phi})F''(\overline{\Phi})}{3(F'(\overline{\Phi}))^2+2F(\overline{\Phi})K(\overline{\Phi})}\Big)(\dot{\overline{\Phi}})^2
+\frac{2F'(\overline{\Phi})V(\overline{\Phi})-F(\overline{\Phi})V'(\overline{\Phi})}{3(F'(\overline{\Phi}))^2+2F(\overline{\Phi})K(\overline{\Phi})}\nn\\
&&
-\frac{F'(\overline{\Phi})}{2\big(3(F'(\overline{\Phi}))^2+2F(\overline{\Phi})K(\overline{\Phi})\big)}\frac{\epsilon}{A^3},\label{ddotphi}\\
\frac{\ddot{A}}{A}&=&\frac{F'(\overline{\Phi})}{F(\overline{\Phi})}\frac{\dot{A}}{A}\dot{\overline{\Phi}}
-\frac{\Big(K(\overline{\Phi})F'(\overline{\Phi})^2
+\frac43F(\overline{\Phi})K^2(\overline{\Phi})+2F(\overline{\Phi})K(\overline{\Phi})F''(\overline{\Phi})
-F'(\overline{\Phi})F(\overline{\Phi})K'(\overline{\Phi})\Big)}{2F(\overline{\Phi})\big(3(F'(\overline{\Phi}))^2
+2F(\overline{\Phi})K(\overline{\Phi})\big)}(\dot{\overline{\Phi}})^2\nn\\
&&
+\Big(\frac16-\frac{(F'(\overline{\Phi}))^2}{3(F'(\overline{\Phi}))^2+2F(\overline{\Phi})K(\overline{\Phi})}\Big)
\frac{V(\overline{\Phi})}{F(\overline{\Phi})}
+\frac{F'(\overline{\Phi}))V'(\overline{\Phi})}{2\big(3(F'(\overline{\Phi}))^2
+2F(\overline{\Phi})K(\overline{\Phi})\big)}\nn\\
&&
+\Big(\frac{1}{12}\frac{1}{F(\overline{\Phi})}
+\frac14\frac{\big(F'(\overline{\Phi})\big)^2}{F(\overline{\Phi})\big(3(F'(\overline{\Phi}))^2
+2F(\overline{\Phi})K(\overline{\Phi})\big)}\Big)\frac{\bar{\epsilon}}{A^3},\label{ddotA}
\ea
where
\ba
\overline{\epsilon}=\overline{H}=A^3\bigg[-6F(\overline{\Phi})\Big(\frac{\dot{A}}{A}\Big)^2-6\frac{\dot{A}}{A}F'(\overline{\Phi})\,{\dot{\overline{\Phi}}}
+K(\overline{\Phi})\,(\dot{\overline{\Phi}})^2+V(\overline{\Phi})\bigg].\label{Hamiltdensity}
\ea
Notice that $\overline{\epsilon}$ is a constant, because as mentioned above the physical Hamiltonian density $\overline{H}$ is conserved during the evolution. From Eqs. (\ref{ddotA}) and (\ref{Hamiltdensity}) we read the gauge invariant Friedmann and Raychaudhuri equations as:
\ba
(\frac{\dot{A}}{A})^2=\frac{1}{6}(\rho^{e}_{\Phi}+\tilde{\rho}),\label{Friedmann}\\
2\frac{\ddot{A}}{A}-(\frac{\dot{A}}{A})^2=-\frac{1}{2}(P^{e}_{\Phi}+\tilde{P}),\label{Raychaudhuri}
\ea
where $\rho^{e}$ and $P^{e}$ denote the effective energy density and pressure of the scalar field, while $\tilde{\rho}$ and $\tilde{P}$ represent the correction terms caused by the non-vanishing physical Hamiltonian density. These quantities are defined as
\ba
\rho^{e}_{\Phi}&:=&-\frac{1}{F(\overline{\Phi})}[6\frac{\dot{A}}{A}F'(\overline{\Phi})\dot{\overline{\Phi}}-K(\overline{\Phi})\cdot(\dot{\overline{\Phi}})^2
-V(\overline{\Phi})],\label{rhoe}\nn\\
P^{e}_{\Phi}&:=&-2\frac{F'(\overline{\Phi})}{F(\overline{\Phi})}\frac{\dot{A}}{A}\dot{\overline{\Phi}}
-4\Big(\frac14-\frac{(F'(\overline{\Phi}))^2}{3(F'(\overline{\Phi}))^2+2F(\overline{\Phi})K(\overline{\Phi})}\Big)
\frac{V(\overline{\Phi})}{F(\overline{\Phi})}
-\frac{2F'(\overline{\Phi}))V'(\overline{\Phi})}{\big(3(F'(\overline{\Phi}))^2
+2F(\overline{\Phi})K(\overline{\Phi})\big)}\nn\\
&&
+2\frac{\Big(\frac12K(\overline{\Phi})F'(\overline{\Phi})^2
+F(\overline{\Phi})K^2(\overline{\Phi})+2F(\overline{\Phi})K(\overline{\Phi})F''(\overline{\Phi})
-F'(\overline{\Phi})F(\overline{\Phi})K'(\overline{\Phi})\Big)}{F(\overline{\Phi})\big(3(F'(\overline{\Phi}))^2
+2F(\overline{\Phi})K(\overline{\Phi})\big)}(\dot{\overline{\Phi}})^2,\label{Pe}
\nn\\
\tilde{\rho}&:=&-\frac{\overline{\epsilon}}{F(\overline{\Phi})A^3},\nn\\
\tilde{P}&:=&-\frac{\big(F'(\overline{\Phi})\big)^2}{F(\overline{\Phi})\big(3(F'(\overline{\Phi}))^2
+2F(\overline{\Phi})K(\overline{\Phi})\big)}\frac{\overline{\epsilon}}{A^3}.\nn
\ea
Given the evolution equations for the background variables we can consider their linear perturbations around an FRW background. The evolution equations for the linear perturbations have the following form
\ba
\delta \ddot{\Phi}&=&\Bigg[\bigg[-\frac{6(F'(\overline{\Phi}))^2
+6F(\overline{\Phi})K(\overline{\Phi})}{3(F'(\overline{\Phi}))^2
+2F(\overline{\Phi})K(\overline{\Phi})}\frac{\dot{A}}{A}
-2\Big(\frac{\frac32F'(\overline{\Phi})K(\overline{\Phi})+F(\overline{\Phi})K'(\overline{\Phi})
+3F'(\overline{\Phi})F''(\overline{\Phi})}{3(F'(\overline{\Phi}))^2
+2F(\overline{\Phi})K(\overline{\Phi})}\Big)(\dot{\overline{\Phi}})\bigg]\frac{\partial}{\partial\tau}\nn\\
&&\quad
+\frac{2F(\overline{\Phi})K(\overline{\Phi})+2(F'(\overline{\Phi}))^2}{3(F'(\overline{\Phi}))^2
+2F(\overline{\Phi})K(\overline{\Phi})}\frac{1}{A^2}\delta^{jk}\overline{D}_j \overline{D}_k+\Big(\frac{3F(\overline{\Phi})F'(\overline{\Phi})}{3(F'(\overline{\Phi}))^2
+2F(\overline{\Phi})K(\overline{\Phi})}\Big)'\Big(\frac{\dot{A}}{A}\Big)^2
\nn\\
&&\quad
-\Big(\frac{6(F'(\overline{\Phi}))^2
+6F(\overline{\Phi})K(\overline{\Phi})}{3(F'(\overline{\Phi}))^2
+2F(\overline{\Phi})K(\overline{\Phi})}\Big)'\frac{\dot{A}}{A}\dot{\overline{\Phi}}
-\Big(\frac{\frac32F'(\overline{\Phi})K(\overline{\Phi})
+F(\overline{\Phi})K'(\overline{\Phi})
+3F'(\overline{\Phi})F''(\overline{\Phi})}{3(F'(\overline{\Phi}))^2
+2F(\overline{\Phi})K(\overline{\Phi})}\Big)'(\dot{\overline{\Phi}})^2\nn\\
&&\quad
+\Big(\frac{\frac32F'(\overline{\Phi})V(\overline{\Phi})
-F(\overline{\Phi})V'(\overline{\Phi})}{3(F'(\overline{\Phi}))^2
+2F(\overline{\Phi})K(\overline{\Phi})}\Big)'\Bigg]\delta \Phi\nn\\
&&
+\Bigg[\bigg[-\frac{(F'\overline{(\Phi)})^2+F(\overline{\Phi})K(\overline{\Phi})}{3(F'(\overline{\Phi}))^2
+2F(\overline{\Phi})K(\overline{\Phi})}\frac{1}{A^2}\dot{\overline{\Phi}}
+\frac{F(\overline{\Phi})F'(\overline{\Phi})}{3(F'(\overline{\Phi}))^2
+2F(\overline{\Phi})K(\overline{\Phi})}\frac{\dot{A}}{A^3}\bigg]\frac{\partial}{\partial\tau}\nn\\
&&\quad
+\frac12\cdot\frac{F(\overline{\Phi})F'(\overline{\Phi})}{3(F'(\overline{\Phi}))^2
+2F(\overline{\Phi})K(\overline{\Phi})}(\overline{D}^j\overline{D}^k
-\frac{1}{A^2}\delta^{jk}\overline{D}^i\overline{D}_i)\nn\\
&&\quad
+\bigg[\frac{2(F'(\overline{\Phi}))^2+2F(\overline{\Phi})K(\overline{\Phi})}{3(F'(\overline{\Phi}))^2
+2F(\overline{\Phi})K(\overline{\Phi})}\frac{\dot{A}}{A^3}\dot{\overline{\Phi}}
-\frac{2F(\overline{\Phi})F'(\overline{\Phi})}{3(F'(\overline{\Phi}))^2
+2F(\overline{\Phi})K(\overline{\Phi})}\frac{\dot{A}^2}{A^4}\bigg]\delta^{jk}\Bigg]\delta Q_{jk}\nn\\
&&
+\Bigg[\bigg[\frac{2(F'(\overline{\Phi}))^2+2F(\overline{\Phi})K(\overline{\Phi})}{3(F'(\overline{\Phi}))^2
+2F(\overline{\Phi})K(\overline{\Phi})}\frac{1}{A^2}\dot{\overline{\Phi}}
-\frac{2F(\overline{\Phi})F'(\overline{\Phi})}{3(F'(\overline{\Phi}))^2
+2F(\overline{\Phi})K(\overline{\Phi})}\frac{\dot{A}}{A^3}\bigg]\frac{\partial}{\partial x^j}\Bigg]\delta N_j,\label{ddotphicos}
\ea
\ba
\delta\ddot{Q}_{jk}&=&\Bigg[\bigg[\Big(\frac{2F'(\overline{\Phi})K(\overline{\Phi})}{3(F'(\overline{\Phi}))^2
+2F(\overline{\Phi})K(\overline{\Phi})}\Big)A\dot{A}-2\Big(\frac{2F''(\overline{\Phi})K(\overline{\Phi})+(K(\overline{\Phi}))^2
-F'(\overline{\Phi})K'(\overline{\Phi})}{3(F'(\overline{\Phi}))^2
+2F(\overline{\Phi})K(\overline{\Phi})}\Big)A^2\dot{\overline{\Phi}}\bigg]\delta_{jk}\frac{\partial}{\partial\tau}\nn\\
&&\quad
-\Big(\frac{2(F'(\overline{\Phi}))^3+2F(\overline{\Phi})F'(\overline{\Phi})K(\overline{\Phi})}
{F(\overline{\Phi})(3(F'(\overline{\Phi}))^2
+2F(\overline{\Phi})K(\overline{\Phi}))}\Big)A^2\delta_{jk}\overline{D}^i\overline{D}_i
+2\frac{F'(\overline{\Phi})}{F(\overline{\Phi})}\overline{D}_j\overline{D}_k\nn\\
&&\quad
+\bigg[\Big(\frac{2F(\overline{\Phi})K(\overline{\Phi})}
{3(F'(\overline{\Phi}))^2
+2F(\overline{\Phi})K(\overline{\Phi})}\Big)'(\dot{A})^2
+\Big(\frac{2F'(\overline{\Phi})K(\overline{\Phi})}{3(F'(\overline{\Phi}))^2
+2F(\overline{\Phi})K(\overline{\Phi})}\Big)' A\dot{A}\dot{\overline{\Phi}}\nn\\
&&\qquad
-\Big(\frac{2F''(\overline{\Phi})K(\overline{\Phi})+(K(\overline{\Phi}))^2
-F'(\overline{\Phi})K'(\overline{\Phi})}{3(F'(\overline{\Phi}))^2
+2F(\overline{\Phi})K(\overline{\Phi})}\Big)'A^2(\dot{\overline{\Phi}})^2
+\Big(\frac{K(\overline{\Phi})V(\overline{\Phi})
+F'(\overline{\Phi})V'(\overline{\Phi})}{3(F'(\overline{\Phi}))^2
+2F(\overline{\Phi})K(\overline{\Phi})}\Big)'A^2\bigg]\delta_{jk}\Bigg]\delta \Phi\nn\\
&&
+\Bigg[\bigg[-3\frac{\dot{A}}{A}-\frac{F'(\overline{\Phi})}
{F(\overline{\Phi})}\dot{\overline{\Phi}}\bigg]\frac{\partial}{\partial\tau}+\overline{D}^i\overline{D}_i
+\Big(\frac{6F(\overline{\Phi})K(\overline{\Phi})+6(F'(\overline{\Phi}))^2}{3(F'(\overline{\Phi}))^2
+2F(\overline{\Phi})K(\overline{\Phi})}\Big)(\frac{\dot{A}}{A})^2
+\Big(\frac{6F'(\overline{\Phi})K(\overline{\Phi})+6\frac{(F'(\overline{\Phi}))^3}{F(\overline{\Phi})}}{3(F'(\overline{\Phi}))^2
+2F(\overline{\Phi})K(\overline{\Phi})}\Big)\frac{\dot{A}}{A}\dot{\overline{\Phi}}\nn\\
&&\quad
-\Big(\frac{2F''(\overline{\Phi})K(\overline{\Phi})+(K(\overline{\Phi}))^2
-F'(\overline{\Phi})K'(\overline{\Phi})}{3(F'(\overline{\Phi}))^2
+2F(\overline{\Phi})K(\overline{\Phi})}\Big)(\dot{\overline{\Phi}})^2
+\frac{K(\overline{\Phi})V(\overline{\Phi})+F'(\overline{\Phi})V'(\overline{\Phi})}{3(F'(\overline{\Phi}))^2
+2F(\overline{\Phi})K(\overline{\Phi})}
\bigg]\delta Q_{jk}\nn\\
&&
+\Bigg[\Big(-\frac{F'(\overline{\Phi})^2}{3(F'(\overline{\Phi}))^2
+2F(\overline{\Phi})K(\overline{\Phi})}\frac{\dot{A}}{A}
+\frac{F(\overline{\Phi})K(\overline{\Phi})
+(F'(\overline{\Phi}))^2}{3(F'(\overline{\Phi}))^2
+2F(\overline{\Phi})K(\overline{\Phi})}\frac{F'(\overline{\Phi})}
{F(\overline{\Phi})}\dot{\Phi}\Big)\delta^{mn}\delta_{jk}
\frac{\partial}{\partial\tau}\nn\\
&&\quad
+\Big(2\frac{F'(\overline{\Phi})^2}{3(F'(\overline{\Phi}))^2+2F(\overline{\Phi})K(\overline{\Phi})}(\frac{\dot{A}}{A})^2
-2\frac{F(\overline{\Phi})K(\overline{\Phi})+(F'(\overline{\Phi}))^2}{3(F'(\overline{\Phi}))^2
+2F(\overline{\Phi})K(\overline{\Phi})}\frac{F'(\overline{\Phi})}{F(\overline{\Phi})}
\frac{\dot{A}}{A}\dot{\overline{\Phi}}\Big)\delta^{mn}\delta_{jk}\nn\\
&&\quad
+\frac{F(\overline{\Phi})K(\overline{\Phi})+(F'(\overline{\Phi}))^2}{3(F'(\overline{\Phi}))^2
+2F(\overline{\Phi})K(\overline{\Phi})}A^2\delta_{jk}(\overline{D}^m\overline{D}^n
-\frac{1}{A^2}\delta^{mn}\overline{D}^i\overline{D}_i)
-4(\frac{\dot{A}}{A})^2\delta^{m}_{(j}\delta^{n}_{k)}+\overline{Q}^{mn}\overline{D}_j\overline{D}_k\Bigg]\delta Q_{mn}\nn\\
&&
+\Bigg[4\frac{\dot{A}}{A}\delta^{m}_{(k}\frac{\partial}{\partial\tau}-2\overline{D}^m\overline{D}_{(k}\Bigg]\delta Q_{j)m}
+\Bigg[2\frac{F'(\overline{\Phi})}{F(\overline{\Phi})}\dot{\overline{\Phi}}\Big(\delta^{m}_{(j}\frac{\partial}{\partial x^{k)}}-2\delta_{jk}\delta^{mw}\frac{\partial}{\partial x^w}\Big)
+2\frac{\dot{A}}{A}\delta^{m}_{(j}\frac{\partial}{\partial x^{k)}}+2\delta^{m}_{(j}\frac{\partial}{\partial x^{k)}}\frac{\partial}{\partial\tau}\Bigg]\delta N_m.\label{ddotqcos}\nn\\
\ea
To study the detailed behavior of the perturbation equations (\ref{ddotphicos}) and (\ref{ddotqcos}) we will work in conformal time $x^0=\eta$ for which $\overline{g}_{00}=-A^2$. However due to the dust observers we always have $g_{\tau\tau}=-N^2+Q^{jk}N_jN_k=-1$ where $\tau$ denotes the physical time, meaning that we automatically work with the proper time of the dust. Using the relation $dx^0=\frac{d\tau}{A}$ we obtain
\be
g_{\tau\tau}=\frac{g_{00}}{A^2}, \quad g_{\tau j}=\frac{g_{0j}}{A}=N_j,\quad
g_{jk}=Q_{jk}.
\ee
Following the procedure in standard cosmological perturbation theory (SCPT), we decompose the  spacetime metric $g_{\mu\nu}=\overline{g}_{\mu\nu}+\delta g_{\mu\nu}$ into  tensor, vector and scalar modes, parametrized by ten functions $\phi,B,E,\psi,S_j,F_j,h_{jk}$ where the vector modes $S_j$ and $F_j$ are transversal with respect to the spatial Euclidian metric and the tensor mode $h_{jk}$ is transversal and traceless,
\ba
g_{00}=(-1+2\phi)A^2,\quad
g_{0j}=(B_{,j}+S_j)A^2,\quad
g_{jk}=\delta_{jk}+A^2[2\psi\delta_{jk}+2E_{,jk}+2F_{(j,k)}+h_{jk}].\label{decompmetric}
\ea
Now from $g_{\tau\tau}=-1$, we immediately get $\phi=0$ and therefore with the dust as dynamical observers we will always work in partly synchronous gauge\footnote{Here partly refers to the fact that in SCPT the gauge $\phi=B=0$ is denoted by synchronous gauge, while longitudinal gauge means $B=E=0$.}.  Furthermore since $N_j$ can be expressed in terms of the other phase space variables, the functions $B$ and $S_j$ are not independent variables in our formalism. We remark that each mode in equations (\ref{decompmetric}) is already gauge invariant in our formalism, which is different from the case of SCPT. At the level of linear perturbation theory the individual modes decouple and we can study their evolution separately. Let us denote the derivative with respect to conformal time $\frac{d}{d\eta}$  by an upper prime ${}^{'}$. Substituting the expressions (\ref{decompmetric}) into (\ref{ddotqcos}) we obtain for the tensor modes
\ba
2\mathcal{H}h^{'}_{jk}+\frac{F'(\overline{\Phi})}{F(\overline{\Phi})}\dot{\overline{\Phi}}Ah^{'}_{jk}+h^{''}_{jk}-\triangle h_{jk}=0,\label{eomtensor}
\ea
where $\mathcal{H}:=\frac{A^{'}}{A}$, $\triangle:=\delta^{ij}\overline{D}_i\overline{D}_j$. For the vector modes we consider the decomposition of the shift vector from (\ref{decompmetric}) given by
\ba
\delta N_{j}=A[B_{,j}+S_{j}].\label{decompshift}
\ea
Since $\dot{\delta N_i}=0$ and in the case of the vector modes we also have $B=0$, we obtain
\ba
\frac{dA}{d\eta}S_{j}+A\frac{dS_{j}}{d\eta}=0&\Longrightarrow& {\cal H}S_j+{S}^{'}_j=0.\label{eoms}
\ea
From (\ref{ddotqcos}) we get for the vector mode the following equation
\ba
2F^{''}_{(j,k)}+4\mathcal{H}F^{'}_{(j,k)}+2\frac{F'(\overline{\Phi})}{F(\overline{\Phi})}\dot{\overline{\Phi}}AF^{'}_{(j,k)}
=2\mathcal{H}S_{(j,k)}+2\frac{F'(\overline{\Phi})}{F(\overline{\Phi})}\dot{\overline{\Phi}}AS_{(j,k)}.\label{vector}
\ea
Denoting $V_{j}=S_j-F^{'}_j$ and using Eq. (\ref{eoms}), Eq. (\ref{vector}) simplifies to
\ba
2\mathcal{H}V_{(j,k)}+\frac{F'(\overline{\Phi})}{F(\overline{\Phi})}\dot{\overline{\Phi}}A V_{(j,k)}+V^{'}_{(j,k)}=0.\label{eomvector}
\ea
For the scalar mode equation (\ref{ddotqcos}) leads to
\ba
&&\quad\Big(\mathcal{H}+\big(\ln\sqrt{F(\overline{\Phi})}\big)^{'}\Big)\Big(4\psi
+2\delta\big(\ln\sqrt{F(\overline{\Phi}})\big)\Big)^{'}\delta_{jk}
+2\Big(\psi+\delta\big(\ln\sqrt{F(\Phi})\big)\Big)^{''}\delta_{jk}\nn\\
&&\quad
+4\Big(\mathcal{H}+\big(\ln\sqrt{F(\overline{\Phi}})\big)^{'}\Big)\big(E^{'}_{,jk}\big)+2E^{''}_{,jk}\nn\\
&=&2\Big(\mathcal{H}+2\big(\ln\sqrt{F(\overline{\Phi})}\big)^{'}\Big)B_{,jk}+2\psi_{,jk}+4\Big(\delta \ln\sqrt{ F(\Phi)}\Big)_{,jk}-\delta_{jk}\bigg[\frac12\big(\overline{\Xi}^{'}\delta\overline{\Xi}^{'}
-A^2(\frac{V(\overline{\Xi})}{F(\overline{\Xi})})'_{\Xi}\delta \Xi\big)\bigg],\label{eomscalar}
\ea
where we used the field redefinition
\ba
\frac{d\Xi(\Phi)}{d\Phi}=\sqrt{3(\frac{F'(\Phi)}{F(\Phi)})^2+2\frac{K(\Phi)}{F(\Phi)}}.\label{defxi}
\ea
 Eq. (\ref{eomscalar}) is of the form $f_{,jk}+g\delta_{jk}=0$ for appropriate definitions of the functions $f$ and $g$. Applying the trace to this equation we have $3g+\Delta f=0$. Now operating with $\partial_j\partial_k$ on  $f_{,jk}+g\delta_{jk}$  yields $\Delta(g+\Delta f)=0$. Since due to our boundary conditions there are no zero modes of the Laplacian, we can conclude $f=g=0$ and therefore the $(.)_{jk}$ part and $\delta_{jk}$ part of Eq. (\ref{eomscalar}) are independent of each other. Considering the separated contribution we get
\ba
&&\bigg[4\big[\mathcal{H}+(\ln\sqrt{F(\overline{\Phi})})^{'}\big]E^{'}+2E^{''}-2\Big(\big[\mathcal{H}+(\ln\sqrt{F(\overline{\Phi})})^{'}\big]
+\big(\ln\sqrt{F(\overline{\Phi})}\big)^{'}\Big)B-2\big[\psi+\delta\ln\sqrt{F(\Phi)}\big]\nn\\
&&\quad
-2\delta \ln\sqrt{F(\Phi)}\bigg]_{,jk}=0,\label{eomsc1}\\
&&\bigg[4\big[\mathcal{H}+(\ln\sqrt{F(\overline{\Phi})})^{'}\big]\big[\psi+\delta\ln\sqrt{F(\Phi)}\big]^{'}
-2\big[\mathcal{H}+(\ln\sqrt{F(\overline{\Phi})})^{'}\big]
\big(\delta\ln\sqrt{F(\Phi})\big)^{'}
+2\big[\psi+\delta\ln\sqrt{F(\Phi)}\big]^{''}\bigg]\delta_{jk}\nn\\
&=&-\bigg[\frac12\big(\overline{\Xi}^{'}\delta\Xi^{'}
-A^2(\frac{V(\overline{\Xi})}{F(\overline{\Xi})})'_{\Xi}\delta \Xi\big)\bigg]\delta_{jk}.\label{eomsc2}
\ea
 In SCPT the system does not involve any physical observers \cite{Mukhanov92}, the gauge invariant variables of STT are defined as:
\ba
\Theta_A &=&-\delta\ln\sqrt{F(\Phi)}-\big[\mathcal{H}+(\ln\sqrt{F(\overline{\Phi})})^{'}\big](B-E^{'})-(B-E^{'})^{'},\nn\\
\Theta_B &=&\big[\psi+\delta\ln\sqrt{F(\Phi)}\big]
+\big[\mathcal{H}+(\ln\sqrt{F(\overline{\Phi})})^{'}\big](B-E^{'}),\nn\\
\delta Z&=&\delta\Xi+\overline{\Xi}^{'}(B-E^{'}).\label{newvariables}
\ea
Notice that in Ref. \cite{Mukhanov92}, there is an additional term proportional to $\delta N$ in the definition of $\Theta_A$. However in our model $\delta N$ is not an independent variable and we have
\ba
\delta N=-\frac{\overline{N}^i\overline{N}^j}{2\overline{N}}\delta Q_{jk}+\frac{\overline{N}^j}{\overline{N}}\delta{N_j}.
\ea
Now for an FRW background we have $\overline{N}=1$ and $\overline{N}_j=0$, and thus $\delta N=0$, explaining why no contribution to $\delta N$ occurs in our equations. In the case of the scalar mode we have $\delta N_j=AB_{,j}$ and consequently
\ba
\frac{dA}{d\eta}B_{,j}+A\frac{dB_{,j}}{d\eta}=0&\Longrightarrow& {\cal H}B_{,j}+{B}^{'}_{,j}=0,\label{eomb}
\ea
where the last relation follows from the fact that the Laplacian has no zero modes.
Using Eq. (\ref{eomb}), Eqs. (\ref{eomsc1}) and (\ref{eomsc2}) are simplified to
\ba
&&2[\Theta_A-\Theta_B]_{,jk}=0,\label{equaltheta}\\
&&\Theta^{''}_B+\big[\mathcal{H}+(\ln\sqrt{F(\overline{\Phi})})^{'}\big](\Theta_A+2\Theta_B)^{'}
+\Big(2\big[\mathcal{H}+(\ln\sqrt{F(\overline{\Phi})})^{'}\big]^{'}+\big[\mathcal{H}+(\ln\sqrt{F(\overline{\Phi})})^{'}\big]^2\Big)\Theta_A\nn\\
&=&
-\frac14\big[(\overline{\Xi}^{'})^2\Theta_A+\overline{\Xi}^{'}\delta Z^{'}
-F(\overline{\Xi})A^2(\frac{V(\overline{\Xi})}{F^2(\overline{\Xi})})'_{\Xi}\delta Z\big],\label{eomscalar1}
\ea
 where $(\frac{V(\overline{\Xi})}{F^2(\overline{\Xi})})'_{\Xi}:=d(\frac{V(\overline{\Xi})}{F^2(\overline{\Xi})})/d\Xi$. Note that in \cite{Giesel10s} a slightly different notation was used but the notations can be related to each other by specializing to $F(\Phi)=1$ and identifying the variables $\Phi$ and $\Psi$ in \cite{Giesel10s} with $\Theta_A=\Phi$ and $\Theta_B=\Psi$.
\\
As the next step we derive the equation of motion for the perturbation $\delta \Phi$. Using the field redefinition (\ref{defxi}), the equation of motion for the scalar field perturbations in (\ref{ddotphicos}) can be written in a more compact form as
\ba
\delta \ddot{\Xi}&=&\Big[\dot{\overline{\Xi}}^2\big(\ln\sqrt{F(\overline{\Xi})}\big)''_{\Xi\Xi}-F(\overline{\Xi})
(\frac{V(\overline{\Xi})}{F^2(\Xi)})''_{\Xi\Xi}
-(\frac{V(\overline{\Xi})}{F^2(\overline{\Xi})})'_{\Xi}F'(\overline{\Xi})+3\big(\dot{(\ln\sqrt{F(\overline{\Xi})})}+\frac{\dot{A}}{A}\big)(\ln F(\overline{\Xi}))'_{\Xi}\dot{\overline{\Xi}}\nn\\
&&
-\frac32(\dot{\overline{\Xi}})^2\frac{F''(\overline{\Xi})}{F(\overline{\Xi})}
-3\frac{\dot{A}}{A}\dot{\overline{\Xi}}(\ln F(\overline{\Xi}))'_{\Xi}+\frac{1}{A^2}\delta^{jk}\overline{D}_j \overline{D}_k\Big]\delta \Xi+\Big[-4(\ln\sqrt{F(\Xi)})'_{\Xi}\dot{\overline{\Xi}}-3\frac{\dot{A}}{A}\Big]\delta \dot{\Xi}\nn\\
&&
+\frac{1}{A^2}(\frac{\dot{A}}{A})\dot{\overline{\Xi}}\delta Q_{jj}-\frac12\frac{1}{A^2}\dot{\overline{\Xi}}\delta\dot{Q}_{jj}+\frac{1}{A^2}\dot{\overline{\Xi}}\delta N_{j,j}-\delta \big(\frac12(\frac{1}{F(\Xi)})'_{\Xi}\frac{C}{\sqrt{\det Q}}\big).
\ea
Using the definition (\ref{newvariables}), it simplifies to
\ba
&&\delta Z^{''}+2\big[\mathcal{H}+(\ln\sqrt{F(\overline{\Xi})})^{'}\big]\delta Z-\Delta \delta Z +(\frac{V(\overline{\Xi})}{F^2(\overline{\Xi})})'_{\Xi\Xi}F(\overline{\Xi})A^2\delta Z+\Xi^{'}\Theta^{'}_A-2(\frac{V(\overline{\Xi})}{F^2(\overline{\Xi})})'_{\Xi}F(\overline{\Xi})A^2\Theta^{'}_A+3\overline{\Xi}^{'}\Theta^{'}_B\nn\\
&=&\big(\frac12(\frac{1}{F(\overline{\Xi})})'_{\Xi}\frac{\overline{\epsilon}}{A}(B-E^{'})\big)^{'}
+\frac12(\frac{1}{F(\overline{\Xi})})'_{\Xi}\frac{\overline{\epsilon}}{A}(B-E^{'})^{'}-A^2\delta \big(\frac12(\frac{1}{F(\Xi)})'_{\Xi}\frac{\epsilon}{\sqrt{\det Q}}\big).\label{eomperscalar}
\ea
Note that compared to SCPT in the manifestly gauge invariant formalism we start with STT plus dust and thus increase the number of degrees of freedom by four. These four additional degrees of freedom are used to construct gauge invariant quantities associated with the spatial 3-metric $q_{jk}$ and the scalar field $\phi$, which we denoted by $Q_{jk}$ and $\Phi$ respectively.  In order to compare our framework with the results of SCPT we expressed $\delta Q_{jk}$ in terms of the invariants $\Theta_A,\Theta_B,V_j$ and $h_{jk}$, usually used in SCPT, and also introduced the field redefinition in Eq. (\ref{defxi}) to finally work with $\delta\Xi$ and $\delta Z$ respectively. Now if we properly identify $\Theta_A,\Theta_B,V_j$, $h_{jk}$ and $\delta Z$ with the analogue quantities in the SCPT formalism (see \cite{Giesel10s} for a detailed discussion), we realize that the evolution equations for the seven perturbed invariants $\Theta_A,\Theta_B,V_j$,$h_{jk}$ and $\delta Z$ take a form almost identical to those equations that one obtains in the Lagrangian formalism using SCPT. The almost refers to the fact that in the manifestly gauge invariant formalism the dust as dynamically coupled observers has the effect that the physical Hamiltonian, which generates the evolution of the gauge invariant quantities $\delta Q_{jk}$ and $\delta\Phi$, is no longer a constraint as the case for SCPT. Instead it becomes a constant of motion. This constant of motion reflects the influence of the dust and hence the dynamically coupled observers on the system. In case of Eq. (\ref{eomperscalar}) these are exactly the terms on the right hand side of this equation, which are corrections caused by the non-vanishing Hamiltonian density $\bar{\epsilon}$ compared to the corresponding equation in SCPT. These corrections also encode the fact that in the manifestly gauge invariant model we have seven physical degrees of freedom whereas for STT in SCPT the theory possesses only three physical degrees of freedom. This can be seen from the fact that in the case of SCPT the seven invariants $\Theta_A,\Theta_B,V_j$,$h_{jk}$ and $\delta Z$ are still subject to constraints which reduce their independent number of physical degrees of freedom down to three. In contrast for the manifestly gauge invariant formalism the constraints have already been reduced by constructing the observables $Q_{jk}$ and $\Phi$. In particular in the limit of vanishing influence of the dust, meaning that its energy and momentum density vanishes, the conservation equations turn into constraint equations again and the corrections that occur in the evolution equations for the linear perturbations vanish. Then we obtain an exact match with the results obtained in SCPT. However the whole manifestly gauge invariant formalism is based on the idea that the influence of the dynamically coupled observers cannot totally be neglected and thus small corrections would always be needed to taken into account. As discussed in \cite{Giesel10s} for the dust observer in the late universe these modifications decay as compared to the usual terms.
\\
\\
As mentioned above the constraints in the SCPT become conservation equations in the manifestly gauge invariant framework due to the dynamically coupled observers since here only the total constraints consisting of STT contribution and the dust contribution vanish.  First, we have two conserved quantities associated with the energy and momentum conservation respectively:
\ba
{\epsilon}_j(\sigma)=-{C}_j(\sigma), \qquad {\epsilon}(\sigma)={H}(\sigma).
\ea
In the following we will derive the equations for the perturbed conserved charges $\delta\epsilon_j$ and $\delta\epsilon$ around an FRW background defined through $\delta\epsilon_j=\epsilon_j-\overline{\epsilon}_j$ and $\delta\epsilon=\epsilon-\overline{\epsilon}$  where for FRW $\overline{\epsilon}_j=0$ and $\overline{\epsilon}=\overline{H}$. We obtain for the linear perturbations
\ba
\delta \epsilon_j&=&F(\overline{\Phi})A(\delta \dot{Q}_{jk,k}-\delta \dot{Q}_{kk,j})-2F(\overline{\Phi})\dot{A}(\delta Q_{jk,k}-\delta Q_{kk,j})-A^3F'(\overline{\Phi})(\delta \dot{\Phi})_{,j}-F(\overline{\Phi})A[\Delta\delta N_j-\delta N_{k,kj}]\nn\\
&&
-\big[2\dot{A}A^2F'(\overline{\Phi})+A^3F''(\overline{\Phi})\dot{\overline{\Phi}}-6A^2\dot{A}F'(\overline{\Phi})
+2A^3\dot{\overline{\Phi}}K(\overline{\Phi})\big](\delta \Phi)_{,j},\nn\\\label{perturbmomentum}
\ea
and
\ba
\delta \epsilon &=& \bigg[-6F'(\overline{\Phi})A^3(\frac{\dot{A}}{A})^2
-6F''(\overline{\Phi})A^3\frac{\dot{A}}{A}\dot{\overline{\Phi}}
-6A^3\frac{\dot{A}}{A}F'(\overline{\Phi})\frac{\partial}{\partial \tau}
+A^3K'(\overline{\Phi})(\overline{\Phi})^2+2A^3K(\overline{\Phi})
\overline{\Phi}\frac{\partial}{\partial \tau}\nn\\
&&\quad
+2A^3F'(\overline{\Phi})\overline{D}^i\overline{D}_i
+A^3\frac{V'(\overline{\Phi})}{F(\overline{\Phi})}\bigg]\delta \Phi+\bigg[\frac12\frac{\overline{\epsilon}}{A^2F(\overline{\Phi})} \delta^{jk}
-2\delta^{jk}F(\overline{\Phi})\dot{A}\frac{\partial}{\partial \tau}+4F(\overline{\Phi})\frac{(\dot{A})^2}{A}\delta^{jk}\nn\\
&&\quad
-A^3F(\overline{\Phi})[G^{-1}]^{jkmn}\overline{D}_m\overline{D}_n-F'(\overline{\Phi})[-\frac{\dot{A}}{A^3}\delta^{jk}\dot{\overline{\Phi}}
+\frac{1}{A^2}\delta^{jk}\dot{\overline{\Phi}}\frac{\partial}{\partial \tau}]\bigg]\delta Q_{jk}\nn\\
&&
+\bigg[4F(\overline{\Phi})\dot{A}\delta^{mn}\frac{\partial}{\partial x^n}
+2F'(\overline{\Phi})A\dot{\overline{\Phi}}\delta^{mj}\frac{\partial}{\partial x^j}\bigg]\delta N_m.\label{perturbhamiltonian}
\ea
As the next step  we decompose the vector equation (\ref{perturbmomentum}) into the longitudinal part
$\delta\epsilon^{\|}_j=\Delta^{-1}\delta\epsilon_{k,kj}$, where  $\Delta^{-1}$ is the Green's function of the Laplacian $\Delta$, and the transversal part $\delta\epsilon^{\bot}=\delta\epsilon_{j}-\delta\epsilon^{\|}_j$. Inserting Eqs. (\ref{decompmetric}) and (\ref{decompshift}) into (\ref{perturbmomentum}),
the longitudinal part of Eq. (\ref{perturbmomentum}) gives
\ba
(\psi +\delta \sqrt{F(\Phi)})^{'}_{,j}-\frac34(\frac{F'(\overline{\Phi})}{F(\overline{\Phi})})^2(\delta\Phi)_{,j}
=-\frac{1}{4}(\frac{\delta \epsilon^{\|}_j}{F(\overline{\Phi})A^2}+\overline{\Xi}^{'}\delta\Xi_{,j}),\label{constvector}
\ea
and the transversal part gives
\ba
\Delta V_j=-\frac{\delta\epsilon^{\bot}_j}{F(\overline{\Phi})A^2}.\label{constvector1}
\ea
Written in terms of variables $\Theta_A$, $\Theta_B$ and $\delta Z$, Eq. (\ref{constvector}) becomes
\ba
\Big[\Theta^{'}_B+\big[\mathcal{H}+(\ln\sqrt{F(\overline{\Phi})})^{'}\big]\Theta_A+\frac{1}{4}\overline{\Xi}^{'}\delta Z\Big]_{,j}=-\frac{1}{F(\overline{\Phi})A}\Big[\frac{\delta\epsilon^{\|}_j}{A}-\overline{\epsilon}(B-E^{'}_{,j})\Big].\label{constvector2}
\ea
Finally, Eq. (\ref{perturbhamiltonian}) gives
\ba
&&\Delta \Theta_B-3\big[\mathcal{H}+(\ln\sqrt{F(\overline{\Phi})})^{'}\big]\Theta^{'}_B
+\frac14\big[-A^2\frac{V(\overline{\Xi})}{F(\overline{\Xi})}\Theta_A+\overline{\Xi}^{'}\delta Z^{'}-F(\overline{\Xi})A^2(\frac{V(\overline{\Xi})}{F^2(\overline{\Xi})})'_{\Xi}\delta Z\big]\nn\\
=&&
\frac{1}{4F(\overline{\Phi})A}\bigg[\delta\epsilon-\overline{\epsilon}\Big(\delta\ln\sqrt{F(\Phi)}+3(\psi +\delta \sqrt{F(\Phi)})+\Delta E-2(B-E^{'})^{'}+\big[\mathcal{H}+(\ln\sqrt{F(\overline{\Phi})})^{'}\big](B-E^{'})\Big)\bigg].\nn\\\label{constscalarmode}
\ea
Up to now we have derived the equations of motion for the background variables in (\ref{ddotphi}), (\ref{Friedmann}) and (\ref{Raychaudhuri}) and the perturbed variables in (\ref{eomtensor}), (\ref{eomvector}), (\ref{equaltheta}), (\ref{eomscalar1}), (\ref{eomperscalar}) and (\ref{constvector1}), (\ref{constvector2}), (\ref{constscalarmode}) in manifestly gauge invariant formalism in the case of STT.
Again we find if we choose $F(\Phi)=1$ and $K(\Phi)=\frac{1}{2}$ the STT + dust reduces to GR + scalar field + dust in \cite{Giesel10s}, which allows us to test the correctness of our equations here.

\section{Gauge Invariant Hamiltonian Formulation in Einstein Frame}
 It is well known that in SCPT of STT  the Jordan frame is related to the Einstein frame by a conformal transformation and can therefore be transformed into each other by appropriate field redefinitions. However, it is not clear whether this kind of equivalence will still hold in the manifestly gauge invariant formalism since such a transformation need to be formulated at the level of the reduced phase space here. In the first part of this section we will prove this equivalence. In the second part, using this equivalence, we  extend our results to slightly different reference systems.

\subsection{Conformal Equivalence}
Under the conformal transformation $\tilde{g}_{ab}=F(\phi)g_{ab}$ and the field redefinition $\frac{d\xi}{d\phi}= \sqrt{3(\frac{F'(\phi)}{F(\phi)})^2+2\frac{K(\phi)}{F(\phi)}}$,
the action (\ref{actionjordan}) in Jordan frame is transformed into the following action in Einstein frame,
\ba
S\, _{(\xi ,\tilde{g},dust)}=\int d^4x\sqrt{|\det (\tilde{g})|}\left[\tilde{R}^{(4)}-\frac12\tilde{g}^{\mu\nu}(\tilde{\triangledown} _\mu\xi) \tilde{\triangledown} _\nu\xi -\frac{V(\xi)}{F^2(\xi)} )\right]-\frac{1}{2}\int d^4x\sqrt{\det (\tilde{g})}\frac{\rho}{F(\xi)} \left[\tilde{g}^{\mu \nu }U_{\mu }U_{\nu }+\frac{1}{F(\xi)}\right],\nn\\\label{actionEinstein}
\ea
where $\tilde{g}_{ab}$ is now taken as the basic variable,
  $\tilde{R}^{(4)}$ stands for the  scalar curvature of the Einstein frame metric $\tilde{g}_{ab}$, $\tilde{g}^{ab}$ is the inverse of $\tilde{g}_{ab}$, $\tilde{\triangledown}$ is the covariant derivative compatible with $\tilde{g}_{ab}$, $V(\xi):=V(\phi(\xi))$ and $F(\xi):=F(\phi(\xi))$. Notice that the definition for the dust variables $\rho$ and $U_{\mu}$ are those in Jordan frame. This means that the reference system remains unchanged.

 Hamiltonian analysis of action (\ref{actionEinstein}) is performed in appendix A, where we get the following physical Hamiltonian,
\ba
\tilde{\textbf{H}}(\tau)=\int_{\mathcal{S}}d^3\sigma \tilde{H}(\tau,\sigma),\qquad \tilde{H}(\tau, \sigma)=\sqrt{F(\Xi)}\sqrt{\tilde{C}^2-\tilde{Q}^{ij}\tilde{C}_i\tilde{C}_j}(\tau, \sigma),
\ea
with
\ba
\tilde{C}_j(\tau,\sigma )&=&\left[-2\tilde{Q}_{jl}\tilde{D}_k\tilde{P}^{kl}+\tilde{\Pi} \tilde{D}_j\Xi\right](\tau ,\sigma ),\label{CjEinstein}\\
\tilde{C}(\tau,\sigma)&=&\Bigg[\frac{1}{\sqrt{\det \tilde{Q}}}\big(\tilde{Q}_{i m}\tilde{Q}_{j n}-\frac{1}{2}\tilde{Q}_{i j}\tilde{Q}_{m n}\big)\tilde{P}^{i j}\tilde{P}^{m n}-\sqrt{\det \tilde{Q}}\tilde{R}^{(3)}\nn\\
&&
+\frac12\frac{\tilde{\Pi}^2}{\sqrt{\det \tilde{Q}}}+\frac12\sqrt{\det\tilde{Q}}\big(\tilde{Q}^{i j}(\tilde{D}_i\Phi)(\tilde{D}_j\Phi)+\frac{V(\Xi)}{F^2(\Xi)}\big)\Bigg](\tau ,\sigma).\label{CEinstein}
\ea
The elementary Poisson brackets read
\ba
\{\tilde{P}^{jk}(\sigma),\tilde{Q}_{mn}(\sigma')\}=\delta ^j_{(m}\delta^k_{n)}\delta (\sigma,\sigma'),\qquad\{\tilde{\Pi}(\sigma),\Xi(\sigma' )\}=\delta (\sigma,\sigma').
\ea
The evolution of a general observable is given by
\ba
\frac{d O}{d\tau}&=&\{\tilde{\textbf{H}}, O\}\nn\\
&=&\int_{\mathcal{S}}d^3\sigma
\frac{F(\sigma)}{\tilde{H}(\sigma)}\bigg(\tilde{C}(\sigma)\{\tilde{C}(\sigma),O\}-\tilde{Q}^{jk}(\sigma)\tilde{C}_j(\sigma)\{\tilde{C}_k (\sigma), O\}+\frac12\tilde{Q}^{jm}(\sigma)\tilde{Q}^{kn}(\sigma)\tilde{C}_j(\sigma)\tilde{C}_k(\sigma)\{\tilde{Q}_{jk}(\sigma), O\}\bigg)\nn\\
&&\qquad
+\frac{\tilde{H}(\sigma)}{2F(\sigma)}\{F(\sigma),O\}\nn\\
&=&\int_{\mathcal{S}}d^3\sigma \bigg(\tilde{N}(\sigma)\{\tilde{C}(\sigma),O\}+\tilde{N}^j\{\tilde{C}_j(\sigma), O\}+\frac{\tilde{H}(\sigma)}{2F(\sigma)}\tilde{N}^i(\sigma)\tilde{N}^j(\sigma)\{\tilde{Q}_{ij}(\sigma),O\}
+\frac{\tilde{H}(\sigma)}{2F(\sigma)}\{F(\sigma),O\}\bigg),
\ea
where the dynamical shift vector and lapse function are defined as
\ba
\tilde{N}_j\equiv-\frac{F\tilde{C}_j}{\tilde{H}}, \qquad \tilde{N}\equiv\frac{F\tilde{C}}{\tilde{H}}=\sqrt{F+\tilde{Q}^{ij}\tilde{N}_i\tilde{N}_j}.\label{eomofO}
\ea
By calculating their Poisson brackets with $\tilde{\textbf{H}}$, we easily verify that $\tilde{\epsilon}_j:=-\tilde{C}_j(\sigma)$ and $\tilde{\epsilon}:=\tilde{H}(\sigma)$ still compose a pair of conserved quantities. Hence $\frac{\tilde{N}_j}{F(\Xi)}:=-\frac{\tilde{C}_j(\sigma)}{\tilde{H}(\sigma)}$ also remains a constant during the evolution.

Simply following the procedures in section two, we get the second order equations of motion for $\Xi$ and $\tilde{Q}_{jk}$ as
\ba
\ddot{\Xi}&=&\Big[\frac{\dot{\tilde{N}}}{\tilde{N}}-\frac{(\sqrt{\det \tilde{Q}})^{\cdot}}{\sqrt{\det \tilde{Q}}}+\frac{\tilde{N}}{\sqrt{\det \tilde{Q}}}\Big(\mathcal{L}_{\overset{\rightharpoonup }{\tilde{N}}}\frac{\sqrt{\det \tilde{Q}}}{\tilde{N}}\Big)\Big](\dot{\Xi}-\mathcal{L}_{\overset{\rightharpoonup }{\tilde{N}}}\Xi)+\tilde{Q}^{jk}\Xi_{,k}\Big[\frac{\tilde{N}}{\sqrt{\det \tilde{Q}}}\Big([\tilde{N}\sqrt{\det\tilde{Q}}]_{,j}-\frac{\tilde{H}F'(\Xi)}{2F(\Xi)}\Big)\Big]\nn\\
&&\quad
+\tilde{N}^2\Big[\tilde{D}^i\tilde{D}_i\Xi+[\tilde{Q}^{jk}]_{,j}\Xi_{,k}-\frac12 (\frac{V(\Xi)}{(F(\Xi))^2})\Big]+2\big(\mathcal{L}_{\overset{\rightharpoonup }{\tilde{N}}}\dot{\Xi}\big)+\big(\mathcal{L}_{\dot{\overset{\rightharpoonup}{N}}}\Xi\big)-\big(\mathcal{L}_{\overset{\rightharpoonup }{\tilde{N}}}(\mathcal{L}_{\overset{\rightharpoonup}{\tilde{N}}}\Xi)\big),\label{ddotxi}\\
\ddot{\tilde{Q}}_{jk}&=&\left[\frac{\dot{\tilde{N}}}{\tilde{N}}-\frac{\Big(\sqrt{\det \tilde{Q}}\Big)^{\cdot}}{\sqrt{\det \tilde{Q}}}+\frac{\tilde{N}}{\sqrt{\det \tilde{Q}}}\Big(\mathcal{L}_{\overset{\rightharpoonup}{\tilde{N}}}(\frac{\sqrt{\det \tilde{Q}}}{\tilde{N}})\Big)\right]\Big(\dot{\tilde{Q}}_{jk}-(\mathcal{L}_{\overset{\rightharpoonup }{\tilde{N}}}\tilde{Q})_{jk}\Big)\nn\\
&&
+\tilde{Q}^{m n}\Big(\dot{\tilde{Q}}_{mj}-(\mathcal{L}_{\overset{\rightharpoonup }{\tilde{N}}}\tilde{Q})_{mj}\Big)\Big(\dot{\tilde{Q}}_{nk}-(\mathcal{L}_{\overset{\rightharpoonup}{\tilde{N}}}\tilde{Q})_{nk }\Big)+\tilde{N}^2\Big(\Xi _{,j}\Xi _{,k}-2\tilde{R}_{jk}\Big)\nn\\
&&
+2\tilde{N} \tilde{D}_j \tilde{D}_k\tilde{N}+2\big(\mathcal{L}_{\overset{\rightharpoonup }{\tilde{N}}}\dot{\tilde{Q}}\big)_{jk}+\big(\mathcal{L}_{\dot{\overset{\rightharpoonup}{N}}}\tilde{Q} \big)_{jk}-\big(\mathcal{L}_{\overset{\rightharpoonup}{\tilde{N}}}(\mathcal{L}_{\overset{\rightharpoonup}{\tilde{N}}}\tilde{Q})\big)_{jk}\nn\\
&&
-\frac{\tilde{N}\tilde{H}(\sigma)}{\sqrt{\det \tilde{Q}}F(\Xi)}\tilde{G}_{jkmn}\tilde{N}^m\tilde{N}^n+\tilde{Q}_{j k}\Big[-\frac{\tilde{N}^2}{2\sqrt{\det \tilde{Q}}}\tilde{C}+\tilde{N}^2\frac{V(\Xi)}{(F(\Xi))^2}\Big].\label{ddottildeq}
\ea
 It is not difficult to check that Eqs. (\ref{ddotxi}) and (\ref{ddottildeq}) exactly reproduce Eqs. (\ref{phisecond}) and (\ref{qsecond}) after substituting $\tilde{Q}_{jk}\rightarrow F(\Phi)Q_{jk}$, $\frac{d\Xi(\Phi)}{d\Phi}\rightarrow\sqrt{3(\frac{F'(\Phi)}{F(\Phi)})^2+2\frac{K(\Phi)}{F(\Phi)}}$. Thus the evolution equations in Einstein frame can be related to those in Jordan frame through the conformal transformation and field redefinition. Since the equivalence holds for general variables, obviously it also holds for the linear perturbed variables under the assumption that the transformations between the two frames are non-singular everywhere. Thus we conclude that the Jordan and Einstein frames are still equivalent to each other in the gauge invariant formulation.
\subsection{Generalizing to Different Reference Systems}
 In above sections we chose the dust particles as the observers, whose equations of motion satisfy $g^{\mu\nu}U_{\mu}U_{\nu}=-1$. An interesting question is whether one can choose different observers, for example the ones satisfying $g^{\mu\nu}U_{\mu}U_{\nu}=-X(\phi)$ with $X(\phi)$ an arbitrary (positive) function. To answer this question, we first generalize the action (\ref{actionEinstein}) to
\ba
S\, _{(\xi ,\tilde{g},dust)}=\int d^4x\sqrt{|\det (\tilde{g})|}\left[\tilde{R}^{(4)}-\frac12\tilde{g}^{\mu\nu}(\tilde{\triangledown}_{\mu}\xi) \tilde{\triangledown}_{\nu}\xi -\frac{V(\xi)}{F^2(\xi)} )\right]-\frac{1}{2}\int d^4x\sqrt{|\det(\tilde{g})|}\frac{\rho}{J(\xi)}\left[\tilde{g}^{\mu \nu}U_{\mu }U_{\nu }+\frac{1}{L(\xi)}\right],\nn\\\label{modactEinstein}
\ea
 where $J(\xi)$ and $L(\xi)$ are arbitrary positive functions. Since it has been shown in the last subsection that the two frames are equivalent, we choose Einstein frame in which the equations look simpler. The physical Hamiltonian reads (see Appendix A)
\ba
\tilde{\textbf{H}}(\tau)=\int_{\mathcal{S}}d^3\sigma \tilde{H}(\tau,\sigma),\qquad \tilde{H}(\tau, \sigma)=\sqrt{L(\Xi)}\sqrt{\tilde{C}^2-\tilde{Q}^{ij}\tilde{C}_i\tilde{C}_j}(\tau, \sigma),
\ea
where expressions for $\tilde{C}_j$ and $\tilde{C}$ are the same as Eqs. (\ref{CjEinstein}) and (\ref{CEinstein}). The dynamical shift vector and lapse function are defined by
\ba
\tilde{N}_j\equiv-\frac{L\tilde{C}_j}{\tilde{H}}, \qquad \tilde{N}\equiv\frac{L\tilde{C}}{\tilde{H}}=\sqrt{L+\tilde{Q}^{ij}\tilde{N}_i\tilde{N}_j}.\label{eomofO}
\ea
In FRW background, we have $\tilde{\overline{N}}_j=0$ and $\tilde{\overline{N}}=\sqrt{L(\overline{\Xi})}$. The equations of motion of background variables read
\ba
\ddot{\overline{\Xi}}&=&\dot{(\ln\sqrt{L(\Xi)})}\dot{\overline{\Xi}}-3(\frac{\dot{\tilde{A}}}{\tilde{A}})\dot{\overline{\Xi}}
-L(\overline{\Xi})(\frac{V(\overline{\Xi})}{F^2(\overline{\Xi})})'_{\Xi}
-\frac{L(\overline{\Xi})'_{\Xi}}{2\sqrt{L(\overline{\Xi})}}\frac{\tilde{\bar{\epsilon}}}{\tilde{A}^3},\label{eomxil0}\\
3(\frac{\dot{\tilde{A}}}{\tilde{A}})^2&=&\frac{1}{4}\big(\dot{\overline{\Xi}}^2+\frac{L(\overline{\Xi})V(\overline{\Xi})}{F^2(\overline{\Xi})}\big)
-\frac{\tilde{\bar{\epsilon}}}{2\sqrt{L(\overline{\Xi})}\tilde{A}^3},\label{constrfrwl0}\\
3(\frac{\ddot{\tilde{A}}}{\tilde{A}})&=&-\frac{1}{4}\big[\frac{1}{2}(\dot{\overline{\Xi}}+\frac{L(\overline{\Xi})V(\overline{\Xi})}{F^2(\overline{\Xi})}
+\frac{3}{2}(\dot{\overline{\Xi}}^2-\frac{L(\overline{\Xi})V(\overline{\Xi})}{F^2(\overline{\Xi})}))\big]+\frac{\tilde{\bar{\epsilon}}}
{4\sqrt{L(\overline{\Xi})}\tilde{A}^3},
\label{friedmannl0}
\ea
where we denote $\tilde{Q}_{jk}\equiv\tilde{A}^2\delta_{jk}$ and $\tilde{\bar{\epsilon}}\equiv\tilde{H}$. The reader can easily check that the above equations reproduce the results in Eqs. (\ref{ddotphi}), (\ref{Friedmann}) and (\ref{Raychaudhuri}) by setting $L(\overline{\Xi})=F(\overline{\Xi})$ and replacing $\tilde{A}$ with $\sqrt{F(\overline{\Xi})}A$. For the linear perturbation equations, using the following decomposition
\ba
\delta \tilde{N}=\frac{L'(\Xi)}{2\sqrt{L(\overline{\Xi})}}\delta \Xi,\qquad
\delta \tilde{N}_j=\sqrt{L(\overline{\Xi})}\tilde{A}(S_i+B_{,i}), \qquad \delta \tilde{Q}_{jk}:=\tilde{A}^2[2\tilde{\psi}\delta_{jk}+2E_{,jk}+2F_{(j,k)}+h_{jk}],
\ea
the evolution equations of the tensor and vector modes in the conformal time frame $\frac{d}{d\eta}=\frac{\tilde{A}}{\tilde{\overline{N}}}\frac{d}{d\tau}$ are
\ba
2\mathcal{\tilde{H}}h^{'}_{jk}+h^{''}_{jk}-\Delta h_{jk}=0,\\
\Delta V_j=-\frac{\delta\tilde{\epsilon}^{\bot}_j}{\tilde{A}^2},\qquad 2\mathcal{\tilde{H}}V_j+V^{'}_{j,k}=0,
\ea
where $\mathcal{\tilde{H}}:=\frac{\tilde{A}^{'}}{\tilde{A}}$. The scalar mode contribution gives
\ba
&&\Big[\tilde{\Theta}^{'}_B+\mathcal{\tilde{H}}\tilde{\Theta}_A+\frac{1}{4}\overline{\Xi}'\delta Z\Big]_{,j}=-\frac{1}{\tilde{A}}\Big[\frac{\delta \tilde{\epsilon}^{\|}_j}{\tilde{A}}-\frac{\tilde{\overline{\epsilon}}}{\sqrt{L(\overline{\Xi})}}(B-E'_{,j})\Big],\\
&&\Delta \tilde{\Theta}_B-3\mathcal{\tilde{H}}\tilde{\Theta}^{'}_B
+\frac14\big[-\tilde{A}^2\frac{V(\overline{\Xi})}{F^2(\overline{\Xi})}\tilde{\Theta}_A+\overline{\Xi}^{'}\delta Z^{'}-\tilde{A}^2(\frac{V(\overline{\Xi})}{F^2(\overline{\Xi})})^{'}_{\Xi}\delta Z\big]\nn\\
&=&\frac{1}{4\sqrt{L(\overline{\Xi})}\tilde{A}}\bigg[\delta \tilde{\epsilon}-\tilde{\overline{\epsilon}}\big(\delta\ln\sqrt{F(\Phi)}+3\tilde{\psi}+\Delta E-2(B-E^{'})^{'}+\mathcal{\tilde{H}}(B-E^{'})\big)\bigg],\\
&&\tilde{\Theta}^{''}_B+\mathcal{\tilde{H}}(\tilde{\Theta}_A+2\tilde{\Theta}_B)+(2\mathcal{\tilde{H}}^{'}+\mathcal{\tilde{H}}^2)\tilde{\Theta}_A
=-\frac14\big[(\overline{\Xi}^{'})^2\tilde{\Theta}_A+\overline{\Xi}^{'}\delta Z^{'}-\tilde{A}^2(\frac{V(\overline{\Xi})}{F^2(\overline{\Xi})})'_{\Xi}\delta Z\big],
\ea
where $\tilde{\Theta}_A:=-\delta \ln\sqrt{ L(\Xi)}-\mathcal{\tilde{H}}(B-E^{'})-(B-E^{'})^{'}$, $\tilde{\Theta}_B:=\tilde{\psi}+\mathcal{\tilde{H}}(B-E^{'})$, $\delta Z:=\delta\Xi+\overline{\Xi}^{'}(B-E^{'})$. It is easy to see that the above definitions are the same as those in (\ref{newvariables}) when $L(\Xi)=F(\Phi(\Xi))$. The evolution of the perturbed gravitational scalar field reads
\ba
&&\delta Z^{''}+2\mathcal{\tilde{H}}\delta Z-\Delta \delta Z +(\frac{V(\overline{\Xi})}{F^2(\overline{\Xi})})'_{\Xi\Xi}\tilde{A}^2\delta Z+\overline{\Xi}^{'}\tilde{\Theta}^{'}_A-2(\frac{V(\overline{\Xi})}{F^2(\overline{\Xi})})'_{\Xi}\tilde{A}^2\tilde{\Theta}^{'}_A
+3\overline{\Xi}^{'}\tilde{\Theta}^{'}_B\nn\\
&=&\big(-\frac{L(\overline{\Xi})'_\Xi}{2L^{\frac{3}{2}}(\overline{\Xi})}\frac{\tilde{\epsilon}}{\tilde{A}}(B-E^{'})\big)^{'}
-\frac{L(\overline{\Xi})'_\Xi}{2L^{\frac{3}{2}}(\overline{\Xi})}\frac{\tilde{\epsilon}}{\tilde{A}}(B-E^{'})^{'}
-\frac{\tilde{A}^2}{L(\overline{\Xi})}\delta \big(\frac{L(\Xi)'_\Xi}{2\sqrt{L(\Xi)}}\frac{\tilde{\epsilon}}{\sqrt{\det \tilde{Q}}}\big).
\ea
Again we can easily match these equations with the ones in Jordan frame. We find that the different choice with $\tilde{g}^{\mu\nu}U_{\mu}U_{\nu}=-\frac{1}{L(\Xi)}$ leads to the change of dynamical lapse function $\tilde{N}$ and the correction terms.

Now we compare our  manifestly gauge invariant cosmological perturbation theory (MGICPT) with the standard cosmological perturbation theory (SCPT). For the convenience of readers, we list all the equations in table 1. We denote the non gauge invariant background variables in SCPT with lowercase letters and the gauge invariant linearly perturbed variables with a hat. Since the Jordan and Einstein frames are equivalent in our manifestly gauge invariant Hamiltonian formalism as well as in the standard formalism, we write all equations in Einstein frame where they take a concise form. In Einstein frame $\tilde{A}:=\sqrt{F(\overline{\Xi})}A$, $\mathcal{\tilde{H}}:=\frac{d\tilde{A}/d\eta}{\tilde{A}}=\frac{1}{\sqrt{F(\overline{\Xi})}}\frac{d\tilde{A}}{d\tau}$, $\tilde{a}:=\sqrt{F(\overline{\xi})}a$, $\tilde{h}:=\frac{d\tilde{a}/d\eta}{\tilde{a}}=\frac{1}{\sqrt{F(\overline{\xi})}}\frac{d\tilde{a}}{dt}$, while the definitions for $\Theta_A$, $\Theta_B$ and $\delta Z$ are given in Eq. (\ref{newvariables}). From the comparison we see that these equations match each other precisely provided that the Hamiltonian density $\overline{\epsilon}$ goes to zero.

\begin{table}
\caption{\large\textbf{Comparison between MGICPT and SCPT of STT in Einstein frame}}
\begin{tabular}{|c|l|l|l|}\hline
\backslashbox{\textbf{E.O.M.}}{}
  & Manifestly Gauge Invariant Formalism & Standard Formalism\\\hline
\multirow{2}*{} &
$\overline{\Xi}^{''}=-2\mathcal{\tilde{H}}\overline{\Xi}^{'}-\tilde{A}^2
\big(\frac{V(\overline{\Xi})}{F^2(\overline{\Xi})}\big)'_\Xi
-\frac{F(\overline{\Xi})'_\Xi}{2F^{\frac32}(\overline{\Xi})}\frac{\bar{\epsilon}}{\tilde{A}},$
&$\overline{\xi}^{''}=-2\tilde{h}\overline{\xi}^{'}-\tilde{a}^2
\big(\frac{V(\overline{\xi})}{F^2(\overline{\xi})}\big)'_\xi,
$\\
\textbf{Background}
 &$-6\mathcal{\tilde{H}}^2+\frac12(\overline{\Xi}^{'})^2+\tilde{A}^2\frac{V(\overline{\Xi})}{F^2(\overline{\Xi})}
=\frac{\bar{\epsilon}}{F^{\frac12}(\overline{\Xi})\tilde{A}},$
&
$-6\tilde{h}^2+\frac12(\overline{\xi}^{'})^2+\tilde{a}^2\frac{V(\overline{\xi})}{F^2(\overline{\xi})}
=0,$
\\
\textbf{Variables}
&
$2\mathcal{\tilde{H}}^{'}+4\mathcal{\tilde{H}}^2
-\tilde{A}^2\frac{V(\overline{\Xi})}{F^2(\overline{\Xi})}=\frac{-\bar{\epsilon}}{2F^{\frac12}(\overline{\Xi})\tilde{A}},$
&
$2\tilde{h}^{'}+4\tilde{h}^2-\tilde{a}^2\frac{V(\overline{\xi})}{F^2(\overline{\xi})}=0,$\\\hline
\textbf{Tensor-Mode}&$2\mathcal{\tilde{H}}h^{'}_{jk}+h^{''}_{jk}-\Delta h_{jk}=0,$
&$2\tilde{h}\hat{h}^{'}_{ab}+\hat{h}^{''}_{ab}-\Delta \hat{h}_{ab}=0,$\\\hline
\multirow{2}*{\textbf{Vector-Mode}}
&$\Delta V_j=-\frac{\delta\epsilon^{\bot}_j}{\tilde{A}^2},$
&$\Delta \hat{V}_a=0,$\\
&$2\mathcal{\tilde{H}}\hat{V}_j+\hat{V}^{'}_{j,k}=0,$
&$2\tilde{h}\hat{V}_a+\hat{V}^{'}_{a,b}=0,$\\\hline
\multirow{6}*{\textbf{Scalar-Mode}}
&\quad$\Big[\Theta^{'}_B+\mathcal{\tilde{H}}\Theta_A+\frac{1}{4}\overline{\Xi}^{'}\delta Z\Big]_{,j}$
&$\hat{\Theta}^{'}_B+\tilde{h}\hat{\Theta}_A+\frac{1}{4}\overline{\xi}'\delta \hat{Z}=0, $\\
&$=-\frac{1}{\tilde{A}}\Big[\frac{\delta \epsilon^{\|}_j}{\tilde{A}}-F^{-\frac{1}{2}}(\overline{\Xi})\overline{\epsilon}(B-E^{'}_{,j})\Big],$
&\\
&\quad$\Delta \Theta_B-3\mathcal{\tilde{H}}\Theta^{'}_B+\frac14\big[-\tilde{A}^2\frac{V(\overline{\Xi})}{F^2(\overline{\Xi})}\Theta_A$
&\quad$\Delta \hat{\Theta}_B-3\tilde{h}\hat{\Theta}^{'}_B+\frac14\big[-\tilde{a}^2\frac{V(\overline{\xi})}{F^2(\overline{\xi})}\hat{\Theta}_A$\\
&$ +\overline{\Xi}^{'}\delta Z^{'}-\tilde{A}^2(\frac{V(\overline{\Xi})}{F^2(\overline{\Xi})})'_{\Xi}\delta Z\big]$
&$ +\overline{\xi}^{'}\delta \hat{Z}^{'}-\tilde{a}^2(\frac{V(\overline{\xi})}{F^2(\overline{\xi})})'_{\xi}\delta \hat{Z}\big]$\\
&$=\frac{1}{4F^{\frac12}(\overline{\Xi})\tilde{A}}\bigg[\delta \epsilon-\overline{\epsilon}(-\delta\ln\sqrt{F(\Xi)}$
&$=0,$\\
&$+3\tilde{\psi}+\Delta E)+2\bar{\epsilon}(B-E^{'})^{'}-\mathcal{\tilde{H}}(B-E^{'})\bigg],$
&\\
&\quad$\Theta^{''}_B+\mathcal{\tilde{H}}(\Theta_A+2\Theta_B)^{'}+(2\mathcal{\tilde{H}}^{'}+\mathcal{\tilde{H}}^2)\Theta_A$
&\quad$\hat{\Theta}^{''}_B+\tilde{h}(\hat{\Theta}_A+2\hat{\Theta}_B)^{'}+(2\tilde{h}^{'}+\tilde{h}^2)\hat{\Theta}_A$
\\
&$=-\frac14\big[(\overline{\Xi}^{'})^2\Theta_A+\overline{\Xi}^{'}\delta Z^{'}
-\tilde{A}^2(\frac{V(\overline{\Xi})}{F^2(\overline{\Xi})})'_{\Xi}\delta Z\big],$
&$=-\frac14\big[(\overline{\xi}^{'})^2\hat{\Theta}_A+\overline{\xi}^{'}\delta \hat{Z}^{'}
-\tilde{a}^2(\frac{V(\overline{\xi})}{F^2(\overline{\xi})})'_{\xi}\delta \hat{Z}\big],$\\
&
&\\\hline
\multirow{2}*{\textbf{Perturbed}}
&$\quad\delta Z^{''}+2\mathcal{\tilde{H}}\delta Z-\Delta \delta Z +(\frac{V(\overline{\Xi})}{F^2(\overline{\Xi})})'_{\Xi\Xi}\tilde{A}^2\delta Z$
&$\quad\delta \hat{Z}^{''}+2\tilde{h}\delta \hat{Z}-\Delta \delta \hat{Z} +(\frac{V(\overline{\xi})}{F^2(\overline{\xi})})'_{\xi\xi}\tilde{a}^2\delta \hat{Z}$\\
\textbf{Gravitational}
&\quad$+\Xi^{'}\Theta^{'}_A-2(\frac{V(\overline{\Xi})}{F^2(\overline{\Xi})})'_{\Xi}\tilde{A}^2\Theta^{'}_A+3\overline{\Xi}^{'}\Theta^{'}_B$
&$\quad+\xi^{'}\hat{\Theta}^{'}_A-2(\frac{V(\overline{\xi})}{F^2(\overline{\xi})})'_{\xi}\tilde{a}^2\hat{\Theta}^{'}_A+3\xi^{'}\hat{\Theta}^{'}_B$\\
\textbf{Scalar Field}
&$=\big(\frac12(\frac{1}{F(\overline{\Xi})})'_{\Xi}\frac{\sqrt{F}\overline{\epsilon}}{\tilde{A}}(B-E^{'})\big)^{'}
+\frac12(\frac{1}{F(\overline{\Xi})})'_{\Xi}\frac{\sqrt{F}\overline{\epsilon}}{\tilde{A}}(B-E^{'})^{'}$
&$=0.$\\
&$-\frac{\tilde{A}^2}{F(\Xi)}\delta \big(\frac12(\frac{1}{F(\Xi)})'_{\Xi}\frac{\overline{\epsilon}}{\sqrt{\det Q}}\big).$
&\\
\hline
\end{tabular}
\end{table}
\newpage
\section{Concluding Remarks}

In this paper we extended the manifestly gauge invariant Hamiltonian formalism introduced in  \cite{Giesel10} to the case of STT. Our results are summarized as follows:
First, we derived the Hamiltonian equations of motion in the manifestly gauge invariant framework using the Brown-Kuch{\v r}-dust as reference fields in the Jordan frame. Afterwards we used these equations to derive the evolution equations for the linear perturbations of the 3-metric and the gravitational scalar field on a general relativistic spacetime background. These are the generalization of the results obtained in \cite{Giesel10} to the case of STT. Then we applied our general result to the special case of a flat FRW background. This allowed to compare the results obtained in our formalism to the results from standard cosmological perturbation theory (SCPT). Likewise to the analysis in \cite{Giesel10} where gravity and a minimally coupled scalar field were considered, also in the case of the STT we have shown that our results and the one from SCPT exactly agree in the limit when the energy momentum and its perturbation of the dust vanishes. However, taking the idea of the manifestly gauge invariant framework seriously this limit is rather artificial since the dynamically coupled observer will always have an imprint on the system and here these are exactly the computed corrections compared to the results of SCPT. 
Second, we analyzed the question whether the Jordan frame and the Einstein frame are still equivalent to each other in the manifestly gauge invariant Hamiltonian formalism by showing their equations of motions can be related to each other through conformal transformations. Third, we generalized our results from the original reference system to a slightly different reference system in the Einstein frame and we showed that these equations are consistent with the ones in the Jordan frame.
\\
\\
There are several possible directions to extend our results. First, since we have derived the linear perturbation equations upon a general background, the direct extension is to choose a background different from the flat FRW case and study the cosmological perturbations in STT. For instance, the homogeneous but anisotropic Bianchi models are interesting candidates. Second, since STT can include a large variety of modified gravity models, we can choose some specific functions of $F(\Phi)$, $K(\Phi)$ and $V(\Phi)$ to study the detailed physical predictions in the inflation period and compare them with the predictions of SCPT. Third, our formalism has the advantage over the procedure used in the context of standard cosmological perturbations theory that we can directly get the gauge invariant perturbed variables from the full metric to \emph{any} order without worrying about the difficulty in constructing the higher order gauge invariant variables as in the standard way, thus it will be very convenient to use our formalism to study the non-linear effects such as the non-Gaussianity in the primordial perturbations.

\section*{ACKNOWLEDGMENTS}

Y.H. thanks the Chinese Scholarship Council (CSC) for financial support.  K.G. thanks the Emerging Field Project ``Quantum Geometry" of the FAU Erlangen-N\"urnberg for financial support. This work is also supported in part by the NSFC (Grant Nos. 11235003 and 11475023) and the Research Fund for the Doctoral Program of Higher Education of China.

 \appendix
\section{ Lagrangian and Hamiltonian Analysis of the Action in Einstein Frame}
In this appendix, we will perform the Lagrangian and Hamiltonian analysis of STT in Einstein frame and derive its physical Hamiltonian. The action which we consider is the following,
\ba
S\, _{(\xi ,\tilde{g},dust)}=\int d^4x\sqrt{|\det (\tilde{g})|}\left[\tilde{R}^{(4)}-\frac12\tilde{g}^{\mu\nu}(\tilde{\triangledown} _{\mu}\xi)\tilde{\triangledown} _{\nu}\xi -\frac{V(\xi)}{F^2(\xi)})\right]-\frac{1}{2}\int d^4x\sqrt{|\det (\tilde{g})|}\frac{\rho}{J(\xi)} \left[\tilde{g}^{\mu \nu}U_{\mu }U_{\nu }+\frac{1}{L(\xi)}\right].\nn\\\label{actionE}
\ea
\subsection*{1. Lagrangian Analysis of the Dust Action}
Recall that the dust velocity field is defined by $U_{\mu}=-T_{,\mu}+W_jS^j_{,\mu}$, and the energy-momentum tensor of the dust reads
\ba
T^{dust}_{\mu\nu}=-\frac{2}{\sqrt{|\det (\tilde{g})|}}\frac{\delta S_{dust}}{\delta \tilde{g}^{\mu\nu}}=\frac{\rho}{J(\xi)}U_{\mu }U_{\nu}-\frac{\rho}{2J(\xi)}\tilde{g}_{\mu\nu}[\tilde{g}^{\alpha\beta}U_{\alpha}U_{\beta}+\frac{1}{L(\xi)}].
\ea
The variation of action (\ref{actionE}) with respect to $\rho$ gives
\ba
\tilde{g}^{\alpha\beta}U_{\alpha}U_{\beta}+\frac{1}{L(\xi)}=0,
\ea
thus the energy-momentum tensor reduces to
\ba
T_{\mu\nu}=\frac{\rho}{J(\xi)} U_{\mu }U_{\nu},
\ea
 which coincides with that of a perfect fluid with energy density $\frac{\rho}{J(\xi)}$ and zero pressure. Similarly, the variations of action (\ref{actionE}) with respect to $W_j$, $T$ and $S^j$ yields the following equations,
\ba
\mathcal{L}_{U^{\mu}}T=\frac{1}{L(\xi)},\qquad
\tilde{\nabla}_{\mu}[\frac{\rho}{J(\xi)}U^{\mu}]=0,\qquad
\tilde{\nabla}_{\mu}[\frac{\rho}{J(\xi)}U^{\mu}W_j]=0.
\ea
With these three equations, the geodesic equation in Einstein frame is modified to
\ba
U^{\mu}\tilde{\nabla}_{\mu}U_{\nu}=-\frac{1}{2}\frac{\tilde{\nabla}_{\nu}L(\xi)}{L^2(\xi)},
\ea
thus unlike in Jordan frame, in Einstein frame $T$ no longer defines the proper time along the dust flow lines and there is a correction term proportional to the gradient of $\xi$ in the geodesic equation. The conservation of the energy momentum tensor transforms as
\ba
\tilde{\nabla}^{\mu}T_{\mu\nu}=-\frac{1}{2}\frac{\rho}{J(\xi)}\tilde{\nabla}_{\nu}\big(\frac{1}{L^2(\xi)}\big).\label{dustconserv}
\ea
The total energy momentum tensor of the scalar field and dust reads
 \ba
 T^{tot}_{\mu\nu}:=-\frac{2}{\sqrt{|\det (\tilde{g})|}}\frac{\delta S_{\xi+dust}}{\delta \tilde{g}^{\mu\nu}}=(\tilde{\nabla}_\mu\xi)\tilde{\nabla}_\nu\xi+\tilde{g}_{\mu\nu}
 \big(-\frac12\tilde{g}^{\alpha\beta}(\tilde{\nabla}_\alpha\xi)\tilde{\nabla}_\beta\xi-\frac{V(\xi)}{F^2(\xi)}\big)
 +\frac{\rho}{J(\xi)} U_{\mu }U_{\nu}.
 \ea
 The variations of action (\ref{actionE}) with respect to the scalar $\xi$ gives
 \ba
 \tilde{\nabla}^{\mu}\tilde{\nabla}_\mu\xi-\big(\frac{V(\xi)}{F^2(\xi)}\big)_{,\xi}
 -\frac{1}{2}\frac{\rho}{J(\xi)}(\frac{1}{L^2(\xi)})_{,\xi}=0.\label{conserveq}
 \ea
 Substituting Eq. (\ref{dustconserv}) into
 \ba
 \tilde{\nabla}^{\mu}T^{tot}_{\mu\nu}=(\tilde{\nabla}_\nu\xi)\tilde{\nabla}^{\mu}\tilde{\nabla}_\mu\xi
 -(\frac{V(\xi)}{F^2(\xi)})_{,\xi}\tilde{\nabla}_{\nu}\xi+\tilde{\nabla}^{\mu}(\frac{\rho}{J(\xi)} U_{\mu }U_{\nu}),
 \ea
we get
 \ba
 \tilde{\nabla}^{\mu}T^{tot}_{\mu\nu}=\big[\tilde{\nabla}^{\mu}\tilde{\nabla}_\mu\xi-\big(\frac{V(\xi)}{F^2(\xi)}\big)_{,\xi}
 -\frac{1}{2}\frac{\rho}{J(\xi)}(\frac{1}{L^2(\xi)})_{,\xi}\big]\tilde{\nabla}_\nu\xi=0.
 \ea
 Hence the total energy momentum tensor is conserved in Einstein frame.
\subsection*{2. Hamiltonian Analysis}
 Although the Hamiltonian analysis for the action (\ref{actionE}) with  $J(\xi)=L(\xi)=1$ was already done in Ref. \cite{Giesel10}, we draw a word sketch of the Hamiltonian analysis for the generalized case when $J(\xi)$ and $L(\xi)$ are arbitrary functions. All the notations remain the same as those in \cite{Giesel10}.

By 3+1 decomposition, the dust action can be written as
\ba
S_{dust}=-\frac12\int dt\int dx^3\sqrt{det(q)}n\frac{\rho}{J(\xi)}\big(-U^2_{n}+q^{ab}U_aU_b+\frac{1}{L(\xi)}\big).
\ea
The conjugate momentum of $T$ and $S^j$ read
\ba
P:&=&\frac{\delta S_{dust}}{\delta T_{,t}}=-\sqrt{det(q)}\frac{\rho}{J(\xi)}U_n,\label{momemtump}\\
P_{j}:&=&\frac{\delta S_{dust}}{\delta S^{j}_{,t}}=\sqrt{det(q)}\frac{\rho}{J(\xi)}U_nW_j.\label{momentumpj}
\ea
equations (\ref{momemtump}) and (\ref{momentumpj}) impose the primary constraint
\ba
Z_j:=P_j+W_jP=0.
\ea
There are also four other primary constraints,
\ba
I:=Z:=\frac{\delta S_{dust}}{\delta \rho_{,t}}=0,\quad I^j:=Z^j:=\frac{\delta S_{dust}}{\delta W_{j,t}}=0,\quad p:=z:=\frac{\delta S}{\delta n_{t}}=0,\quad p_a:=z_a:=\frac{\delta S}{\delta n^a_{,t}}=0.
\ea
The resulted Hamiltonian constraint of the coupled system of action (\ref{actionE}) is given by
\ba
\tilde{c}^{tot}=\tilde{c}^{geo}+\tilde{c}^{scalar}+\tilde{c}^{dust},
\ea
where
\ba
\tilde{c}^{geo}&\equiv&\frac{1}{\sqrt{\det(\tilde{q})}}\big[\tilde{q}_{ac}\tilde{q}_{bd}-\frac12\tilde{q}_{ab}\tilde{q}_{cd}\big]\tilde{p}^{ab}\tilde{p}^{cd}
-\sqrt{\det(\tilde{q})}\tilde{R}^{(3)},\\
\tilde{c}^{scalar}&\equiv&\frac12\Big[\frac{\tilde{\pi}^2}{\sqrt{\det(\tilde{q})}}+\sqrt{\det(\tilde{q})}\big(\tilde{q}^{ab}\xi_a\xi_b+\frac{V(\xi)}{F^2(\xi)}\big)\Big],\\
\tilde{c}^{dust}&\equiv&\frac12\Big[\frac{P^2J(\xi)}{\rho \sqrt{\det(\tilde{q})}}+\sqrt{\det(\tilde{q})}\frac{\rho}{J(\xi)}(\tilde{q}^{ab}U_aU_b+\frac{1}{L(\xi)})\Big]
\ea
with $U_a=-T_{,a}+W_jS^j_{,a}$. Note that $\{\tilde{p}^{cd},\tilde{q}_{ab},\}$, $\{\tilde{\pi},\xi\}$ and $\{P,T\}$ are three pairs of conjugate variables. The diffeomorphism constraint is given by
\ba
\tilde{c}^{tot}_a=\tilde{c}^{geo}_a+\tilde{c}^{scalar}_a+\tilde{c}^{dust}_a,
\ea
where
\ba
\tilde{c}^{geo}_a\equiv-2\tilde{q}_{ac}\tilde{D}_b\tilde{p}^{bc},\quad \tilde{c}^{scalar}_a\equiv\tilde{\pi}\xi_{,a},\quad \tilde{c}^{dust}_a\equiv-PU_a.
\ea
Now the total Hamiltonian is defined as
\ba
H=\int dx^3\big(\mu^jZ_j+\mu Z+\mu_jZ^j+\nu z +\mu^az_a+n\tilde{c}^{tot}+n^a\tilde{c}^{tot}_a\big),
\ea
where $\mu^j,\mu,\mu_j,\nu,\mu^a$ are Lagrangian multipliers. Now we consider the condition that the
primary constraints should be maintained during evolution. Explicitly we have
\ba
z_{,t}&=&\{H, p\}=-\tilde{c}^{tot},\label{zt}\\
z_{a,t}&=&\{H,p_a\}=-\tilde{c}^{tot}_a,\label{zat}\\
Z_{,t}&=&\{H, I\}=\frac{n}{2}\Big[-\frac{P^2J(\xi)}{\rho^2\sqrt{\det(\tilde{q})}}
+\frac{\sqrt{\det(\tilde{q})}}{J(\xi)}(\tilde{q}^{ab}U_aU_b+\frac{1}{L(\xi)})\Big]=:\tilde{c},\label{zt}\\
Z^j_{,t}&=&\{H, I^j\}=-\mu^jP+n^aPS^j_{,a}-n\frac{\rho}{J(\xi)}\sqrt{\det(\tilde{q})}q^{ab}U_aS^j_{,b},\label{Zjt}\\
Z_{j,t}&=&\{H, Z_j\}=\mu_j P-n^aPW_{j,a}+n\frac{\rho}{J(\xi)}\tilde{q}^{ab}U_bW_{j,a}.\label{ZJt}
\ea
We can solve the constraints (\ref{Zjt}) and (\ref{ZJt}) by fixing the Lagrangian multipliers $\mu^j$ and $\mu_j$. The new constraint (\ref{zt}) is of second-class, by asking the conservation of this constraint, we can fix the Lagrangian multiplier $\mu$. It is also easy to check that both $\tilde{c}^{tot}$ and $\tilde{c}^{tot}_a$ are conserved during the evolution. After solving all the second-class constraints, we get
\ba
W_j=-\frac{P_j}{P},\quad I^j=0,\quad I=0,\quad \rho=\frac{PJ(\xi)}{\sqrt{\det(\tilde{q})}}\Big(\sqrt{q^{ab}U_aU_b+\frac{1}{L(\xi)}}\Big)^{-1}.
\ea
Finally, in the reduced phase space with $z=z_a=0$, we are left with only two constraints
\ba
\tilde{c}^{tot}=\tilde{c}^{geo}+\tilde{c}^{scalar}+\tilde{c}^{dust},\qquad \tilde{c}^{tot}_a=\tilde{c}^{geo}_a+\tilde{c}^{scalar}_a+\tilde{c}^{dust}_a,
\ea
where
\ba
\tilde{c}^{dust}&=&P\sqrt{\tilde{q}^{ab}U_aU_b+\frac{1}{L(\xi)}},\label{cdust}\\
\qquad \tilde{c}^{dust}_a&=&PT_{,a}+P_jS^j_{,a}=-PU_a.\label{cdusta}
\ea
Substituting Eq. (\ref{cdusta}) into (\ref{cdust}) and using $\tilde{c}^{dust}_a=-(\tilde{c}^{geo}+\tilde{c}^{scalar})$, we can rewrite the constraints as
\ba
\tilde{c}^{tot}&=&P+h, \qquad h=\sqrt{L(\xi)}\sqrt{\tilde{c}^2-\tilde{q}^{ab}\tilde{c}_a\tilde{c}_b},\\
\tilde{c}^{tot}_j&=&P_j+h_j,\qquad h_j=-hS^{a}_jT_{,a}+S^a_j\tilde{c}_a,
\ea
where $\tilde{c}:=\tilde{c}^{geo}+\tilde{c}^{scalar}, \tilde{c}_a:=\tilde{c}^{geo}_a+\tilde{c}^{scalar}_a$ and $S^a_jS^j_{,b}=\delta^a_b$.

\end{document}